\documentclass[aps,prd,twocolumn,amsmath,amssymb]{revtex4-1}
\usepackage{mathtools}
\usepackage{graphicx}
\usepackage{subfigure}
\usepackage{epstopdf}
\usepackage{color}
\usepackage{multirow}
\usepackage{setspace}
\usepackage{overpic}
\usepackage[bookmarksnumbered, pdfstartview=FitH,colorlinks,urlcolor=blue, citecolor=blue,linkcolor=blue] {hyperref}
\usepackage{lineno}
\usepackage{bm}
\usepackage{rotating}
\usepackage[utf8]{inputenc}
\usepackage{morefloats}
\usepackage{hyperref}
\usepackage{braket}

\let\oldequation\equation
\let\oldendequation\endequation

\renewenvironment{equation}
  {\linenomathNonumbers\oldequation}
  {\oldendequation\endlinenomath}

\newcommand{\klnu}{D \to \bar K\ell^+\nu_{\ell}}
\newcommand{\kenu}{D^0\to K^-e^+\nu_e}
\newcommand{\kmunu}{D^0\to K^-\mu^+\nu_\mu}
\newcommand{\ksenu}{D^+\to \bar K^0e^+\nu_e}
\newcommand{\ksmunu}{D^+\to \bar K^0\mu^+\nu_\mu}
\newcommand{\koenu}{D^+\to \bar K^0e^+\nu_e}
\newcommand{\komunu}{D^+\to \bar K^0\mu^+\nu_\mu}
\newcommand{\ffK}{f_+(0)}

\newcommand{\BESIIIorcid}[1]{\href{https://orcid.org/#1}{\hspace*{0.1em}\raisebox{-0.45ex}{\includegraphics[width=1em]{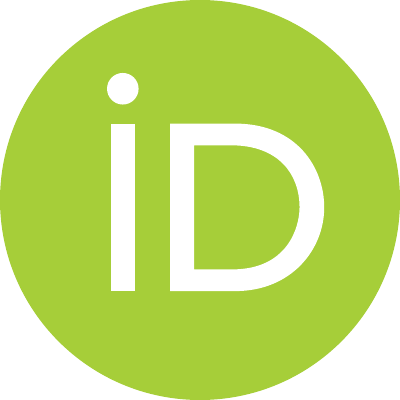}}}}

\begin{document}

%\linenumbers

\title{\boldmath Precise measurements of $D^0 \to K^-\ell^+\nu_\ell$ and $D^+ \to \bar K^0\ell^+\nu_\ell$ decays}
%% Saved at => 2025-10-23
\author{
M.~Ablikim$^{1}$\BESIIIorcid{0000-0002-3935-619X},
M.~N.~Achasov$^{4,c}$\BESIIIorcid{0000-0002-9400-8622},
P.~Adlarson$^{81}$\BESIIIorcid{0000-0001-6280-3851},
X.~C.~Ai$^{87}$\BESIIIorcid{0000-0003-3856-2415},
C.~S.~Akondi$^{31A,31B}$\BESIIIorcid{0000-0001-6303-5217},
R.~Aliberti$^{39}$\BESIIIorcid{0000-0003-3500-4012},
A.~Amoroso$^{80A,80C}$\BESIIIorcid{0000-0002-3095-8610},
Q.~An$^{77,64,\dagger}$,
Y.~H.~An$^{87}$\BESIIIorcid{0009-0008-3419-0849},
Y.~Bai$^{62}$\BESIIIorcid{0000-0001-6593-5665}
% ...
\\
\vspace{0.2cm}
(BESIII Collaboration)\\
\vspace{0.2cm} {\it
$^{1}$ Institute of High Energy Physics, Beijing 100049, People's Republic of China\\
% ...
$^{4}$ Budker Institute of Nuclear Physics SB RAS (BINP), Novosibirsk 630090, Russia\\
% ...
$^{31}$ INFN Sezione di Ferrara, (A)INFN Sezione di Ferrara, I-44122, Ferrara, Italy; (B)University of Ferrara, I-44122, Ferrara, Italy\\
% ...
$^{39}$ Johannes Gutenberg University of Mainz, Johann-Joachim-Becher-Weg 45, D-55099 Mainz, Germany\\
% ...
$^{62}$ Southeast University, Nanjing 211100, People's Republic of China\\
% ...
$^{64}$ State Key Laboratory of Particle Detection and Electronics, Beijing 100049, Hefei 230026, People's Republic of China\\
% ...
$^{77}$ University of Science and Technology of China, Hefei 230026, People's Republic of China\\
% ...
$^{80}$ University of Turin and INFN, (A)University of Turin, I-10125, Turin, Italy; (B)University of Eastern Piedmont, I-15121, Alessandria, Italy; (C)INFN, I-10125, Turin, Italy\\
$^{81}$ Uppsala University, Box 516, SE-75120 Uppsala, Sweden\\
% ...
$^{87}$ Zhengzhou University, Zhengzhou 450001, People's Republic of China\\
% ...
$^{\dagger}$ Deceased\\
% ...
$^{c}$ Also at the Novosibirsk State University, Novosibirsk, 630090, Russia\\
% ...
}
}
%% ends here %%

\begin{abstract} 

Using $e^+e^-$ collision data corresponding to an integrated
luminosity of 20.3 fb$^{-1}$, collected at the center-of-mass energy
of 3.773 GeV with the BESIII detector, we present precise measurements
of $D^0 \to K^-\ell^+\nu_\ell$ and $D^+ \to \bar K^0\ell^+\nu_\ell$
($\ell=e,\mu$) decays. The branching fractions of $\kenu$, $\kmunu$,
$\koenu$, and $\komunu$ are measured to be $(3.548\pm0.006_{\rm
stat}\pm0.017_{\rm syst}) \%$, $(3.445\pm0.007_{\rm
stat}\pm0.017_{\rm syst}) \%$, $(8.928\pm0.025_{\rm
stat}\pm0.050_{\rm syst}) \%$, and $(8.770\pm0.029_{\rm
stat}\pm0.053_{\rm syst}) \%$, respectively. The partial decay rates
of these four decays are measured with improved precision, and their
forward-backward asymmetries are determined for the first time. By
performing a simultaneous fit to the measured partial decay rates and
the forward-backward asymmetries of $\klnu$, a search for a possible
scalar current contribution in the $c\to s\ell^+\nu_{\ell}$ transition
is performed.  The results are ${\rm
Re}(c_S^{\mu})=0.017\pm0.008_{\rm stat}\pm0.006_{\rm syst}$ and
${\rm Im}(c_S^{\mu})=\pm(0.077\pm0.011_{\rm stat}\pm0.009_{\rm
syst})$, corresponding to a difference from the SM with a significance
of $2.3\sigma$. The product of the form factor $\ffK$ and the modulus
of the $c\to s$ Cabibbo-Kobayashi-Maskawa matrix element $|V_{cs}|$ is
determined to be $\ffK|V_{cs}|=0.7183\pm0.0007_{\rm
stat}\pm0.0014_{\rm syst}$. With the inputs
$|V_{cs}|=0.97349\pm0.00016$ from the Standard Model global fit or
$\ffK=0.7452\pm0.0031$ from the lattice quantum chromodynamics
calculation, we derive $\ffK=0.7383\pm0.0007_{\rm stat}\pm0.0014_{\rm
syst}$ and $|V_{cs}|=0.9639\pm0.0009_{\rm stat}\pm0.0018_{\rm
syst}\pm0.0040_{\rm LQCD}$.  Combining with the results of the
semi-muonic and semi-electronic decays, lepton flavor universality is
tested via the ratios of the decay rates between semi-muonic and
semi-electronic decays in the full and different momentum transfer ranges.
\end{abstract}

\maketitle

\section{Introduction}

Precise measurements of semileptonic decays of charmed mesons are
important for our understanding of the weak and strong interactions in
the charm sector. Their decay dynamics are sensitive to the magnitude
of the Cabibbo-Kobayashi-Maskawa (CKM) matrix element $|V_{cs(d)}|$ as
well as to the corresponding hadronic transition form factors.  As the
dominant semileptonic $D^{0(+)}$ decay modes, the $\klnu$ channels are
of special interest to investigate both theoretically and
experimentally.  Calculations on the hadronic form
factor at zero-momentum transfer
$\ffK$~\cite{Lubicz:2017syv,Chakraborty:2021qav,Parrott:2022rgu,FermilabLattice:2022gku,Wu:2006rd,Verma:2011yw,Ivanov:2019nqd,Faustov:2019mqr,Ke:2023qzc}
have been performed in various theoretical approaches, including
lattice quantum
chromodynamics~(LQCD)~\cite{Lubicz:2017syv,Chakraborty:2021qav,Parrott:2022rgu,FermilabLattice:2022gku},
QCD light-cone sum rules (LCSR)~\cite{Wu:2006rd}, covariant
light-front quark model (LFQM)~\cite{Verma:2011yw}, the covariant
confined quark model (CCQM)~\cite{Ivanov:2019nqd}, and the
relativistic quark model (RQM)~\cite{Faustov:2019mqr}. A precisely
measured $\ffK$ enables a more stringent test of the different
theoretical predictions, and the accurately determined $|V_{cs}|$ is
important for the test of the unitarity of the CKM matrix.

Lepton flavor universality~(LFU), as predicted by the Standard
Model~(SM), requires identical couplings between the gauge bosons and
the three lepton families. Any violation of LFU could reveal potential
new physics (NP) beyond the SM. In recent years, several differences
from LFU have been observed in the $b\to
c\tau^+\nu_{\tau}$~\cite{BaBar:2012obs,LHCb:2023zxo,LHCb:2023uiv,LHCb:2024jll,Belle:2015qfa} and
$b\to s
\ell^+\ell^-$~\cite{LHCb:2013ghj,Belle:2016fev,CMS:2017rzx,ATLAS:2018gqc,LHCb:2024onj}
transitions. While some of these violations disappear after further update measurements, some anomalies remain. Various NP scenarios, including two-Higgs-doublet
models~\cite{Iguro:2022uzz,Blanke:2022pjy} and leptoquark
models~\cite{Sakaki:2013bfa,Becirevic:2018afm}, as well as model
independent investigation~\cite{Iguro:2024hyk}, suggest that these
differences may originate from scalar currents in the weak
interaction. Therefore, it is important to study the $c\to
s\ell^+\nu_\ell$ transition in the (semi-)leptonic decays of charmed
mesons, as highlighted in
Refs.~\cite{Barranco:2013tba,Fajfer:2015ixa,Barranco:2016njc,Zhang:2018jtm}. With
large branching fractions~(BFs) and low backgrounds, the semileptonic
decays $\klnu$ offer a unique platform to access potential scalar
current contributions in the $c\to s\ell^+\nu_\ell$ transition. The
corresponding effective Lagrangian is written as~\cite{Fajfer:2015ixa}
\begin{equation} \label{eq:eft_lagr} \mathcal{L}_{\rm eff} = -\frac{4
G_{F}}{\sqrt{2}} V^*_{cs} \sum_{\ell=e,\mu,\tau}\sum_{i} c^{\ell}_{i}
\mathcal{O}_{i}^\ell +{\rm H.C.}, \end{equation} where the only
allowed operator in the SM is $\mathcal{O}^\ell_{\rm
SM}=(\bar{s}\gamma_{\mu}P_{L}c)(\bar{\nu}_\ell \gamma^\mu P_L\ell)$
with coefficient $c^\ell_{\rm SM}=1$. A potential right(left)-handed
scalar current is described by the NP operator
$\mathcal{O}^\ell_{R(L)}=(\bar{s}P_{R(L)}c)(\bar{\nu}_\ell P_R\ell)$
with complex Wilson coefficient $c^\ell_{R(L)}$. Here, $\gamma_{\mu}$
indicates the Dirac matrix, and $P_{L,R}$ are the chirality projection
operators. Non-zero values of $c^\ell_{R(L)}$ will cause differences
from the SM in several observables, such as full or partial decay
rates, and forward-backward asymmetries. Therefore, the precise
measurements of these variables are important for searching for the
scalar current in the $c\to s\ell^+\nu_\ell$ transition.

Experimentally, the branching fractions and dynamics of the $\klnu$
decays have been extensively studied by
BES~\cite{BES:2004rav,BES:2004obp,BES:2006kzp},
BaBar~\cite{BaBar:2007zgf}, Belle~\cite{Belle:2006idb},
CLEO~\cite{CLEO:2005rxg,CLEO:2005cuk,CLEO:2007ntr,CLEO:2009svp}, and
BESIII~\cite{BESIII:2021mfl,BESIII:2015tql,BESIII:2018ccy,BESIII:2017ylw,BESIII:2016hko,BESIII:2015jmz,BESIII:2016gbw,BESIII:2024slx}.
However, experimental studies of the scalar current and the
forward-backward asymmetry in the $\klnu$ transition have not been
reported to date. In this paper, we provide the complete experimental details and results for the precise measurements of the partial decay rates and forward-backward asymmetries in the $\klnu$ transition, from which the first constraint on the scalar current~\cite{PRL_draft} is derived. These analyses are based on $e^+e^-$ collision data, which corresponds to a total integrated luminosity of 20.3 fb$^{-1}$ collected with the BESIII detector at the center-of-mass energy
$\sqrt{s}=3.773$ GeV~\cite{BESIII:2024lbn} during 2010-2011 and
2022-2024. In addition, we present updated measurements of the
branching fractions of $D\to K\ell^+\nu_\ell$, the form factor
$f_+(0)$, and the modulus of the CKM matrix element $|V_{cs}|$, as
well as a test of LFU using the $\klnu$ decays. Compared to our
previous measurements~\cite{BESIII:2024slx}, which were based on a 7.9
fb$^{-1}$ data sample taken during 2010-2011 and 2022, all results
reported here are obtained with improved precision and thereby
supersede those in Ref.~\cite{BESIII:2024slx}. Throughout this paper,
charge conjugated modes are always included.

\section{BESIII detector and Monte Carlo simulations}

The BESIII detector~\cite{BESIII:2009fln} records symmetric $e^+e^-$
collisions provided by the BEPCII storage ring~\cite{Yu:2016cof}, and
has collected large data samples in the center-of-mass energy range from
1.84 to 4.95~GeV~\cite{BESIII:2020nme,Li:2021iwf}, with a peak
luminosity of $1.1 \times 10^{33}\;\text{cm}^{-2}\text{s}^{-1}$
achieved at $\sqrt{s} = 3.773~\text{GeV}$.  The cylindrical core of
the BESIII detector covers 93\% of the full solid angle and consists
of a helium-based
 multilayer drift chamber~(MDC), a time-of-flight system~(TOF), and a
 CsI(Tl) electromagnetic calorimeter~(EMC), which are all enclosed in
 a superconducting solenoidal magnet providing a 1.0~T magnetic
 field. The solenoid is supported by an octagonal flux-return yoke
 with resistive-plate muon counters~(MUC) interleaved
 with steel. %The main function of the MUC is to separate muons from charged pions, other hadrons and backgrounds based on their hit patterns in the instrumented flux-return yoke.
 The charged-particle momentum resolution at $1~{\rm
 GeV}/c$ is $0.5\%$, and the ${\rm d}E/{\rm d}x$ resolution is $6\%$
 for electrons from Bhabha scattering. The EMC measures photon
 energies with a resolution of $2.5\%$ ($5\%$) at $1$~GeV in the
 barrel (end-cap) region. The time resolution in the plastic
 scintillator TOF barrel region is 68~ps, while that in the end-cap
 region was 110~ps. The end-cap TOF system was upgraded in 2015 using
 multi-gap resistive plate chamber technology, providing a time
 resolution of 60~ps, which benefits 85\% of the data used in this analysis~\cite{Li:2017jpg,Guo:2017sjt,Cao:2020ibk}.

Simulated samples produced with {\sc geant4}-based~\cite{GEANT4:2002zbu} Monte
Carlo (MC) software, which includes the geometric
description~\cite{Huang:2022wuo} of the BESIII detector and the
detector response, are used to determine detection efficiencies and to
estimate backgrounds. The simulation models the beam energy spread and
initial state radiation (ISR) in the $e^+e^-$ annihilations with the
generator {\sc kkmc}~\cite{Jadach:1999vf}.  Signal MC samples of the decays $\klnu$ are simulated with a specific two-parameter series
expansion model~\cite{Becher:2005bg}, which is described in Sec.~\ref{th_formula}.  The background is studied using
an inclusive MC sample that includes $D\bar D$ pairs originating from
the decay of the $\psi(3770)$ with quantum coherence for the neutral $D$ channels considered, the non-$D\bar D$ decays of the $\psi(3770)$,
the ISR production of the charmonium states, and the continuum
processes. These processes are also generated with {\sc kkmc}.  The
known decay modes are modeled by {\sc evtgen}~\cite{Lange:2001uf,Ping:2008zz} with
branching fractions taken from the Particle Data
Group~(PDG)~\cite{ParticleDataGroup:2024cfk}, while the remaining unknown charmonium decays are modeled with {\sc lundcharm}~\cite{Chen:2000tv,Yang:2014vra}. Final state radiation from charged final-state particles is incorporated using {\sc photos} version 2.15~\cite{Golonka:2005pn}. %The signal MC samples of $\klnu$ are generated based on the series expansion model described in Sec.~\ref{th_formula}. 

\section{Double Tag Method}

At $\sqrt s=3.773$ GeV, the $D$ and $\bar D$ mesons are produced in
pairs via the $e^+e^-\to \psi(3770)\to D\bar D$ process, where $D$
stands for $D^0$ or $D^+$. This property allows for the absolute
branching fraction measurement with the well established double-tag
(DT) method~\cite{MARK-III:1985hbd}.  In this method, the single-tag
(ST) candidate events are selected by reconstructing a $\bar D^0$ in
six hadronic decay modes $\bar D^0 \to K^+\pi^-$, $K^+\pi^-\pi^0$,
$K^+\pi^-\pi^-\pi^+$, $K^0_S\pi^+\pi^-$, $K^+\pi^-\pi^0\pi^0$, and
$K^+\pi^-\pi^-\pi^+\pi^0$, or a $D^-$ in six hadronic decay modes
$D^-\to K^{+}\pi^{-}\pi^{-}$, $K^0_{S}\pi^{-}$,
$K^{+}\pi^{-}\pi^{-}\pi^{0}$, $K^0_{S}\pi^{-}\pi^{0}$,
$K^0_{S}\pi^{+}\pi^{-}\pi^{-}$, and $K^{+}K^{-}\pi^{-}$.  These
inclusively selected candidates are referred to as ST $\bar D$ mesons.
In events containing an ST $\bar{D}$ meson, candidates for the signal
decays $\klnu$ are selected to form DT events, where $\bar{K}^0$ is
reconstructed via $K^0_S\to\pi^+\pi^-$ and the lepton $\ell$ is either a
positron or a muon. The branching fraction of the signal decay is
determined as \begin{equation} \label{eq:bf} {\mathcal B}_{\rm
sig}=N_{\mathrm{DT}}/(N_{\mathrm{ST}}^{\rm tot}\cdot
\bar\varepsilon_{\rm sig}\cdot\mathcal{B}_{\rm sub}), \end{equation}
where $N^{\rm tot}_{\rm ST}=\sum_{i}{N^i_{\rm ST}}$ and $N_{\rm DT}$
are the total ST and DT yields summed over all tag modes. The weighted
signal efficiency is defined as \begin{equation}
\bar{\varepsilon}_{\rm sig}=\sum_{i}\frac{N^i_{\rm ST}}{N^{\rm
tot}_{\rm ST}}\frac{\varepsilon^i_{\rm DT}}{\varepsilon^i_{\rm ST}}
\end{equation} where $\varepsilon^i_{\rm ST}$ and $\varepsilon^i_{\rm
DT}$ are the ST and DT efficiencies for the $i$-th tag mode,
respectively. The factor $\mathcal{B}_{\rm sub}$ accounts for the
branching fractions of intermediate decays, with $\mathcal{B}_{\rm
sub}=\mathcal{B}(\bar{K}^0\to K^0_S) \cdot
\mathcal{B}(K^0_S\to\pi^+\pi^-)$~\cite{ParticleDataGroup:2024cfk} for
$D^+$ modes and $\mathcal{B}_{\rm sub}=1$ for $D^0$ modes.

\section{\boldmath Selection of single tag $\bar D$ mesons}

For each charged track other than those used for $K^0_S$ reconstruction, the distances of closest approach to the interaction point~(IP) in the plane perpendicular to the $z$-axis, $|V_{xy}|$, and along the $z$-axis, $|V_z|$, are required to be within 1~cm and 10~cm, respectively, where the $z$-axis is defined as the symmetry axis of the MDC. The polar angle $\theta$ of the track with respect to the $z$-axis is required to satisfy $|\cos\theta|<0.93$. Particle identification~(PID) for charged tracks is performed by combining the $\mathrm{d}E/\mathrm{d}x$ and TOF information to calculate the confidence levels for the pion and kaon hypotheses. Each charged track is assigned the particle type with the higher probability.

The $K_S^0$ candidates are reconstructed via $K_S^0\to \pi^+\pi^-$
from pairs of oppositely charged tracks, satisfying
$|\cos\theta|<0.93$ and $|V_z|<$ 20~cm, with no requirement on
$|V_{xy}|$ nor PID. The two tracks are constrained to originate from a
common vertex, and the decay length must be greater than twice its
resolution. The quality of the vertex fit is required to satisfy
$\chi^2 < 100$, and the invariant mass of the $\pi^+\pi^-$ system is
required to fall within $(0.487,0.511)$~GeV/$c^2$.

The $\pi^0$ candidates are reconstructed via the $\pi^0\to\gamma\gamma$ decay using the EMC showers. For each shower, 
%the EMC time difference from the event start time is required to be within $(0,700)$~ns. 
the EMC time must be within $(0,700)$~ns of the event start time.
The deposited energy in the EMC is required to be greater than 25~MeV in the barrel region ($|\cos\theta|<0.80$) and 50~MeV in the end-cap region ($0.86<|\cos\theta|<0.92$). The opening angle between the EMC shower and the nearest charged track must be greater than $10^{\circ}$. The invariant mass of the photon pair is required to fall within $(0.115,0.150)$~GeV$/c^{2}$. A mass-constrained~(1C) fit to the known $\pi^{0}$ mass~\cite{ParticleDataGroup:2024cfk} is imposed on the photon pair, and the corresponding $\chi^2$ is required to be less than 50. The four-momentum of the $\pi^0$ updated by the 1C fit is used in the subsequent analysis.

For the two-body tag mode $\bar D^0\to K^+\pi^-$, the backgrounds
originating from cosmic rays, Bhabha and dimuon events are
suppressed using the procedure defined in
Ref.~\cite{BESIII:2014rtm}. The TOF time difference between the two
charged tracks is required to be less than 5~ns. In addition, each
event must contain either at least one EMC shower with energy greater
than 50~MeV or at least one additional good charged track detected in
the MDC.

To identify the ST $\bar D$ mesons, two kinematic variables are defined: the energy difference $\Delta E\equiv E_{\bar D}-E_{\mathrm{beam}}$
and the beam-constrained mass $M_{\rm BC}\equiv\sqrt{E_{\mathrm{beam}}^{2}/c^{4}-|\vec{p}_{\bar D}|^{2}/c^{2}}$, where, $E_{\mathrm{beam}}$ is the beam energy and $(E_{\bar D},\vec{p}_{\bar D})$ is the four-momentum of the ST $\bar D$ meson in the $e^+e^-$ center-of-mass frame. If multiple $\bar D$ candidates are reconstructed in the same tag mode, the one with the smallest $|\Delta E|$ is kept for
further analysis. The multiple combinations rates are less than 10\% for the two- and three-body decays without $\pi^0$, while higher for the other decay modes. Good consistency between data and MC simulations on the multiple combinations rates is confirmed. To suppress the combinatorial backgrounds in the
$M_{\rm BC}$ distribution, tag-mode-dependent $\Delta E$ requirements are applied on the ST candidates.  The $\Delta E$ requirements and the corresponding ST efficiencies estimated with the inclusive MC sample are summarized in Table~\ref{ST:realdata}.

\begin{figure*}[htbp]\centering
\includegraphics[width=0.98\linewidth]{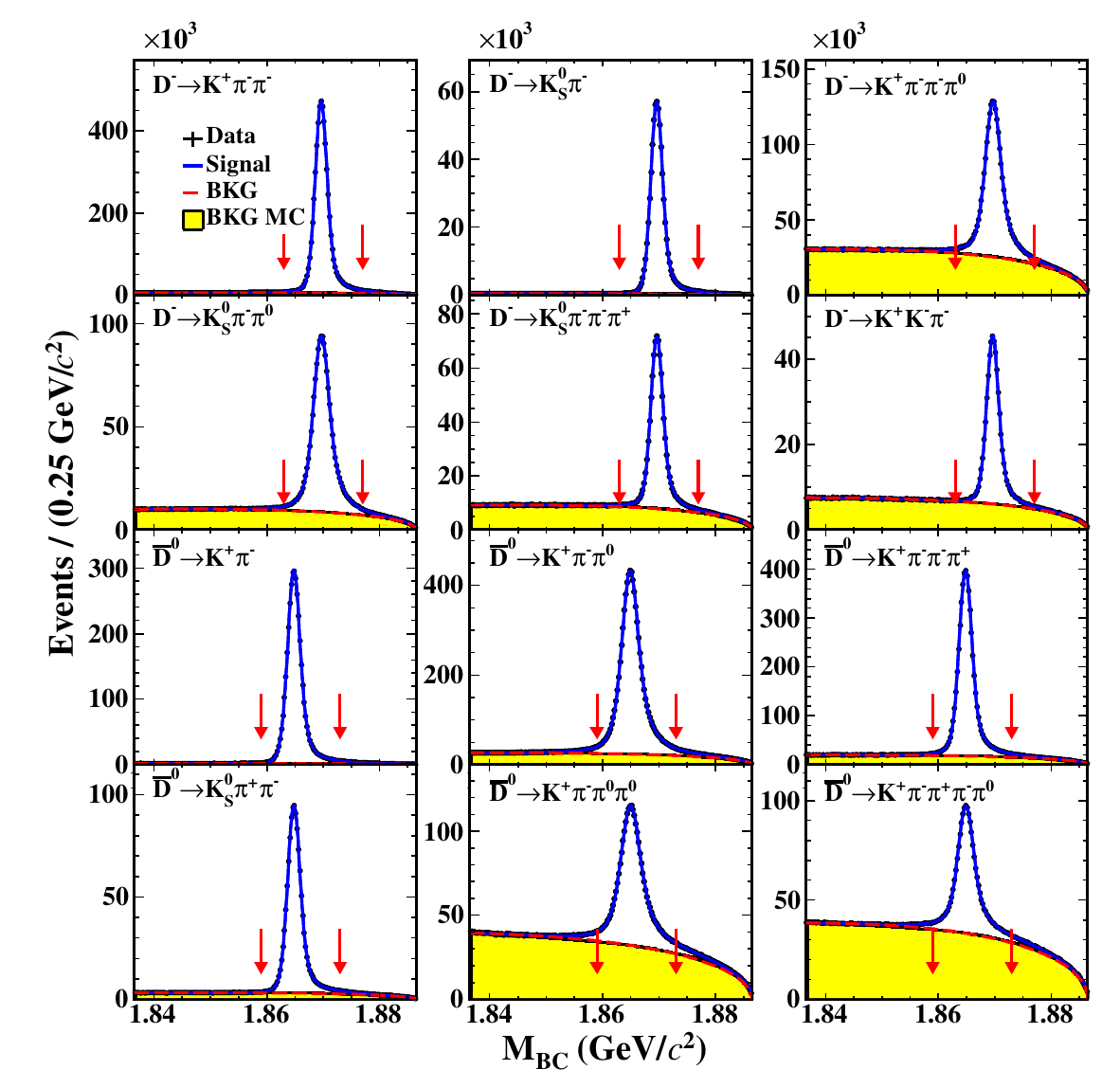}
\caption{The $M_{\rm BC}$ distributions of the ST $\bar D$
  candidates in data, with fit results overlaid.  The points with error bars are data, the blue curves are the best fits, and the red dashed curves are the fitted ARGUS functions. The pairs of red arrows show the
  $M_{\rm BC}$ signal windows, and the yellow filled histograms are the combinatorial background from the inclusive MC simulation.
\label{fig:datafit_Massbc}}
\end{figure*}

The ST yield of each tag mode is obtained from a binned maximum
likelihood fit to the corresponding $M_{\rm BC}$ distribution. The
signal is described by the MC-simulated shape, where only events with
the opening angle between the generated and reconstructed momentum
directions of ST daughter particles smaller than 15$^\circ$ are
retained. The MC shape is then convolved with a double-Gaussian
function to account for the resolution difference between data and MC
simulation, and an additional single-Gaussian function is included to
describe ISR effects. The background is modeled by an ARGUS
function~\cite{ARGUS:1990hfq} with the endpoint fixed at the $E_{\rm
beam}$. The potential peaking backgrounds such as doubly Cabibbo-suppressed $\bar{D}^0$ decays are subtracted from the ST yields. The fit results for various tag modes are shown in
Fig.~\ref{fig:datafit_Massbc}.  To further suppress the combinatorial
backgrounds, $M_{\rm BC}$ is required to be within $(1.859,1.873)$
GeV/$c^2$ for $\bar D^0$ tags and $(1.863,1.877)$ GeV/$c^2$ for $D^-$
tags. Summing over all the tag modes, the total ST yields of $\bar
D^0$ and $D^-$ mesons ($N_{\rm ST}^{\rm tot}$) are determined to be
$(20,229.8\pm5.5_{\rm stat})\times 10^3$ and $(10,646.9\pm3.8_{\rm
stat})\times10^3$, respectively.

\begin{table}
\renewcommand{\arraystretch}{1.2}
\centering
\caption {The $\Delta E$ requirements, the ST $\bar D$ yields
  in data($N_{\rm ST}^i$) and the ST
  efficiencies of the tag mode $i$ ($\varepsilon_{\rm ST}^i$), where the uncertainties are statistical only.  \label{ST:realdata}}
%\scalebox{0.87}{
\resizebox{0.49\textwidth}{!}{
\begin{tabular}{c|c|c|c|c}
\hline
\hline
 $D$ & Tag mode & $\Delta E$~(MeV)  &  $N_{\rm ST}^i~(10^3)$  &  $\varepsilon_{\rm ST}^i~(\%)    $       \\\hline
\multirow{6}{*}{$D^0$}
&$K^+\pi^-$                  &  $(-27,+27)$ & $3725.9\pm2.0$&$65.09\pm0.01$\\
&$K^+\pi^-\pi^0$             &  $(-62,+49)$ & $7420.0\pm3.2$&$35.53\pm0.01$\\
&$K^+\pi^-\pi^-\pi^+$        &  $(-26,+24)$ & $4987.2\pm2.5$&$40.70\pm0.01$\\
&$K^0_S\pi^+\pi^-$           &  $(-24,+24)$ & $1168.7\pm1.2$&$37.81\pm0.01$\\
&$K^+\pi^-\pi^0\pi^0$        &  $(-68,+53)$ & $1771.8\pm2.2$&$15.11\pm0.01$\\
&$K^+\pi^-\pi^-\pi^+\pi^0$   &  $(-57,+51)$ & $1152.8\pm1.8$&$16.19\pm0.01$\\
\hline
\multirow{6}{*}{$D^-$}
&$K^+\pi^-\pi^-$                  &  $(-25,+24)$ & $5552.7\pm2.5$&$51.01\pm0.01$\\
&$K^{0}_{S}\pi^{-}$               &  $(-25,+26)$ & $~656.4\pm0.8$&$51.41\pm0.02$\\
&$K^{+}\pi^{-}\pi^{-}\pi^{0}$     &  $(-57,+46)$ & $1723.7\pm1.8$&$24.40\pm0.01$\\
&$K^{0}_{S}\pi^{-}\pi^{0}$        &  $(-62,+49)$ & $1442.3\pm1.5$&$26.42\pm0.01$\\
&$K^{0}_{S}\pi^{-}\pi^{-}\pi^{+}$ &  $(-28,+27)$ & $~790.6\pm1.1$&$29.61\pm0.01$\\
&$K^{+}K^{-}\pi^{-}$              &  $(-24,+23)$ & $~481.2\pm0.9$&$40.87\pm0.01$\\
\hline
\hline
          \end{tabular}
          }

          \end{table}

\section{Selection of double tag events}
\label{sec:double_tag}

The DT events for $\kenu$, $\kmunu$, $\koenu$, and $\komunu$ are
selected from the remaining tracks in the presence of the ST $\bar D$
candidates. The selection criteria for $K^-$ and $K_{S}^{0}$ are the same as those used in the ST selection. The positron and muon candidates are identified using PID information based on the TOF, $\mathrm{d}E/\mathrm{d}x$, and EMC, from which the confidence levels $\mathcal{L}_{e}$, $\mathcal{L}_{\mu}$, $\mathcal{L}_{K}$, and $\mathcal{L}_{\pi}$ are calculated. The positron candidates must satisfy $\mathcal{L}_e > 0.8\cdot (\mathcal{L}_e+\mathcal{L}_\pi+\mathcal{L}_K)$ and $\mathcal{L}_e > 0.001$. The muon candidates are required to satisfy $\mathcal{L}_\mu > \mathcal{L}_e$ and $\mathcal{L}_\mu >0.001$, and the deposited energy of muon in the EMC is required to be within $(0.1,0.3)$ GeV.

The backgrounds due to hadronic $D$ decays are suppressed by requiring no additional good charged tracks on the signal side ($N_{\rm extra}^{\rm trk}=0$). The hadronic backgrounds with a $\pi^0$ are rejected by requiring the maximum energy of extra photons ($E_{\text{extra~}\gamma}^{\rm max}$) to be less than 0.25 GeV. To veto backgrounds associated with the misidentification between $\pi^+$ and $\ell^+$, the invariant mass of the $\bar{K}\ell^+$ system is required to be less than 1.83, 1.56, 1.84, and 1.59 GeV/$c^2$ for $\kenu$, $\kmunu$, $\ksenu$, and $\ksmunu$, respectively.

Since the neutrino is not detectable by the BESIII detector, the
kinematic variable $U_{\mathrm{miss}}\equiv
E_{\mathrm{miss}}-|\vec{p}_{\mathrm{miss}}|c$ is defined to determine
the DT yields. The missing four-momentum $(E_{\rm miss},\vec{p}_{\rm
miss})$ in the $e^+e^-$ center-of-mass frame is calculated as
$E_{\mathrm{miss}}\equiv E_{\mathrm{beam}}-E_{\bar K}-E_{\ell^+}$ and
$\vec{p}_{\mathrm{miss}}\equiv\vec{p}_{D}-\vec{p}_{\bar
K}-\vec{p}_{\ell^+}$. To improve the $U_{\mathrm{miss}}$ resolution,
$\vec{p}_{D}$ is evaluated as $\vec{p}_{D} = -\hat{p}_{\bar D}
\sqrt{E_{\mathrm{beam}}^{2}/c^{2}-m_{\bar D}^{2} c^{2} }$, where
$\hat{p}_{\bar D}$ is the unit vector along the momentum direction of
the ST $\bar D$ meson, $m_{\bar D}$ is the known $\bar D$
mass~\cite{ParticleDataGroup:2024cfk}, and $(E_{\bar
K(\ell^+)},\vec{p}_{\bar K(\ell^+)})$ is the measured four-momentum of
the $\bar K(\ell^+)$ candidate.

\section{Branching fractions}

\subsection{Branching fraction results}
\label{bf_result}

\begin{figure*}[htbp]
\begin{center}
\subfigure{\includegraphics[width=0.8\linewidth]{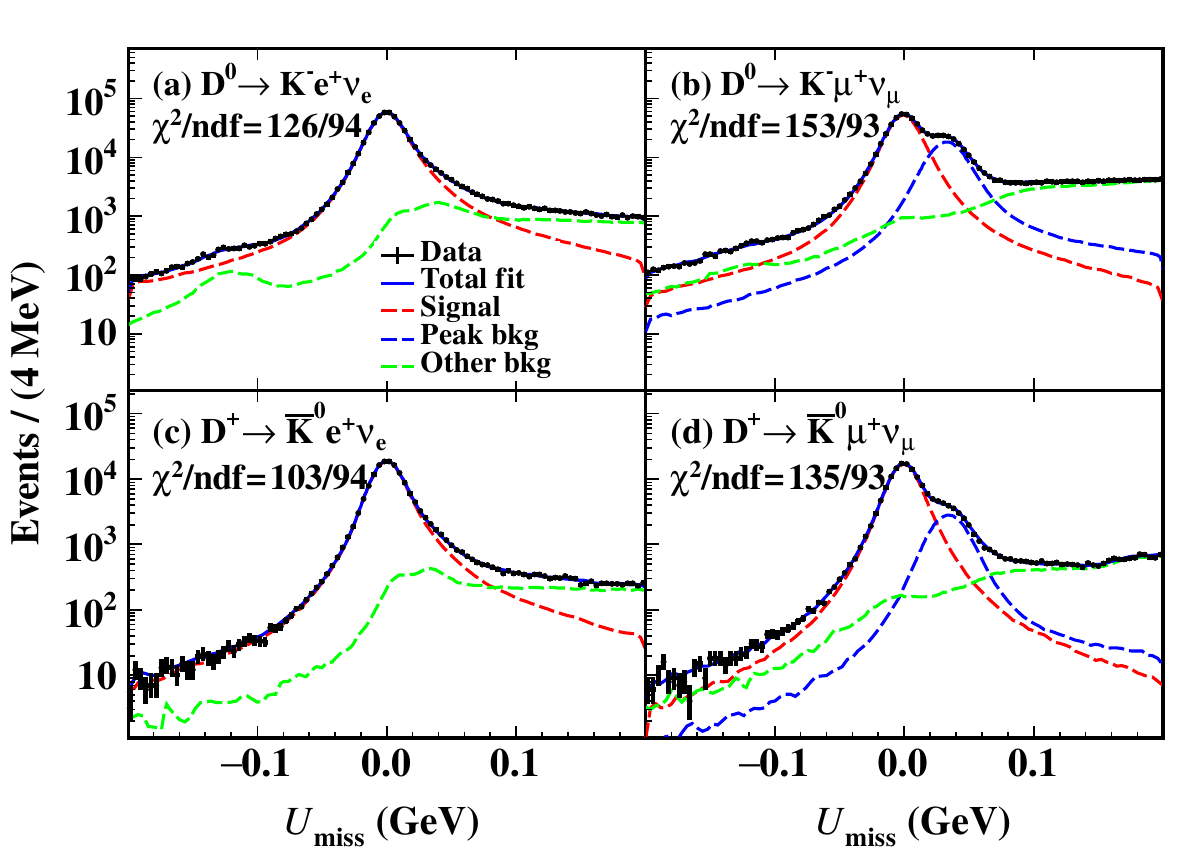}}

\caption{$U_{\rm miss}$ distributions of the accepted candidate events for $\klnu$ in data, shown on a logarithmic scale with fit results overlaid. The points with error bars are data. The blue solid lines denote the total fits. The red, blue, and black dashed lines show the signal, peaking background, and combinatorial background contributions, respectively.
\label{fit_umiss}
}
\end{center}
\end{figure*}

After applying all the selection criteria, the $U_{\rm miss}$ distributions of the accepted candidates for $D\to \bar K\ell^+\nu_\ell$ in data are shown in Fig.~\ref{fit_umiss}. Based on the inclusive MC simulation, the remaining backgrounds are dominated by the misidentification of $\mu^+$ or $\pi^+$ as $e^+$ for positron channels; misidentification of $\pi^+$ as $\mu^+$ for muon channels, and events with a missing $\pi^0$ for all signal decays. The normalized yields and fractions of main background components are listed in Table~\ref{umissbkg}.

The DT signal yields~($N_{\rm DT}$) in data are determined using
binned maximum likelihood fits to the corresponding $U_{\rm miss}$
distributions. The signal is modeled using MC simulated shape
convolved with a double Gaussian function with free parameters to
accounts for resolution difference between data and MC simulation. For
the muon channels, the dominant peaking backgrounds $D\to \bar K
\pi^+\pi^0$ are described using the corresponding MC-simulated shape
convolved with the same Gaussian function as that for the signal. The
other background components are combined into a combinatorial
background, whose shape is obtained from the inclusive MC sample. The
yields of the signal, the peaking background, and the combinatorial
background are allowed to float in the fits.

The DT efficiencies are obtained with the signal MC samples. Table~\ref{tab:DT_efficiency} summarizes the DT efficiencies and signal efficiencies for different signal decays in each tag mode and the weighted signal efficiencies ($\bar \varepsilon_{\rm sig}$). The data-MC differences on the $\bar \varepsilon_{\rm sig}$ associated with the tracking and PID have been corrected based on the studies of control samples, as described in Sec.~\ref{section_b}.

\begin{table}[htbp]
\caption{Main background sources for each signal decay mode, including the normalized yields ($N_{\rm bkg}$) and the corresponding fractions~($f_{\rm bkg}$) to the total background, as estimated from the inclusive MC sample.}
\label{umissbkg}
\centering
%\resizebox{0.48\textwidth}{!}{
\begin{tabular}{|c|l|c|c|}
\hline
\hline
Signal decay & Background source & $N_{\rm bkg}$ & $f_{\rm bkg}(\%)$ \\ \hline
\multirow{4}{*}{$\kenu$}
&$D^0 \to K^{*-}e^+\nu_e$      & 24816 & 43.4\\
&$D^0 \to K^{-}\mu^+\nu_\mu$   & 14859 & 26.0\\
&$D^0 \to K^{-}\pi^+\pi^0$     & 7740 & 13.5\\
%&$D^0 \to \pi^{-}e^+\nu_e$     & 1183 & 2.1\\
\hline
\multirow{3}{*}{$\kmunu$}
&$D^0 \to K^{-}\pi^+\pi^0$        & 194586 & 57.5\\
&$D^0 \to K^{-}\pi^+\pi^0\pi^0$   & 67142 & 19.8\\
&$D^0 \to K^{*-}\mu^+\nu_\mu$     & 21061  & 6.2\\
\hline
\multirow{3}{*}{$\ksenu$}
&$D^+ \to \bar K^{*0}e^+\nu_e$       & 7249 &  50.8\\
&$D^+ \to \bar K^{0}\mu^+\nu_\mu$    & 4309 &  30.2\\
&$D^+ \to K_S^0\pi^+\pi^0$           & 740  &  5.2\\
\hline
\multirow{3}{*}{$\ksmunu$}
&$D^+ \to K_S^0\pi^+\pi^0$            & 28379 &  55.8\\
&$D^+ \to \bar K^{*0}\mu^+\nu_\mu$    & 6514 &  12.8\\
&$D^+ \to  K_S^0\pi^+\pi^0\pi^0$      & 5179 &  10.2\\
\hline
\hline
\end{tabular}
%	}
\end{table}

\begin{table*}[htbp]
\centering\linespread{1.1}
\caption{The DT efficiencies $\varepsilon_{\rm DT}$, signal
  efficiencies $\varepsilon$ for different signal decays in each
  tag mode, and the weighted signal efficiencies $\bar
  \varepsilon_{{\rm sig}}$. The listed efficiencies are all in unit of \%.
  For the $D^+$ signal decays, the efficiencies also include the branching
  fraction of $\bar K^0\to \pi^+\pi^-$. The uncertainties are
  statistical only.}  \small
\label{tab:DT_efficiency}
\resizebox{1.0\textwidth}{!}{
\begin{tabular}{l|cccc|l|ccccc }\hline \hline
\multicolumn{5}{c|}{$D^0$ decay} &         \multicolumn{5}{c}{$D^+$ decay} \\ \hline

Tag mode &$\varepsilon_{{\rm DT}, K^-e^+\nu_e}$&$\varepsilon_{K^-e^+\nu_e}$&$\varepsilon_{{\rm DT}, K^-\mu^+\nu_\mu}$&$\varepsilon_{K^-\mu^+\nu_\mu}$ &
Tag mode  &$\varepsilon_{{\rm DT}, \bar K^0e^+\nu_e}$&$\varepsilon_{\bar K^0e^+\nu_e}$&$\varepsilon_{{\rm DT}, \bar K^0\mu^+\nu_\mu}$&$\varepsilon_{\bar K^0\mu^+\nu_\mu}$ \\ \hline
$\bar{D}^{0} \to K^{+} \pi^{-}$ & $42.34 \pm 0.02$ & $65.04 \pm 0.04$ & $36.25 \pm 0.02$ & $55.69 \pm 0.04$ & $D^{-} \to K^{+} \pi^{-} \pi^{-}$ & $22.83 \pm 0.01$ & $44.75 \pm 0.02$ & $22.83 \pm 0.01$ & $44.75 \pm 0.02$   \\  
$\bar{D}^{0} \to K^{+} \pi^{-} \pi^{0}$ & $24.04 \pm 0.01$ & $67.67 \pm 0.04$ & $20.63 \pm 0.01$ & $58.07 \pm 0.03$ & $D^{-} \to K_{S}^{0} \pi^{-}$ & $23.09 \pm 0.03$ & $44.92 \pm 0.06$ & $23.09 \pm 0.03$ & $44.92 \pm 0.06$   \\  
$\bar{D}^{0} \to K^{+} \pi^{+} \pi^{-} \pi^{-}$ & $26.34 \pm 0.01$ & $64.72 \pm 0.04$ & $22.15 \pm 0.01$ & $54.42 \pm 0.04$ & $D^{-} \to K^{+} \pi^{-} \pi^{-} \pi^{0}$ & $10.78 \pm 0.01$ & $44.17 \pm 0.04$ & $10.78 \pm 0.01$ & $44.17 \pm 0.04$   \\  
$\bar{D}^{0} \to K_{S}^{0} \pi^{+} \pi^{-}$ & $24.48 \pm 0.03$ & $64.73 \pm 0.08$ & $20.67 \pm 0.03$ & $54.68 \pm 0.08$ & $D^{-} \to K_{S}^{0} \pi^{-} \pi^{0}$ & $11.72 \pm 0.01$ & $44.37 \pm 0.05$ & $11.72 \pm 0.01$ & $44.37 \pm 0.05$   \\  
$\bar{D}^{0} \to K^{+} \pi^{-} \pi^{0} \pi^{0}$ & $10.93 \pm 0.01$ & $72.35 \pm 0.09$ & $9.42 \pm 0.01$ & $62.33 \pm 0.08$ & $D^{-} \to K_{S}^{0} \pi^{-} \pi^{-} \pi^{+}$ & $12.60 \pm 0.02$ & $42.55 \pm 0.06$ & $12.60 \pm 0.02$ & $42.55 \pm 0.06$   \\  
$\bar{D}^{0} \to K^{+} \pi^{+} \pi^{-} \pi^{-} \pi^{0}$ & $11.46 \pm 0.01$ & $70.81 \pm 0.10$ & $9.77 \pm 0.01$ & $60.32 \pm 0.09$ & $D^{-} \to K^{+} K^{-} \pi^{-}$ & $18.11 \pm 0.03$ & $44.30 \pm 0.07$ & $18.11 \pm 0.03$ & $44.30 \pm 0.07$   \\
\hline
$\bar \varepsilon_{{\rm sig}}$& &$68.25\pm0.02$& &$57.96\pm0.02$& $\bar \varepsilon_{{\rm sig}}$& &$45.39\pm0.02$& &$38.74\pm0.02$\\
\hline \hline
        \end{tabular}
        }
\end{table*}

Combining the signal yields in data~$N_{\rm DT}$, the weighted signal efficiencies~$\bar \varepsilon_{\rm sig}$, and the ST yields in data, the branching fractions of $\kenu$, $\kmunu$, $\ksenu$, and $\ksmunu$ are calculated using Eq.~(\ref{eq:bf}) and are summarized in Table~\ref{bf_klnu}. Input and output checks have been performed with MC simulation to ensure the analysis workflow are correct.

\begin{table*}[htp]
\centering
\setlength\tabcolsep{7pt}
\caption{The signal yields in data $N_{\rm DT}$, the weighted signal efficiency $\bar \varepsilon_{\rm sig}$, and the branching fractions $\mathcal B_{\rm sig}$ for the four signal decay modes. For $\mathcal B_{\rm sig}$, the first and second uncertainties are statistical and systematic, respectively.  For other quantities, the uncertainties are statistical only.  }
\label{bf_klnu}
\begin{tabular}{lccc}
\hline\hline
Signal decay&$N_{\rm DT}$ & $\bar \varepsilon_{{\rm sig}}~(\%)$ &$\mathcal B_{\rm sig}~(\%)$\\ \hline
$D^0\to K^-e^+\nu_{e}$&$489~811\pm770$&$68.25\pm0.02$&$3.548\pm0.006\pm0.017$\\ \hline
$D^0\to K^-\mu^+\nu_{\mu}$&$403~893\pm830$&$57.96\pm0.02$&$3.445\pm0.007\pm0.017$\\ \hline
$D^+\to \bar K^{0}e^+\nu_{e}$&$149~286\pm413$&$45.39\pm0.02$&$8.928\pm0.025\pm0.050$\\ \hline
$D^+\to \bar K^{0}\mu^+\nu_{\mu}$&$125~156\pm409$&$38.74\pm0.02$&$8.770\pm0.029\pm0.053$\\ \hline \hline
\end{tabular}
\end{table*}

\subsection{Systematic uncertainties on branching fractions}
\label{section_b}

Table~\ref{table:bf_systot} summarizes the sources of the systematic
uncertainties in the branching fraction measurements, the detailed
estimations of which are described below. The uncertainties caused
by ST $\bar D$ yields, tracking and PID of charged particles, $K^0_S$
reconstruction, $E_{\rm extra~\gamma}^{\rm max}$ and $N_{\rm
extra}^{\rm trk}$ requirements, and quoted branching fractions are
correlated among the corresponding decay channels in the subsequent
fits and calculations of LFU.

\paragraph{\boldmath \bf ST $\bar D$ yields}

The systematic uncertainty of ST $\bar D$ yields is estimated by varying the signal and background shapes in the fits to the $M_{\rm BC}$ spectra. The alternative signal shape is obtained by varying the parameters of Gaussian functions, while the alternative background shape is obtained by varying the endpoint by $\pm0.1$ MeV. The quadratic sum of these two sources results in a 0.3\% variation, which is taken as the systematic uncertainty on $N_{\rm ST}$.

\paragraph{\boldmath \bf $K^-$ tracking and PID}

The tracking and PID efficiencies of the $K^-$ are studied with a
control sample of hadronic $D\bar D$ events, where $D^0$ decays into
$K^-\pi^+$, $K^-\pi^+\pi^+\pi^-$, $D^+$ decays into $K^- \pi^+\pi^+$,
and $\bar{D}$ decays into the corresponding charge conjugated
channels. The correction factors are $0.997\pm0.001$ and
$1.001\pm0.001$ for the tracking and PID, respectively. Based on these
values, the signal efficiencies are corrected, and the uncertainties of
correction factors are taken as a source of systematic uncertainty.

\paragraph{\boldmath \bf $K^0_S$ reconstruction}
\label{sec:kso}

The uncertainty in the reconstruction efficiency of $K^0_S\to\pi^+\pi^-$ includes two sources. The tracking efficiencies of the $\pi^\pm$ are studied with the same control sample as for the $K^-$, with the correction factor determined to be $0.996\pm0.001$. Based on this value, the signal efficiencies are corrected, and the uncertainties of the correction factor are taken as a source of the systematic uncertainty. The efficiencies due to the $K^0_S$ mass window and $K^0_S$ decay vertex fit are investigated using the hadronic $D\bar D$ events, where $D^0$ decaying into $K^0_S\pi^+\pi^-$, $K^0_S\pi^+\pi^-\pi^0$, $K^0_S\pi^0$, and $D^+$ decaying into $K^0_S\pi^+$, $K^0_S\pi^+\pi^0$, and $K^0_S\pi^+\pi^+\pi^-$. The $\bar{D}$ is reconstructed via the tag modes $\bar{D}^0\to K^+ \pi^-,K^+\pi^-\pi^0, K^+\pi^-\pi^-\pi^+$ and $D^-\to K^+\pi^-\pi^-$. The correction factors are determined to be $0.41\%$ for $\ksenu$ and $0.45\%$ for $\ksmunu$, respectively, which are taken as the systematic uncertainties.
%In the further fits to the partial decay rates, these uncertainties are correlated for $\ksenu$ and $\ksmunu$.

\paragraph{\boldmath \bf $\ell^+$ tracking and PID}
\label{sec:electron}

The tracking and PID efficiencies of $e^+$ and $\mu^+$ are studied by using the control samples of $e^+e^-\to \gamma e^+e^-$ and $e^+e^-\to \gamma \mu^+\mu^-$, respectively. The correction factors are obtained to be $0.999\pm0.001$ for $e^+$ tracking and $0.983\pm0.001$ for $e^+$ PID, as well as $1.001\pm0.001$ for $\mu^+$ tracking and $0.985\pm0.001$ for $\mu^+$ PID. The signal efficiencies are corrected by these factors, and their uncertainties are assigned as sources of systematic uncertainty.
%In the further fits to the partial decay rates, these uncertainties are correlated for electron and muon channels, respectively.

\paragraph{\bf MC model}

The uncertainty due to the hadronic transition form factors used in the MC generation is estimated by varying corresponding parameters by $\pm1\sigma$.
%In the further fits to the partial decay rates, these uncertainties are independent for each signal decay.

\paragraph{\boldmath \bf $M_{\bar K \ell^+}$ requirement}
\label{m}

The requirements on $M_{\bar K\ell^+}$ result in negligible efficiency loss for the positron channels, and thereby their corresponding uncertainties are ignored. For muon channels, the reliability of MC modeling on the $M_{K\ell}$ requirement is estimated with positron channels as a control sample. By applying the same $M_{K\ell}$ requirements of muon channels to the positron ones, no significant variation in the measured branching fractions is observed. Therefore, this source of uncertainty is neglected.

\paragraph{\boldmath \bf $E_{\rm extra~\gamma}^{\rm max}$ and $N_{\rm extra}^{\rm trk}$ requirements}
\label{em}

The uncertainties associated with the $E_{\rm extra~\gamma}^{\rm max}$ and $N_{\rm extra}^{\rm trk}$ requirements are studied with the control samples of hadronic $D\bar D$ events, where both $D$ and $\bar D$ decay into one of the ST hadronic final states used in this analysis. The efficiency differences between data and MC simulation are assigned as the systematic uncertainty.
%In the further fits to the partial decay rates, these uncertainties are correlated for all four decay channels.

\paragraph{\boldmath \bf $U_{\rm miss}$ fit}

The systematic uncertainty due to the $U_{\rm miss}$ fit includes two sources. Given the resolution difference between data and MC simulation has been taken into account by convolving a Gaussian function to the simulated signal shapes, this contribution is neglected. The systematic uncertainty due to the background shape is estimated by varying the relative fractions of backgrounds from $e^+e^-\to q\bar q$ and the dominant background channels in the inclusive MC sample within the uncertainties of their input branching fractions. The variations of measured branching fractions are taken as the corresponding systematic uncertainties.
%In the further fits to the partial decay rates, these uncertainties are independent for all four signal channels.

\paragraph{\bf MC statistics}

The relative uncertainties on the signal efficiencies are assigned as the systematic uncertainties due to the limited statistics of the MC samples.
%In the further fits to the partial decay rates, these uncertainties are independent for the four signal decays.

\paragraph{\bf Quoted branching fractions}

For the $\ksenu$ and $\ksmunu$ decays, the uncertainty due to the quoted
branching fraction of $K_{S}^{0}\to \pi^+\pi^-$ is 0.07\%~\cite{ParticleDataGroup:2024cfk}.
%In the further fits to the partial decay rates, these uncertainties are correlated for $\ksenu$ and $\ksmunu$.

\begin{table*}[htbp]
\caption{Relative systematic uncertainties (in \%) in the measurements of the branching fractions.
\label{table:bf_systot}}
\centering
%\scalebox{1.0}{
\begin{tabular}{lcccc}
\hline
\hline

Source                              & $\kenu$ & $\kmunu$ & $\ksenu$ & $\ksmunu$\\\hline
$N_{\rm ST}^{\rm tot}$                       &0.30 &0.30 &0.30 &0.30   \\
$K^{-}$ tracking                   &0.10 &0.10 &--  &--   \\
$K^{-}$ PID                        &0.10 &0.10 &--  &--   \\
$K_S^0$ reconstruction              &--  &--  &0.41 &0.45 \\
$\ell^{+}$ tracking                &0.10 &0.10 &0.10 &0.10  \\
$\ell^{+}$ PID                     &0.10 &0.10 &0.10 &0.10  \\
MC model                            &0.12 &0.15 &0.04 &0.15   \\
$M_{\bar K \ell}$ requirement       &--   &--   &--   &--   \\
$E_{{\rm extra}~\gamma}^{\rm max}$ and $N_{\rm extra}^{\rm trk}$ requirement   &0.10 &0.10 &0.10 &0.10  \\
$U_{\rm miss}$ fit                  &0.27 &0.29 &0.08 &0.09  \\
MC statistics                       &0.03 &0.03 &0.04 &0.04  \\
Quoted branching fractions          &--   &--   &0.07 &0.07  \\

\hline

Total                               &0.47 &0.51 &0.56 &0.60  \\

\hline\hline

\end{tabular}
%	}
\end{table*}

\section{Hadronic transition form factors}

\subsection{Theoretical formula}
\label{th_formula}

The differential decay width $\frac{d\Gamma^{\ell}}{dq^2}$ of the semileptonic decay $D \to \bar K \ell^+\nu_\ell$ can be expressed as~\cite{Fajfer:2015ixa,Faustov:2019mqr}
\begin{equation}
\label{ffk_function1}
\small
\frac{d\Gamma^{\ell}}{dq^2} = \mathcal{N}(q^2)\left(1-\frac{m_\ell^2}{q^2}\right)^2\left[\left(1+\frac{m_\ell^2}{2q^2}\right)|h_0(q^2)|^2 + \frac{3m_\ell^2}{2q^2}|h_t(q^2)|^2\right],
\end{equation}
where $\mathcal{N}(q^2)=\frac{G_F^2|V_{cs}|^2|\textbf{q}|q^2}{96\pi^3 m^2_D}$ contains the Fermi coupling constant $G_F$, the modulus of the Cabibbo-Kobayashi-Maskawa matrix element $|V_{cs}|$, and the mass of $D$ meson $m_{D}$~\cite{ParticleDataGroup:2024cfk}; $q$ is the four momentum of the $\ell\nu_\ell$ system in the $D$ rest frame and $|\textbf{q}|$ is the magnitude of the three momentum of $q$; $m_\ell$ is the mass of the lepton. The hadronic helicity amplitudes $h_{0(t)}(q^2)$ are written as
\begin{equation}
\label{eq:hadron_hel_amp}
\begin{split}
h_0(q^2) &= \frac{2 m_D |\textbf{q}|}{\sqrt{q^2}}f_{+}(q^2),\\
h_t(q^2) &= \frac{m_D^2-m_K^2}{\sqrt{q^2}}f_0(q^2),
\end{split}
\end{equation}
where $f_{+}(q^2)$ and $f_{0}(q^2)$ are the vector and scalar form factors, respectively.

The series expansion~\cite{Becher:2005bg} approach is applied to describe the hadronic transition form factor in this paper, which takes the form
\begin{equation}
\begin{array}{l}
	\displaystyle f_{+,0}\left(q^2\right)=\frac{1}{P\left(q^2\right)\Phi\left(q^2\right)}\sum_{k=0}^{\infty}a_{k}
	\left(t_0\right)\left[z\left(q^2,t_0\right)\right]^k.\\
	\end{array}
\end{equation}
Here, $a_{k}(t_0)$ are the real coefficients, and $P(q^2)=z(q^2,m_{D^{*+}_{s}}^{2})$, where $z(q^2,t_{0})=\frac{\sqrt{t_{+}-q^2}-\sqrt{t_{+}-t_{0}}}{\sqrt{t_{+}-q^2}+\sqrt{t_{+}-t_{0}}}$. The function $\Phi$ is given by
\begin{equation}
\begin{array}{l}
	\displaystyle \Phi(q^2)=\sqrt{\frac{1}{24\pi\chi_{V}}}\left(\frac{t_{+}-q^2}{t_{+}-t_{0}}\right)^{1/4}\left(\sqrt{t_{+}-q^2}+\sqrt{t_{+}}\right)^{-5}\\
	\displaystyle \times\left(\sqrt{t_{+}-q^2}+\sqrt{t_{+}-t_{0}}\right)\left(\sqrt{t_{+}-q^2}+\sqrt{t_{+}-t_{-}}\right)^{3/2}\\
	\displaystyle \times\left(t_{+}-q^2\right)^{3/4},
\end{array}
\end{equation}
where $t_{\pm}=(m_{D}\pm m_{K})^{2}$; $t_{0}=t_{+}(1-\sqrt{1-t_{-}/t_{+}})$; $m_{D}$ and $m_{K}$ are the masses of $D$ and $K$, respectively. The pole mass of the vector form factor $m_{D_s^{*+}}=2112.2$~MeV/$c^2$ is identified with the mass of the lowest lying $c\bar s$ vector meson $D_s^{*+}$~\cite{ParticleDataGroup:2024cfk}. The parameter $\chi_{V}=\frac{3}{32\pi^2m_c^2}$ is obtained from dispersion relations using perturbative QCD~\cite{Boyd:1995sq}, where $m_c=1.27$~GeV/$c^2$ is the $\overline{\rm MS}$ mass of $c$-quark at the energy scale $\mu=m_c$~\cite{ParticleDataGroup:2024cfk}.

Following Ref.~\cite{BESIII:2024slx}, the vector form factor $f_+(q^2)$ is parameterized using a two-parameter series expansion as
\begin{equation}
\begin{array}{l}
	\displaystyle f_{+}\left(q^2\right)=\frac{1}{P\left(q^2\right)\Phi\left(q^2\right)}\left[a_{0}
	\left(t_0\right)+a_{1}\left(t_0\right)z\left(q^2,t_0\right)\right].\\
\end{array}
\end{equation}
By defining $r_1(t_0)=a_1(t_0)/a_0(t_0)$, $f_{+}(q^2)$ is re-written as
\begin{equation}
\label{ffk_function2}
\begin{array}{l}
	\displaystyle f_{+}\left(q^2\right)=\frac{1}{P\left(q^2\right)\Phi\left(q^2\right)}\frac{f_{+}\left(0\right)P\left(0\right)\Phi\left(0\right)}{1+r_{1}
	\left(t_{0}\right)z\left(0,t_{0}\right)}\\
	\displaystyle\times\left(1+r_{1}\left(t_{0}\right)\left[z\left(q^2,t_{0}\right)\right]\right).
	\end{array}
\end{equation}
The scalar form factor $f_{0}(q^2)$ is described using a one-parameter series expansion as
\begin{equation}
f_0(q^2) = \frac{1}{P(q^2)\Phi(q^2)}f_0(0)P(0)\Phi(0),
\label{equation:ff0}
\end{equation}
where the pole mass $m_{D_{s0}^{*}}=2317.8$~MeV/$c^2$ is assigned as the mass of $D_{s0}^{*}(2317)^+$~\cite{ParticleDataGroup:2024cfk}. The usual kinematic constraint $f_+(0)=f_0(0)$~\cite{Fajfer:2015ixa} is applied in this paper.

\subsection{Partial decay rates in data}
\label{sec:ddr_measure}

The form factors are determined by performing a fit to the $q^2$-binned
partial decay rates, which are defined as \begin{equation} \Delta
\Gamma_{i} = \int^{q^2_{{\rm max}(i)}}_{q^2_{{\rm
min}(i)}}{\frac{d\Gamma}{dq^2} dq^2} \end{equation} and determined
with $\Delta\Gamma_{i} = N_{\mathrm{prd}}^{i}/(\tau_{D}\cdot N_{\mathrm{ST}}^{\rm
tot})$. Here, $N_{\mathrm{prd}}^{i}$ is the number of events produced
in the $i$-th $q^{2}$ interval $\left(q^2_{{\rm min}(i)},q^2_{{\rm
max}(i)}\right)$ and $\tau_{D}$ is the $D$-meson
lifetime~\cite{ParticleDataGroup:2024cfk}. The four momentum of neutrino used in both $q^2$ and further $\cos\theta_{W}$ calculation is defined as $p_{\rm \nu}=(E_{\rm miss},E_{\rm miss}\times\hat{\vec{p}}_{\rm miss})$, where $E_{\rm miss}$ is the missing energy and $\hat{\vec{p}}_{\rm miss}$ is the unit vector along the missing momentum direction defined in Sec.~\ref{sec:double_tag}.

The number of events produced in the $i$-th $q^{2}$ interval of the data sample is calculated as
\begin{equation}
	N_{\mathrm{prd}}^{i}=\sum_{j}^{N_{\mathrm{intervals}}}\left(\varepsilon^{-1}\right)_{ij}N_{\mathrm{DT}}^{j},
\end{equation}
where $(\varepsilon^{-1})_{ij}$ is the element of the inverse efficiency matrix, obtained by analyzing the signal MC events. The statistical uncertainty of $N_{\mathrm{prd}}^{i}$ is calculated as
\begin{equation}
\sigma^2\left(N_{\mathrm{prd}}^{i}\right)=\sum_{j}^{N_{\mathrm{intervals}}}\left(\varepsilon^{-1}\right)_{ij}^2\sigma^2_{\rm stat}\left(N_{\mathrm{DT}}^{j}\right),
\end{equation}
where $\sigma_{\rm stat}(N_{\mathrm{DT}}^{j})$ is the statistical uncertainty of $N_{\mathrm{DT}}^{j}$. The efficiency matrix element $\varepsilon_{ij}^\alpha$ of the tag mode $\alpha$ is given as
\begin{equation}
\label{eq:eff_matrix}
	\varepsilon_{ij}^{\alpha}=\frac{N_{ij}^{\mathrm{rec}}}{N_{j}^{\mathrm{gen}}}\cdot \frac{1}{\varepsilon_{\mathrm{ST}}^{\alpha}},
\end{equation}
where $N_{ij}^{\mathrm{rec}}$ is the number of events generated in the $j\text{-}$th $q^{2}$ interval and reconstructed in the $i$-th $q^{2}$ interval; $N_{j}^{\mathrm{gen}}$ is the number of generated events in the $j\text{-}$th $q^{2}$ interval; $\varepsilon_{\mathrm{ST}}^\alpha$ is the ST efficiency for the tag mode $\alpha$. The efficiency matrix elements are further weighted by the ST yields in data as
\begin{equation}
\varepsilon_{ij}=\sum_{\alpha=1}^{6}\frac{N_{\rm {ST}}^{\alpha}\varepsilon_{ij}^\alpha}{N_{\rm {ST}}^{\rm tot}},
\end{equation}
and corrected for the data-MC difference due to tracking and PID. The detailed efficiency matrices are shown in Tables~\ref{kenu_effmatrix}$-$\ref{ksmunu_effmatrix} in the \hyperref[appendix]{Appendix}.

\begin{figure*}[htbp]
\begin{center}
\subfigure{\includegraphics[width=0.98\textwidth]{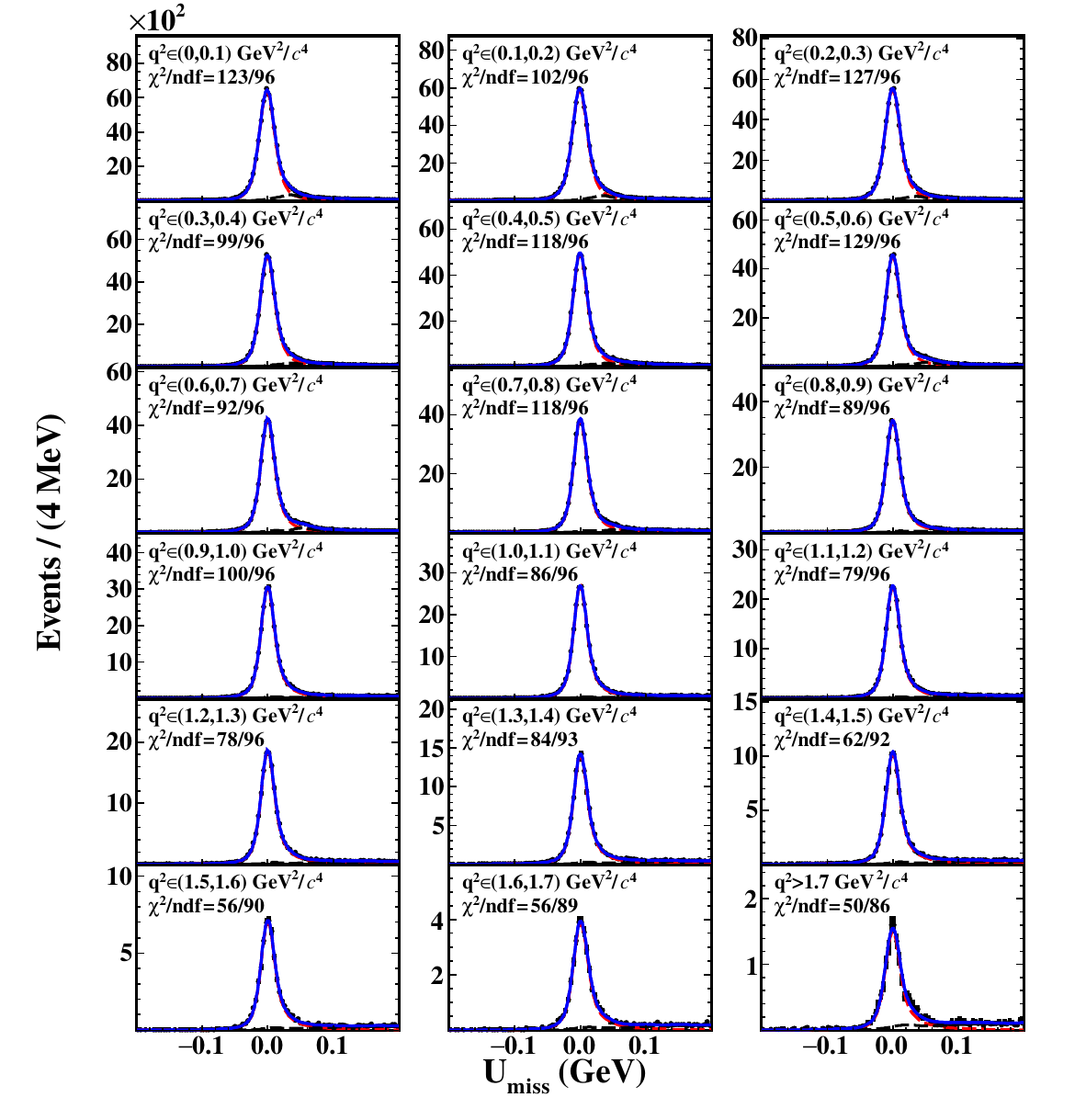}}
\caption{The $U_{\rm miss}$ distributions of the accepted candidate events for $\kenu$ in individual $q^2$ intervals in data, with fit results overlaid. The points with error bars are data, the blue solid curves are the fit results, the red dashed curves are the signal shapes, and
the black dashed curves are the fitted combinatorial background shapes.
\label{kenu_umissq2}
}
\end{center}
\end{figure*}

For each signal decay, the DT yield observed in the $j$-th
reconstructed $q^{2}$ interval $N_{\mathrm{DT}}^{j}$ is obtained by
fitting the corresponding $U_{\rm miss}$ distribution. The fitting
method is the same as that described in Sec.~\ref{bf_result}. The fit
results for the $U_{\rm miss}$ distributions are shown in
Fig.~\ref{kenu_umissq2} for the signal decay $\kenu$, while those for
the other three decays are presented in
Figs.~\ref{kmunu_umissq2}$-$\ref{ksmunu_umissq2} of
the \hyperref[appendix]{Appendix}.

Table~\ref{tab:kenu_decayrate} summarizes the $q^2$ ranges, the fitted DT yields $N_{\rm DT}$, the corresponding produced yields $N_{\rm prd}$, and the $q^2$-binned decay rates $\Delta\Gamma$ for the decay $\kenu$. The results for $\kmunu$, $\ksenu$, and $\ksmunu$ decays are presented in Tables~\ref{tab:kmunu_decayrate}$-$\ref{tab:ksmunu_decayrate} in the \hyperref[appendix]{Appendix}.

\begin{table}[htbp]
\caption{The fitted DT yields $N_{\rm DT}^i$, the produced yields $N_{\rm prd}^i$ and the $q^2$-binned partial decay rates $\Delta\Gamma$
  of $\kenu$ in different $q^2$ intervals of data, where the uncertainties are statistical only. \label{tab:kenu_decayrate}}
%\centering \resizebox{0.49\textwidth}{!}{
\centering{
\begin{tabular}{cccc}
\hline
\hline
$q^2~({\rm GeV^2}/c^4)$&$N_{\rm DT}^i$&$N_{\rm prd}^i$&$\Delta\Gamma~({\rm ns^{-1}})$\\ \hline
$(0.0,0.1)$&  $55232\pm258$&$76678\pm383$&$9.240\pm0.046$ \\
$(0.1,0.2)$&  $51026\pm248$&$72259\pm401$&$8.707\pm0.048$ \\
$(0.2,0.3)$&  $47871\pm241$&$68637\pm403$&$8.271\pm0.049$ \\
$(0.3,0.4)$&  $44255\pm229$&$63851\pm393$&$7.694\pm0.047$ \\
$(0.4,0.5)$&  $41687\pm222$&$60472\pm386$&$7.287\pm0.046$ \\
$(0.5,0.6)$&  $38575\pm214$&$56279\pm376$&$6.782\pm0.045$ \\
$(0.6,0.7)$&  $35282\pm204$&$51461\pm360$&$6.201\pm0.043$ \\
$(0.7,0.8)$&  $31878\pm193$&$46750\pm343$&$5.633\pm0.041$ \\
$(0.8,0.9)$&  $28543\pm180$&$42021\pm321$&$5.063\pm0.039$ \\
$(0.9,1.0)$&  $25497\pm170$&$37895\pm304$&$4.566\pm0.037$ \\
$(1.0,1.1)$&  $22333\pm158$&$33297\pm283$&$4.012\pm0.034$ \\
$(1.1,1.2)$&  $18934\pm146$&$28601\pm263$&$3.446\pm0.032$ \\
$(1.2,1.3)$&  $15810\pm133$&$24352\pm243$&$2.934\pm0.029$ \\
$(1.3,1.4)$&  $12230\pm118$&$19229\pm218$&$2.317\pm0.026$ \\
$(1.4,1.5)$&  $8939\pm101$&$14485\pm192$&$1.745\pm0.023$ \\
$(1.5,1.6)$&  $6152\pm84$&$10471\pm164$&$1.262\pm0.020$ \\
$(1.6,1.7)$&  $3563\pm65$&$6541\pm135$&$0.788\pm0.016$ \\
$>1.7$&  $1425\pm43$&$3471\pm115$&$0.418\pm0.014$ \\
\hline
\hline
\end{tabular}
	}
\end{table}

\subsection{Fit to the partial decay rates}

A fit to the $q^2$-binned partial decay rates $\Delta\Gamma$ is
performed using the least $\chi^{2}$ method, with the objective
function constructed as \begin{equation} \label{eq:chi}
\chi^{2}=\sum_{i,j=1}^{N_{\mathrm{intervals}}}\left(\Delta\Gamma_{i}^{\mathrm{msr}}-\Delta\Gamma_{i}^{\rm
fit}\right)
(C^{-1})_{ij}\left(\Delta\Gamma_{j}^{\mathrm{msr}}-\Delta\Gamma_{j}^{\rm
fit}\right).  \end{equation} Here, $\Delta\Gamma_{i}^{\mathrm{msr}}$ is the measured $\Delta\Gamma_{i}$ and $\Delta\Gamma_i^{\rm fit}$ is the
fitted decay rate obtained by integrating Eq.~\ref{ffk_function1} in
the $i$-th interval. The covariance matrix is defined as $C_{ij} =
C_{ij}^{\mathrm{stat}}+C_{ij}^{\mathrm{syst}}$, which includes both
the statistical and systematic uncertainties, and takes the
correlations between different $q^{2}$ intervals into
consideration. The elements of the statistical covariance matrix are
defined as \begin{equation} C_{ij}^{\rm stat} =\left
(\frac{1}{\tau_{D}N_{\mathrm{ST}}^{\rm
tot}}\right)^{2}\sum_{\alpha}(\varepsilon^{-1})_{i\alpha}(\varepsilon^{-1})_{j\alpha}\left(\sigma\left(N_{\mathrm{DT}}^{\alpha}\right)\right)^{2},
\end{equation} where $\sigma(N_{\mathrm{DT}}^{\alpha})$ is the
statistical uncertainty of the signal yield observed in the
$\alpha$-th interval. The detailed statistical covariance matrices of
$\kenu$, $\kmunu$, $\koenu$ and $\komunu$ decays are listed in
Tables~\ref{tab:kenu_statmatrix}$-$\ref{tab:ksmunu_statmatrix} of the
\hyperref[appendix]{Appendix}.

\subsection{Systematic uncertainties on partial decay rates}
\label{sec:sys_ddr}

The sources of systematic uncertainties in the measurement of $q^2$-binned decay rates are discussed below. Table~\ref{tab:kenu_sysq2} summarizes the systematic uncertainties of $\Delta\Gamma$ in different $q^2$ intervals for $\kenu$, while those for $\kmunu$, $\ksenu$, and $\ksmunu$ are presented in Tables~\ref{tab:kmunu_sysq2}$-$\ref{tab:ksmunu_sysq2} of the \hyperref[appendix]{Appendix}. The systematic covariance matrices for all signal decays are also available in Tables~\ref{tab:kenu_sysmatrix}$-$\ref{tab:ksmunu_sysmatrix} of the \hyperref[appendix]{Appendix}.

\paragraph{\boldmath \bf ST $\bar D$ yields}

The systematic uncertainties due to the ST yields are fully correlated across all the $q^2$ intervals. The corresponding systematic covariance matrix is calculated as
\begin{equation}
C_{ij}^{\mathrm{syst}}\left(N_{\rm ST}\right)=\Delta\Gamma_{i}\Delta\Gamma_{j}\left(\frac{\sigma\left(N_{\rm ST}\right)}{N_{\rm ST}}\right)^2,
\end{equation}
where $\sigma(N_{\rm ST})/N_{\rm ST}$ is the relative uncertainty of ST yield.

\paragraph{\boldmath \bf $D$ lifetime}

The systematic uncertainties associated with the $D$ meson lifetime are fully correlated across all the $q^2$ intervals. The corresponding systematic covariance matrix is calculated as
\begin{equation}
C_{ij}^{\mathrm{syst}}\left(\tau_{D}\right)=\sigma\left(\Delta\Gamma_{i}\right)\sigma\left(\Delta\Gamma_{j}\right),
\end{equation}
where $\sigma(\Delta\Gamma_{i})=\sigma \tau_{D}\cdot\Delta\Gamma_{i}$ and $\sigma \tau_{D}$ is the uncertainty of the $D$ lifetime~\cite{ParticleDataGroup:2024cfk}.

\paragraph{\bf MC statistics}

The systematic uncertainties of MC statistics are described with the covariance matrix as
\begin{equation}
\begin{array}{l}
\displaystyle C_{ij}^{\mathrm{syst}}\left(\rm MC^{stat}\right)=\left(\frac{1}{\tau_{D}N_{\mathrm{ST}}}\right)^{2}\\
\displaystyle\times \sum_{\alpha\beta}N_{\mathrm{DT}}^{\alpha}N_{\mathrm{DT}}^{\beta}\mathrm{Cov}\left(\left(\varepsilon^{-1}\right)_{i\alpha},\left(\varepsilon^{-1}\right)_{j\beta}\right),
\end{array}
\end{equation}
where $N_{\mathrm{DT}}^{\alpha(\beta)}$ is the DT yield observed in the interval $\alpha(\beta)$, and the covariances of the inverse efficiency matrix elements are given by
\begin{equation}
\begin{array}{l}
\mathrm{Cov}\left(\left(\varepsilon^{-1}\right)_{i\alpha},\left(\varepsilon^{-1}\right)_{j\beta}\right)=\\ \sum\limits_{mn}\left(\left(\varepsilon^{-1}\right)_{im}\left(\varepsilon^{-1}\right)_{jm}\right)\left[\sigma\left(\varepsilon_{mn}\right)\right]^2\left(\left(\varepsilon^{-1}\right)_{\alpha n}\left(\varepsilon^{-1}\right)_{\beta n}\right).
\end{array}
\end{equation}

\paragraph{\bf MC model}

The systematic uncertainty from the MC model is estimated by varying the parameters of the two-parameter series expansion model by $\pm1\sigma$. Alternative partial decay rates are calculated based on the updated efficiency matrix, and their differences from the nominal ones are taken as the systematic uncertainties. The corresponding covariance matrix is assigned as
\begin{equation}
C_{ij}^{\mathrm{syst}}\left (\mathrm{MC~model}\right )=\delta\left(\Delta\Gamma_{i}\right)\delta\left(\Delta\Gamma_{j}\right),
\end{equation}
where $\delta(\Delta\Gamma_{i})$ denotes the change of the partial decay rate in the $i$-th $q^{2}$ interval.

\paragraph{\bf Tracking and PID}

The systematic uncertainties associated with the $\ell^+$ or $K^-$ tracking and PID efficiencies are estimated by varying the corresponding correction factors within $\pm 1\sigma$. With the new efficiency matrices, alternative partial decay rates are obtained and their differences from the nominal ones are taken as the systematic uncertainties. The related covariance matrix is calculated as
\begin{equation}
C_{ij}^{\mathrm{syst}}\left (\mathrm{Tracking,~PID}\right)=\delta\left(\Delta\Gamma_{i}\right)\delta\left(\Delta\Gamma_{j}\right),
\end{equation}
where $\delta(\Delta\Gamma_{i})$ denotes the change of the partial decay rate in the $i$-th $q^{2}$ interval.

\paragraph{\boldmath \bf $U_{\rm miss}$ fit}

The systematic uncertainties arising from the fit to the $U_{\rm miss}$ distributions in the interval $\alpha$, $\sigma_{\alpha}^{\mathrm{Fit}}$, are estimated in the same approach as described in Sec.~\ref{section_b}. The corresponding covariance matrix is assigned as
\begin{equation}
C_{ij}^{\mathrm{syst}}\left(U_{\rm miss}~\mathrm{fit}\right)=\left(\frac{1}{\tau_{D}N_{\mathrm{ST}}^{\rm tot}}\right)^{2}\sum_{\alpha}\varepsilon_{i\alpha}^{-1}\varepsilon_{j\alpha}^{-1}\left(\sigma_{\alpha}^{\mathrm{Fit}}\right)^{2}.
\end{equation}

\paragraph{\bf Remaining uncertainties}

The remaining systematic uncertainties, including the $E_{\rm extra~\gamma}^{\rm max}$ and $N_{\rm extra}^{\rm trk}$ requirements, $K^0_S$ reconstruction, and quoted branching fractions, are assumed to be fully correlated across $q^{2}$ intervals, and the corresponding systematic covariance matrix is calculated as
\begin{equation}
C_{ij}^{\mathrm{syst}}=\sigma\left(\Delta\Gamma_{i}\right)\sigma\left(\Delta\Gamma_{j}\right),
\end{equation}
where $\sigma(\Delta\Gamma_{i})=\sigma_{\rm syst}\cdot\Delta\Gamma_{i}$.

\begin{table*}[htbp]
\caption{The systematic uncertainties (in \%) of the measured decay rates of $\kenu$ in different
$q^2$ bins.
\label{tab:kenu_sysq2}}
\centering
%\resizebox{1.0\textwidth}{!}{
\begin{tabular}{lcccccccccccccccccc}
\hline
\hline
$i$-th $q^2$ bin&1&2&3&4&5&6&7&8&9&10&11&12&13&14&15&16&17&18\\ \hline
$N_{\rm tag}$&0.30&0.30&0.30&0.30&0.30&0.30&0.30&0.30&0.30&0.30&0.30&0.30&0.30&0.30&0.30&0.30&0.30&0.30\\
$D^0$ lifetime&0.24&0.24&0.24&0.24&0.24&0.24&0.24&0.24&0.24&0.24&0.24&0.24&0.24&0.24&0.24&0.24&0.24&0.24\\
MC stat.&0.08&0.09&0.09&0.09&0.10&0.10&0.10&0.11&0.12&0.12&0.13&0.14&0.16&0.18&0.21&0.25&0.34&0.56\\
$E_{\rm extra \gamma}^{\rm max}$ cut&0.10&0.10&0.10&0.10&0.10&0.10&0.10&0.10&0.10&0.10&0.10&0.10&0.10&0.10&0.10&0.10&0.10&0.10\\
$U_{\rm miss}$ fit&0.19&0.18&0.20&0.17&0.17&0.18&0.18&0.20&0.15&0.14&0.14&0.13&0.15&0.18&0.18&0.17&0.25&0.45\\
$K$ tracking&0.10&0.10&0.10&0.10&0.10&0.10&0.10&0.10&0.10&0.10&0.10&0.10&0.10&0.10&0.10&0.11&0.14&0.27\\
$K$ PID&0.10&0.10&0.10&0.10&0.10&0.10&0.10&0.10&0.10&0.10&0.10&0.10&0.10&0.10&0.10&0.10&0.10&0.10\\
$e$ tracking&0.10&0.10&0.10&0.10&0.10&0.10&0.10&0.10&0.10&0.10&0.10&0.10&0.10&0.10&0.10&0.10&0.10&0.10\\
$e$ PID&0.10&0.10&0.10&0.10&0.10&0.10&0.10&0.10&0.10&0.10&0.10&0.10&0.10&0.10&0.10&0.10&0.10&0.10\\
MC model&0.08&0.21&0.37&0.29&0.23&0.38&0.25&0.06&0.31&0.75&0.12&0.19&0.17&0.44&0.06&0.70&0.27&0.83\\
\hline
Total&0.50&0.53&0.62&0.57&0.54&0.62&0.55&0.50&0.58&0.89&0.50&0.52&0.53&0.68&0.53&0.88&0.68&1.21\\
\hline
\hline
\end{tabular}
%	}
\end{table*}

\subsection{Results based on individual fits}

\begin{figure*}[htbp]
\begin{center}
\includegraphics[width=0.99\textwidth]{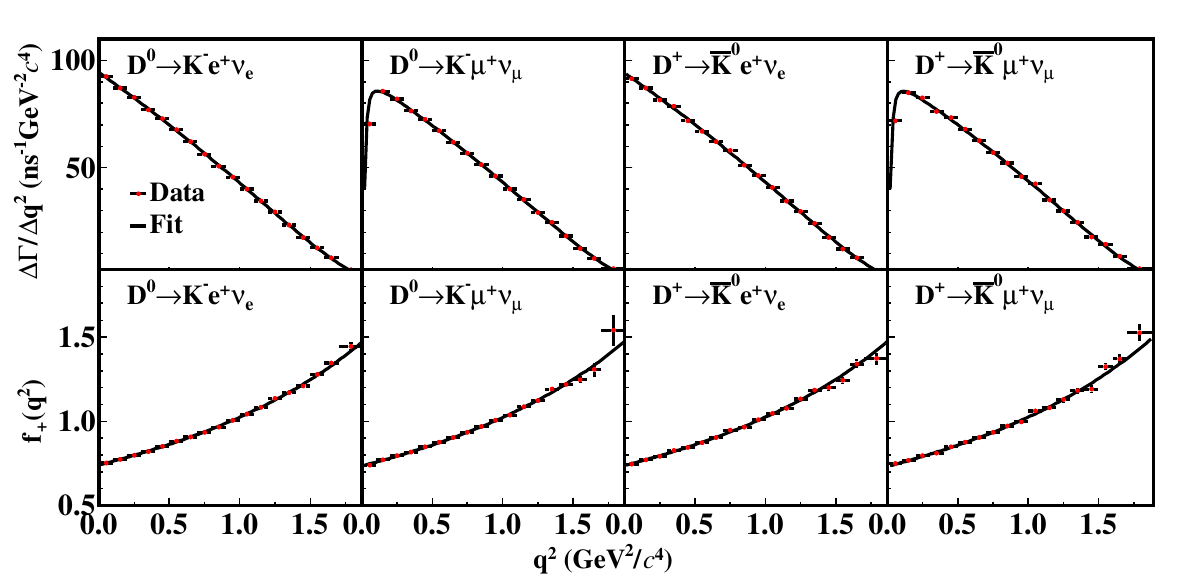}
\caption{Fits to the partial decay rates of $\klnu$ and projections of the form factor as functions of $q^2$, where the red points with error bars are the measured partial decay rates and the solid curves are the best fits. The fit qualities are available in Table~\ref{tab:fit_par}.
\label{fig:ff_klnu}
}
\end{center}
\end{figure*}

Individual fits are performed to the $q^2$-binned partial decay rates of $\kenu$, $\kmunu$, $\ksenu$, and $\ksmunu$ to determine the fit parameters $f_+(0)|V_{cs}|$ and $r_1(t_0)$ defined in Sec.~\ref{th_formula}. The statistical uncertainties of the fitted parameters are obtained from the fits with only statistical covariance matrix included. 
The systematic uncertainties of the fitted parameters are assigned as the quadratic differences between the uncertainties obtained from the fits with the statistical covariance matrix alone and those with both the statistical and systematic covariance matrices included.
The fit projections for all the decays are shown in Fig.~\ref{fig:ff_klnu} and the fitted parameters are listed in Table~\ref{tab:fit_par}.

Furthermore, the projection of form factor $f_{+}$ in each $q^2$
interval is obtained as
\begin{equation}
\label{eq:ffp_extract}
\begin{array}{l}
\displaystyle f_{+}^{\mathrm{data}}(q_i^{2})=
\displaystyle \sqrt{\frac{\left(\Delta\Gamma_i^{\mathrm{measured}}-B\right)\cdot|f_{+}(q^2_i)|^2}{A}},\\
\end{array}
\end{equation}
where
\begin{equation}
\small
\begin{array}{l}
\displaystyle A=\int^{q^2_{{\rm max}(i)}}_{q^2_{{\rm min}(i)}}{\mathcal{N}(q^2)\left(1-\frac{m_\ell^2}{q^2}\right)^2\left[\left(1+\frac{m_\ell^2}{2q^2}\right)|h_0(q^2)|^2\right] dq^2},
\end{array}
\end{equation}
and
\begin{equation}
\small
\begin{array}{l}
\displaystyle B=\int^{q^2_{{\rm max}(i)}}_{q^2_{{\rm min}(i)}}{\mathcal{N}(q^2)\left(1-\frac{m_\ell^2}{q^2}\right)^2\left[ \frac{3m_l^2}{2q^2}|h_t(q^2)|^2\right]dq^2},
\end{array}
\end{equation}
where $q^{2}_{\mathrm{min}(i)}$ and $q^{2}_{\mathrm{max}(i)}$ are the
low and high boundaries of the $i$-th $q^{2}$ bin. The numerical results are summarized in Table~\ref{tab:ffp_data_klnu}.
%The contribution of
%the form factor $f_{0}$ is suppressed by a factor $m^2_l$, which could
%hardly be extracted via a model independent approach as shown in
%Eq.~\ref{eq:ffp_extract} for most of $q^2$ intervals, even in current
%statistics.

\begin{table*}[htbp]
	\caption{ The determined $f_{+}^{\mathrm{data}}(q_i^{2})$ values in different $q^2$ bins of $\klnu$, where the first uncertainties are statistical and the second are systematic.}
	\setlength{\tabcolsep}{6pt}
	\centering
	\scalebox{1.00}{
		\begin{tabular}{c|cccc}
			\hline
			\hline
			$q^2~({\rm GeV}/c^4)$&$\kenu$&$\kmunu$&$\ksenu$ &$\ksmunu$\\ \hline
			$(0.0,0.1)$&$0.754\pm0.002\pm0.002$&$0.741\pm0.004\pm0.003$&$0.748\pm0.003\pm0.002$&$0.749\pm0.006\pm0.004$ \\
			$(0.1,0.2)$&$0.774\pm0.002\pm0.002$&$0.772\pm0.003\pm0.003$&$0.772\pm0.004\pm0.003$&$0.767\pm0.005\pm0.003$ \\
			$(0.2,0.3)$&$0.800\pm0.002\pm0.002$&$0.796\pm0.003\pm0.003$&$0.792\pm0.004\pm0.003$&$0.797\pm0.005\pm0.003$ \\
			$(0.3,0.4)$&$0.821\pm0.003\pm0.002$&$0.817\pm0.003\pm0.002$&$0.826\pm0.004\pm0.003$&$0.812\pm0.005\pm0.003$ \\
			$(0.4,0.5)$&$0.852\pm0.003\pm0.002$&$0.848\pm0.003\pm0.002$&$0.844\pm0.005\pm0.003$&$0.849\pm0.005\pm0.003$ \\
			$(0.5,0.6)$&$0.881\pm0.003\pm0.003$&$0.875\pm0.004\pm0.003$&$0.873\pm0.005\pm0.003$&$0.877\pm0.006\pm0.003$ \\
			$(0.6,0.7)$&$0.907\pm0.003\pm0.003$&$0.903\pm0.004\pm0.003$&$0.905\pm0.006\pm0.004$&$0.904\pm0.006\pm0.004$ \\
			$(0.7,0.8)$&$0.936\pm0.003\pm0.002$&$0.935\pm0.004\pm0.003$&$0.946\pm0.006\pm0.004$&$0.933\pm0.007\pm0.004$ \\
			$(0.8,0.9)$&$0.966\pm0.004\pm0.003$&$0.970\pm0.004\pm0.003$&$0.966\pm0.007\pm0.005$&$0.971\pm0.007\pm0.004$ \\
			$(0.9,1.0)$&$1.006\pm0.004\pm0.004$&$1.006\pm0.005\pm0.003$&$1.010\pm0.007\pm0.006$&$0.997\pm0.008\pm0.005$ \\
			$(1.0,1.1)$&$1.044\pm0.004\pm0.003$&$1.036\pm0.005\pm0.003$&$1.046\pm0.008\pm0.006$&$1.062\pm0.009\pm0.006$ \\
			$(1.1,1.2)$&$1.084\pm0.005\pm0.003$&$1.086\pm0.006\pm0.004$&$1.077\pm0.009\pm0.006$&$1.079\pm0.011\pm0.006$ \\
			$(1.2,1.3)$&$1.136\pm0.006\pm0.003$&$1.125\pm0.007\pm0.004$&$1.134\pm0.010\pm0.006$&$1.130\pm0.013\pm0.008$ \\
			$(1.3,1.4)$&$1.170\pm0.007\pm0.004$&$1.191\pm0.009\pm0.003$&$1.184\pm0.012\pm0.007$&$1.182\pm0.015\pm0.008$ \\
			$(1.4,1.5)$&$1.209\pm0.008\pm0.003$&$1.220\pm0.012\pm0.007$&$1.200\pm0.014\pm0.009$&$1.191\pm0.019\pm0.010$ \\
			$(1.5,1.6)$&$1.277\pm0.010\pm0.006$&$1.246\pm0.012\pm0.018$&$1.243\pm0.018\pm0.011$&$1.325\pm0.018\pm0.011$ \\
			$(1.6,1.7)$&$1.346\pm0.014\pm0.005$&$1.308\pm0.019\pm0.037$&$1.341\pm0.023\pm0.014$&$1.370\pm0.026\pm0.015$ \\
			$>1.7$& $1.444\pm0.024\pm0.009$&$1.542\pm0.038\pm0.085$&$1.373\pm0.035\pm0.012$&$1.526\pm0.044\pm0.027$ \\\hline\hline
		\end{tabular}
	}
	\label{tab:ffp_data_klnu}
\end{table*}

\begin{table*}[htbp]
\centering
\caption{The parameters ($f_+(0)|V_{cs}|$, $r_1(t_0)$) determined from the fits to the partial decay rates of the $\klnu$ decays, where the first and second uncertainties
  are statistical and systematic, respectively. The column labeled $\rho$ gives the correlation
  coefficients of the two parameters, and ndf denotes the number of degrees of freedom.
\label{tab:fit_par}}
\scalebox{1.0}{
\begin{tabular}{|l|c|c|c|c|c|}
 \hline\hline
Case & Signal decay &$f_+(0)|V_{cs}|$ & $r_1(t_0)$ & $\rho$ & $\chi^2/\rm ndf$\\
  \hline
\multirow{4}{*}{Individual fit}&$\kenu$   &$0.7217\pm0.0010\pm0.0017$&$-2.22\pm0.04\pm0.02$&0.38&14.7/16  \\
                               &$\kmunu$  &$0.7159\pm0.0013\pm0.0019$&$-2.33\pm0.05\pm0.02$&0.47&13.1/16  \\
                               &$\ksenu$  &$0.7187\pm0.0017\pm0.0024$&$-2.30\pm0.06\pm0.06$&0.27&14.1/16  \\
                               &$\ksmunu$ &$0.7132\pm0.0021\pm0.0025$&$-2.43\pm0.07\pm0.06$&0.41&18.1/16  \\  \hline
Simultaneous fit               &$\klnu$   &$0.7183\pm0.0007\pm0.0014$&$-2.28\pm0.02\pm0.02$&0.30&72.6/70  \\
\hline\hline
\end{tabular}
}
\end{table*}

%%%%fah
\subsection{Results based on a simultaneous fit}
\label{sec:ddr_smfit}

Considering the correlations between the systematic uncertainties of
the fitted parameters from individual fits, a simultaneous fit is
performed to the $q^2$-binned partial decay rates of $\kenu$,
$\kmunu$, $\ksenu$, and $\ksmunu$ to determine the combined
$\ffK|V_{cs}|$ and $r_1(t_0)$. In the simultaneous fit, the parameters
$f_+(0)|V_{cs}|$ and $r_1(t_0)$ are shared among the four signal
decays. The corresponding $\chi^2$ function is defined in the same way
as in Eq.~(\ref{eq:chi}), where the indices $i,j$ sum over all the 72
$q^2$ intervals. The covariance matrix $C_{ij}$ is redefined as
$C_{ij}=C^{\rm stat}_{ij}+C^{\rm csyst}_{ij}+C^{\rm usyst}_{ij}$,
where $C^{\rm stat}_{ij}$ is the element of statistical covariance
matrix, which is block-diagonal, {\it i.e.}
\begin{equation}
C^{\rm stat}=
\begin{pmatrix}
A&0&0&0\\
0&B&0&0\\
0&0&C&0\\
0&0&0&D
\end{pmatrix}.
\end{equation}
Here $A$, $B$, $C$, and $D$ are the statistical covariance matrices for each signal channel. The elements of the correlated systematic covariance matrix are defined as
\begin{equation}
\label{eq:csys_matrix}
C^{\mathrm{csyst}}_{ij}=\delta(\Delta\Gamma_{i})\delta(\Delta\Gamma_{j}).
\end{equation}
The uncorrelated systematic covariance matrix is also defined in a block-diagonal form as
\begin{equation}
C^{\rm usyst}=
\begin{pmatrix}
a&0&0&0\\
0&b&0&0\\
0&0&c&0\\
0&0&0&d
\end{pmatrix},
\end{equation}
where $a$, $b$, $c$, and $d$ are the uncorrelated systematic covariance matrices obtained from each signal channel. The detailed covariance matrices for the simultaneous fit are presented in Tables~\ref{klnumatrix_cov1}$-$\ref{klnumatrix_cov4} of the \hyperref[appendix]{Appendix}.

\begin{figure*}[htbp]
\begin{center}
\includegraphics[width=1.0\textwidth]{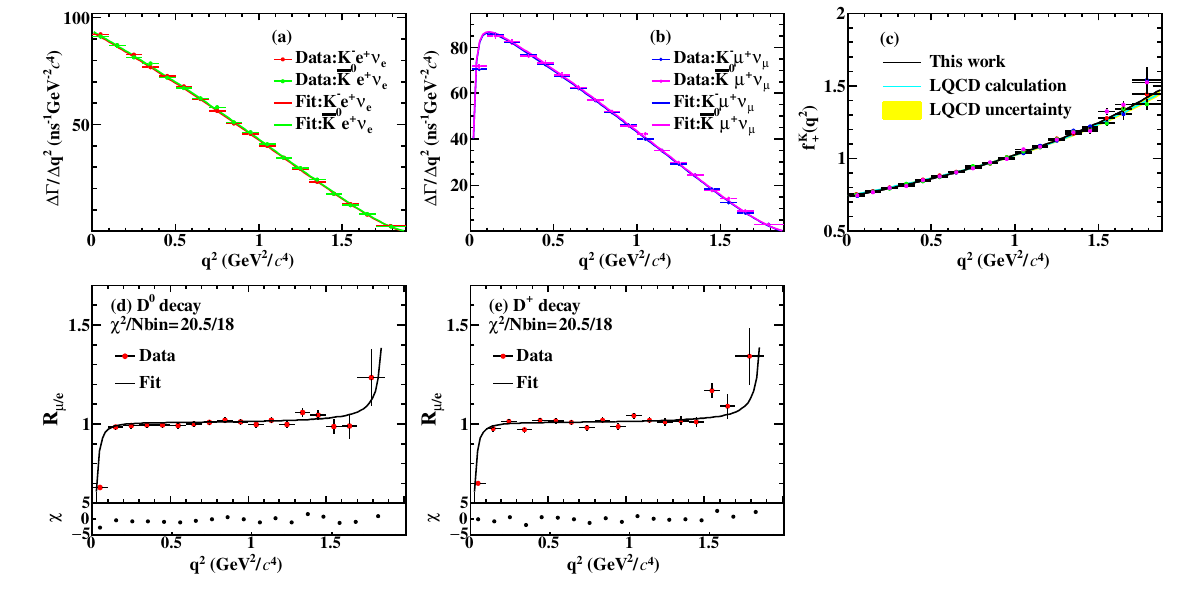}
\caption{(a)--(b) Simultaneous fit to the partial decay rates of $D^0(D^+)\to \bar K\ell^+\nu_\ell$. (c) Hadronic form factor $f_+(q^2)$ as a function of $q^2$ for the four signal modes. (d)--(e) Ratios of differential decay rates of $\kmunu$ over $\kenu$ and $\ksmunu$ over $\ksenu$ in each $q^2$ bin, respectively. The dots with error bars are data, and the solid lines are the results with the parameters of the simultaneous fit. The colors red, green, blue, and purple are for $\kenu$, $\ksenu$, $\kmunu$, and $\ksmunu$, respectively. The fit quality is available in Table~\ref{tab:fit_par}.}
  
  %The bottom sub-figures of (e)(f) show the residual $(R^{\rm dt}-R^{\rm fit})/\sigma_{R^{\rm dt}}$, where $R^{\rm dt}$ and $R^{\rm fit}$ are the ratios of decay rates between muon and positron channels obtained from data sample and calculated based on fit results, respectively.

\label{ff_klnu}
\end{center}
\end{figure*}

The fit projections of the simultaneous fit are shown in
Fig.~\ref{ff_klnu}. The parameters obtained are
$f_+(0)|V_{cs}|=0.7183\pm0.0007_{\rm stat}\pm0.0014_{\rm syst}$ and
$r_1(t_0)=-2.28\pm0.02_{\rm stat}\pm0.02_{\rm syst}$.

\section{Forward-Backward asymmetries}

\subsection{Theoretical formula}

In addition to the partial decay rate, the angular distribution of the
semileptonic decay $\klnu$ is of particular interest because of its
sensitivity to the potential scalar current contribution in the $c\to
s\ell^+\nu_\ell$ transition. The angular observable forward-backward
asymmetry is defined as \begin{equation} A_{\rm
FB}(q^2)=\frac{d\Gamma^{\ell}(\cos\theta_{W}>0)-d\Gamma^{\ell}(\cos\theta_{W}<0)}{d\Gamma^{\ell}(\cos\theta_{W}>0)+d\Gamma^{\ell}(\cos\theta_{W}<0)}.
\end{equation} Here, $\theta_{W}$ is the angle between the lepton
momentum and the direction opposite to the $D$-meson momentum in the
$\ell\nu_\ell$ rest frame. The theoretical expression of $A_{\rm
FB}(q^2)$ is written as~\cite{Fajfer:2015ixa,Faustov:2019mqr}
\begin{equation} \label{eq:afb} \small A_{\rm FB}(q^2) =
\frac{3\mathcal{N}(q^2)}{2}\frac{1}{d\Gamma^{\ell}/dq^2}\left(1-\frac{m_\ell^2}{q^2}\right)^2\frac{m_\ell^2}{q^2}
\Re\left(h_0(q^2)h^*_t(q^2)\right), \end{equation} where the
definitions of variables are exactly the same as those in
Sec.~\ref{th_formula}. In the SM, this asymmetry is proportional to
the squared lepton mass $m^2_{\ell}$, resulting in a trivial
distribution $A^{e}_{\rm FB}(q^2)=0$ for the positron channels in the
full $q^2$ range. For the muon channels, the overall asymmetry is
predicted to be $-0.053$~\cite{Faustov:2019mqr} and
$-0.055\pm0.002$~\cite{Fajfer:2015ixa}.

With the scalar currents included, the hadronic helicity amplitude $h_{t}(q^2)$ shown in Eq.~\ref{eq:hadron_hel_amp} is modified as
\begin{equation}
\label{eq:mod_hel_amp}
h_t(q^2) = \left(1+c_s^\ell\frac{q^2}{m_\ell(m_s-m_c)}\right)\frac{m_D^2-m_K^2}{\sqrt{q^2}}f_0(q^2),
\end{equation}
where $c^\ell_{S}=c^\ell_{R}+c^\ell_{L}$ is the scalar combination of Wilson coefficients defined in Eq.~\ref{eq:eft_lagr}. Hence, any deviation of $A_{\rm FB}$ from the SM prediction may suggest potential scalar current contribution in the decay $\klnu$. In this paper, both the overall and $q^2$-binned asymmetries are determined.

\subsection{Average forward-backward asymmetries}

\begin{figure*}[htbp]
	\begin{center}
		\includegraphics[width=0.99\textwidth]{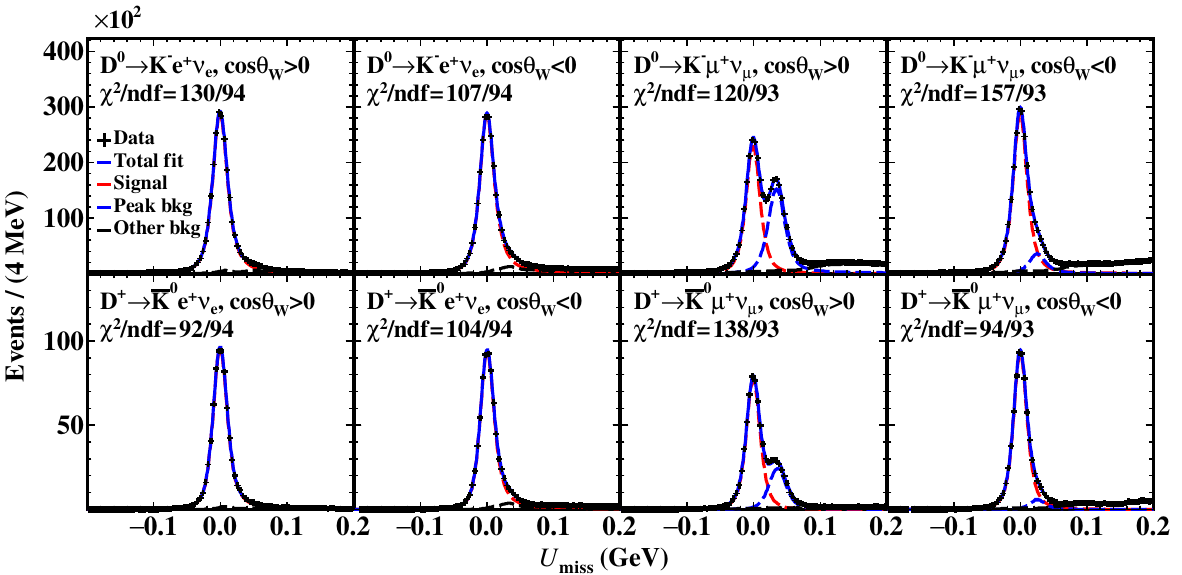}
		\caption{The $U_{\rm miss}$ distributions of the accepted forward and backward candidates of $\klnu$ in data, with fit results overlaid. The points with error bars are data; the blue solid lines denote the total fits; the red, blue, and black dashed lines show the signal, peaking background, and non-peaking background contributions, respectively.
		}
		\label{fig:umissfit_fbasym}
	\end{center}
\end{figure*}

The overall forward-backward~(FB) asymmetry $\braket{A_{\rm FB}}$ over the full region $q^2\in(q^2_{\rm min},q^2_{\rm max})$ is defined as
\begin{equation}
\label{eq:def_afb}
	\braket{A_{\rm FB}} = \int^{q^2_{\rm max}}_{q^2_{\rm min}}{A_{\rm FB}(q^2)\frac{d\Gamma(q^2)}{dq^2}dq^2}\bigg{/}\int^{q^2_{\rm max}}_{q^2_{\rm min}}{\frac{d\Gamma(q^2)}{dq^2}dq^2}
\end{equation}
and measured as
\begin{eqnarray}
\label{eq:cal_afb}
	\braket{A_{\rm FB}} = \frac{N_{\rm prd}(\cos\theta_W>0)-N_{\rm prd}(\cos\theta_W<0)}{N_{\rm prd}(\cos\theta_W>0)+N_{\rm prd}(\cos\theta_W<0)}.
\end{eqnarray}
Here, $N_{\rm prd}$ is the number of produced events obtained with $N_{\mathrm{prd}}^{\alpha}=\sum_{\beta}\left(\varepsilon^{-1}\right)_{\alpha\beta}N_{\mathrm{DT}}^{\beta}$, where the indices $\alpha,\beta$ correspond to the forward or backward regions. Here, DT yields $N_{\mathrm{DT}}$ are determined from the $U_{\rm miss}$ fits as shown in Fig.~\ref{fig:umissfit_fbasym} with fitted yields summarized in Table~\ref{tab:klnu_Nobspro_AFB_average}. The efficiency matrix $\varepsilon_{\alpha\beta}$ is determined in the same approach as Eq.~\ref{eq:eff_matrix} with detailed values shown in Table~\ref{tab:klnu_effmatrix_AFB_average}. The statistical covariance matrices of $N_{\rm prd}$ are constructed with
\begin{equation}
C_{\alpha,\beta}^{\rm stat}=\sum_{\gamma}(\varepsilon^{-1})_{\alpha\gamma}(\varepsilon^{-1})_{\beta\gamma}\left(\sigma\left(N_{\mathrm{DT}}^{\gamma}\right)\right)^{2}
\end{equation}
as summarized in Table~\ref{tab:klnu_stat_covmatrix_AFB_average}.

\begin{table}[htbp]
	\caption{The weighted efficiency matrix of overall asymmetry (in units of \%) for $\klnu$, where the indices $\alpha=1$ and $2$ correspond to the forward and backward events, respectively.}
	\scalebox{0.83}{
	\centering
	\begin{tabular}{ccc|ccc|ccc|ccc}
		\hline
		\hline
		\multicolumn{3}{c|}{$\kenu$} &\multicolumn{3}{c|}{$\kmunu$} &\multicolumn{3}{c|}{$\ksenu$}&\multicolumn{3}{c}{$\ksmunu$}\\\hline
		$\varepsilon_{\alpha\beta}$&1&2 & $\varepsilon_{\alpha\beta}$&1&2 & $\varepsilon_{\alpha\beta}$&1&2 & $\varepsilon_{\alpha\beta}$&1&2\\ \hline
		1&67.41&0.78 & 1&53.62&0.77 & 1&45.29&0.49 & 1&37.07&0.50\\
		2&2.02&66.58 & 2&0.99&60.43 & 2&1.22&44.12 & 2&0.56&39.53\\
		\hline
		\hline			
		\label{tab:klnu_effmatrix_AFB_average}
	\end{tabular}
	}
\end{table}

\begin{table}[htbp]
	\caption{The statistical covariance matrix of $N_{\rm pro}$ for $\klnu$, where the indices $\alpha=1$ and $2$ correspond to the forward and backward events, respectively.}
	\scalebox{0.85}{
	\centering
	\begin{tabular}{ccc|ccc|ccc|ccc}
		\hline
		\hline
		\multicolumn{3}{c|}{$\kenu$} &\multicolumn{3}{c|}{$\kmunu$} &\multicolumn{3}{c|}{$\ksenu$}&\multicolumn{3}{c}{$\ksmunu$}\\\hline
		$\rho_{\alpha\beta}$&1&2 & $\rho_{\alpha\beta}$&1&2 & $\rho_{\alpha\beta}$&1&2 & $\rho_{\alpha\beta}$&1&2\\ \hline
		1&1.00&-0.04 & 1&1.00&-0.03 & 1&1.00&-0.04 & 1&1.00&-0.03\\
		2&-0.04&1.00 & 2&-0.03&1.00 & 2&-0.04&1.00 & 2&-0.03&1.00\\
		\hline
		\hline			
		\label{tab:klnu_stat_covmatrix_AFB_average}
	\end{tabular}
	}
\end{table}

Some of the systematic uncertainties of $N_{\rm prd}$ cancel in the calculation of $A_{\rm FB}$, except those from the $U_{\rm miss}$ fit, $\ell^+$ tracking and PID~(Trk/PID), requirements on $M_{K\ell}$~(negligible), and MC statistics. All of these uncertainties are estimated using the same approach as described in Sec.~\ref{sec:sys_ddr}, and the detailed values are summarized in Table~\ref{tab:klnu_AFB_average_syserr}.

\begin{table}[htbp]
	\caption{The systematic uncertainties~($\times10^{-4}$) of the overall forward-backward asymmetry $\braket{A_{\rm FB}}$.}
	\centering
	\scalebox{0.92}{
		\begin{tabular}{c|c|c|c|c}
			\hline
			\hline
			 Signal decay& $U_{\rm miss}$ fit & $\ell^+$ Trk/PID & MC statistics & Total\\ \hline
			 $\kenu$ & 16.0 & $0.1$& 3.1 & 16.3\\
			 $\kmunu$ & 22.1 & 0.8 & 3.4 & 22.3\\
			 $\ksenu$ & 14.9 & $0.1$ & 3.7 & 15.5\\
			 $\ksmunu$ & 5.1 & 0.9 & 4.1 & 6.4\\
			\hline
			\hline
		\end{tabular}
	}
	\label{tab:klnu_AFB_average_syserr}
\end{table}

With the measured $N_{\rm prd}$, the overall asymmetries $\braket{A_{\rm FB}}$ are determined, and the corresponding uncertainties are estimated with
\begin{equation}
\left[\sigma(\braket{A_{\rm FB}})\right]^2 = \sum_{\alpha,\beta}{\frac{\partial \braket{A_{\rm FB}}}{\partial N_{\rm prd}^{\alpha}}\frac{\partial \braket{A_{\rm FB}}}{\partial N_{\rm prd}^{\beta}}\times C_{\alpha\beta}}.
\end{equation}
Here, the statistical uncertainties are obtained with $C_{\alpha\beta}=C^{\rm stat}_{\alpha\beta}$, while the systematic ones are assigned as the quadratic difference between the uncertainties obtained with $C_{\alpha\beta}=C^{\rm stat}_{\alpha\beta}+C^{\rm syst}_{\alpha\beta}$ and $C_{\alpha\beta}=C^{\rm stat}_{\alpha\beta}$. The results are summarized in Table~\ref{tab:klnu_Nobspro_AFB_average}. No significant deviation from the theoretical predictions~\cite{Faustov:2019mqr,Fajfer:2015ixa} is observed.

\begin{table}[htbp]
	\caption{The fitted signal yields for $\klnu$, and the overall forward-backward asymmetry $\braket{A_{\rm FB}}$, where the first uncertainties are statistical and the second are systematic.}
	\centering
	\scalebox{0.90}{
		\begin{tabular}{c|c|c|c}
			\hline
			\hline
			Signal decay& & $N_{\rm DT}$ &  $\braket{A_{\rm FB}}(\times 10^{-3})$\\ \hline
			\multirow{2}{*}{$\kenu$}&$\cos\theta_l>0$&$244569\pm550$ & \multirow{2}{*}{$+0.2\pm1.6\pm1.6$}\\
			& $\cos\theta_l<0$&$245113\pm545$\\ \hline
			\multirow{2}{*}{$\kmunu$}&$\cos\theta_l>0$&$178328\pm539$ & \multirow{2}{*}{$-59.2\pm2.2\pm2.2$}\\ 
			& $\cos\theta_l<0$&$225896\pm696$\\ \hline
			\multirow{2}{*}{$\ksenu$}&$\cos\theta_l>0$&$75066\pm293$ & \multirow{2}{*}{$+0.0\pm2.9\pm1.5$}\\
			&$\cos\theta_l<0$&$74297\pm294$\\ \hline
			\multirow{2}{*}{$\ksmunu$}&$\cos\theta_l>0$&$57122\pm281$ & \multirow{2}{*}{$-53.1\pm3.5\pm0.7$}\\
			&$\cos\theta_l<0$&$67591\pm329$\\
			\hline
			\hline
		\end{tabular}
	}
	\label{tab:klnu_Nobspro_AFB_average}
\end{table}

\subsection{\boldmath $q^2$-binned forward-backward asymmetries}

\begin{figure*}[htbp]
	\begin{center}	
		\subfigure{\includegraphics[width=0.49\textwidth]{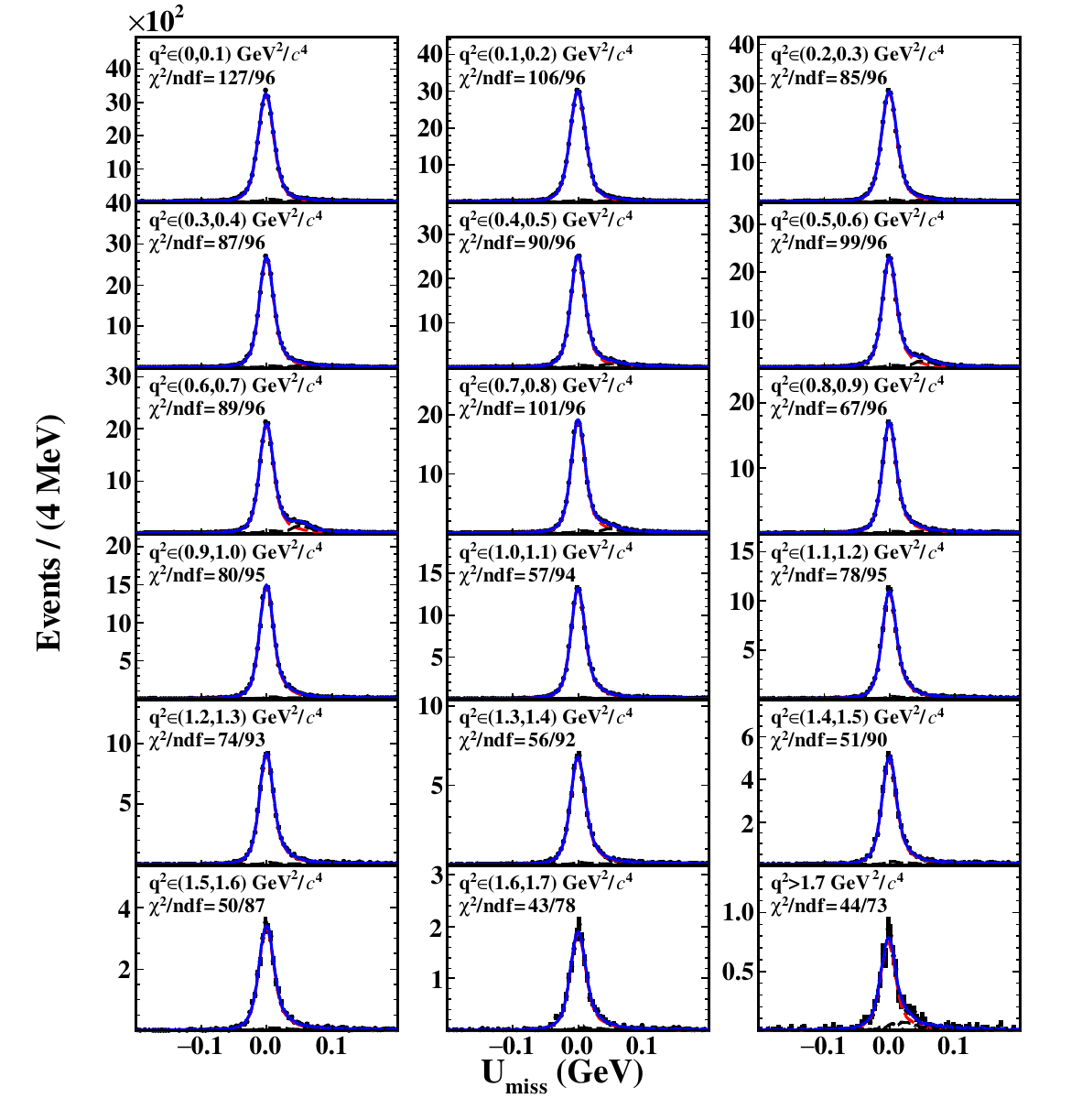}}
		\subfigure{\includegraphics[width=0.49\textwidth]{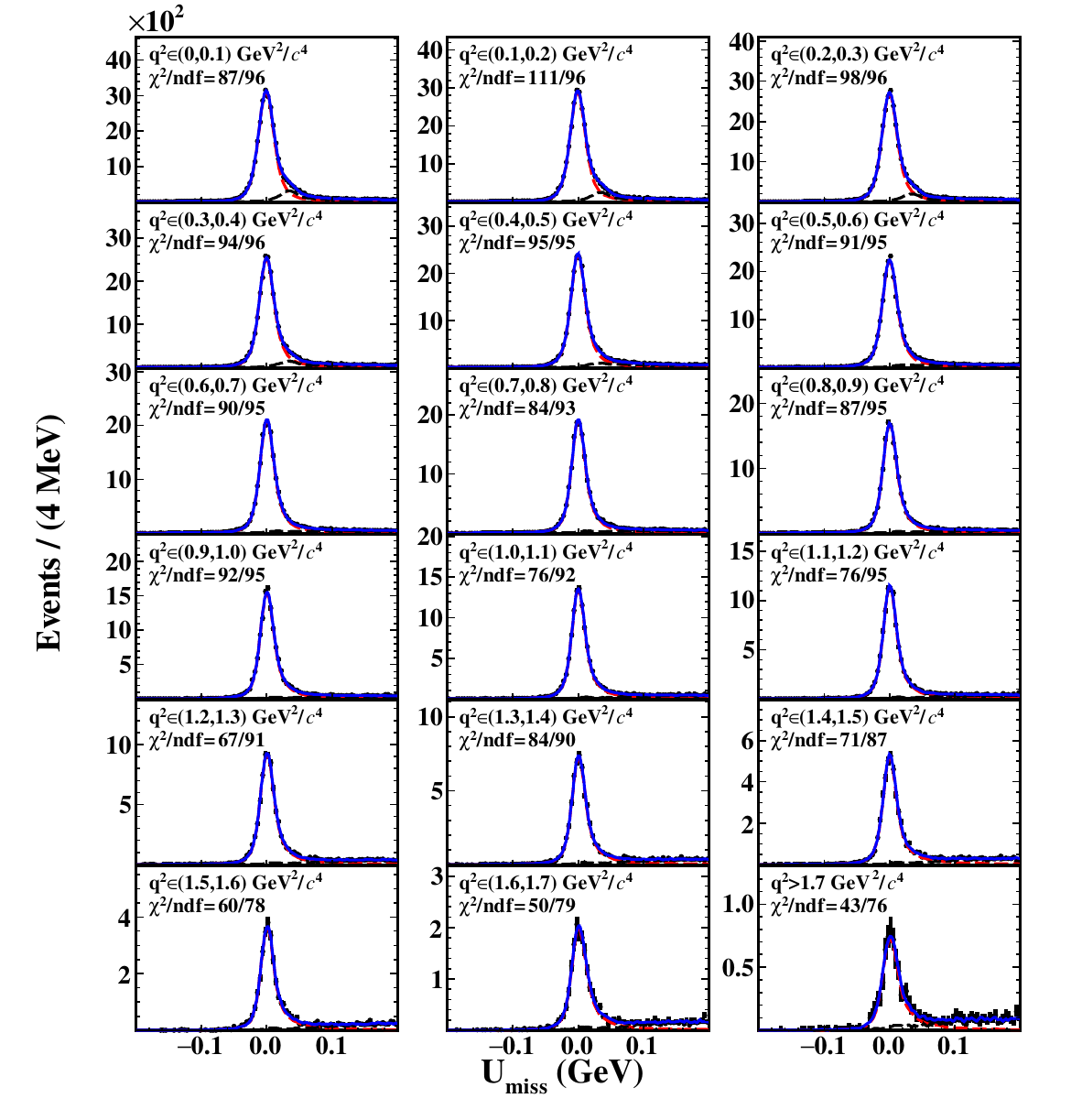}}
		\caption{The $U_{\rm miss}$ distributions of the accepted forward (left) and backward (right) candidate events in different $q^2$ bins for $\kenu$ in data, with fit results overlaid as the blue solid curves. The red dashed curves are the signal shapes, and the black dashed curves are the fitted combinatorial background shapes.}
		\label{kenu_umiss_AFB_q2}
	\end{center}
\end{figure*}

The forward-backward asymmetry in the $i$-th $q^2$ interval is defined as
\begin{equation}
A_{{\rm FB},i}=\int^{q^2_{{\rm max}(i)}}_{q^2_{{\rm min}(i)}}{A_{{\rm FB}}(q^2)\frac{d\Gamma}{dq^2} dq^2}\bigg{/}\int^{q^2_{{\rm max}(i)}}_{q^2_{{\rm min}(i)}}{\frac{d\Gamma}{dq^2} dq^2},
\end{equation}
where the $q^2$ intervals are the same as that of Sec.~\ref{sec:ddr_measure}. Similar to the measurement of the overall asymmetry, the number of produced events in the $i$-th $q^2$ interval is obtained with
\begin{equation}
	N_{\mathrm{prd}}^{i,\alpha}=\sum_{(j,\beta)}\left(\varepsilon^{-1}\right)_{(i,\alpha)(j,\beta)}N_{\mathrm{DT}}^{j,\beta},
\end{equation}
where $N^{\rm prd}$ is the number of produced events, and the indices $\alpha$ and $\beta$ indicate forward and backward categories, respectively. The results of the fits to the $U_{\rm miss}$ distributions of $\kenu$ are shown in Fig.~\ref{kenu_umiss_AFB_q2} with DT yields $N_{\rm DT}$ summarized in Table~\ref{tab:kenu_AFB_q2_stat}. The results for the other three decay modes are provided in Figs.~\ref{kmunu_umiss_AFB_q2}$-$\ref{ksmunu_umiss_AFB_q2} and Tables~\ref{tab:kmunu_AFB_q2_stat}$-$\ref{tab:ksmunu_AFB_q2_stat} in the \hyperref[appendix]{Appendix}. The systematic uncertainties are estimated in the same approach as those of overall asymmetries.

\begin{figure}[htbp]
	\begin{center}
		\includegraphics[width=\linewidth]{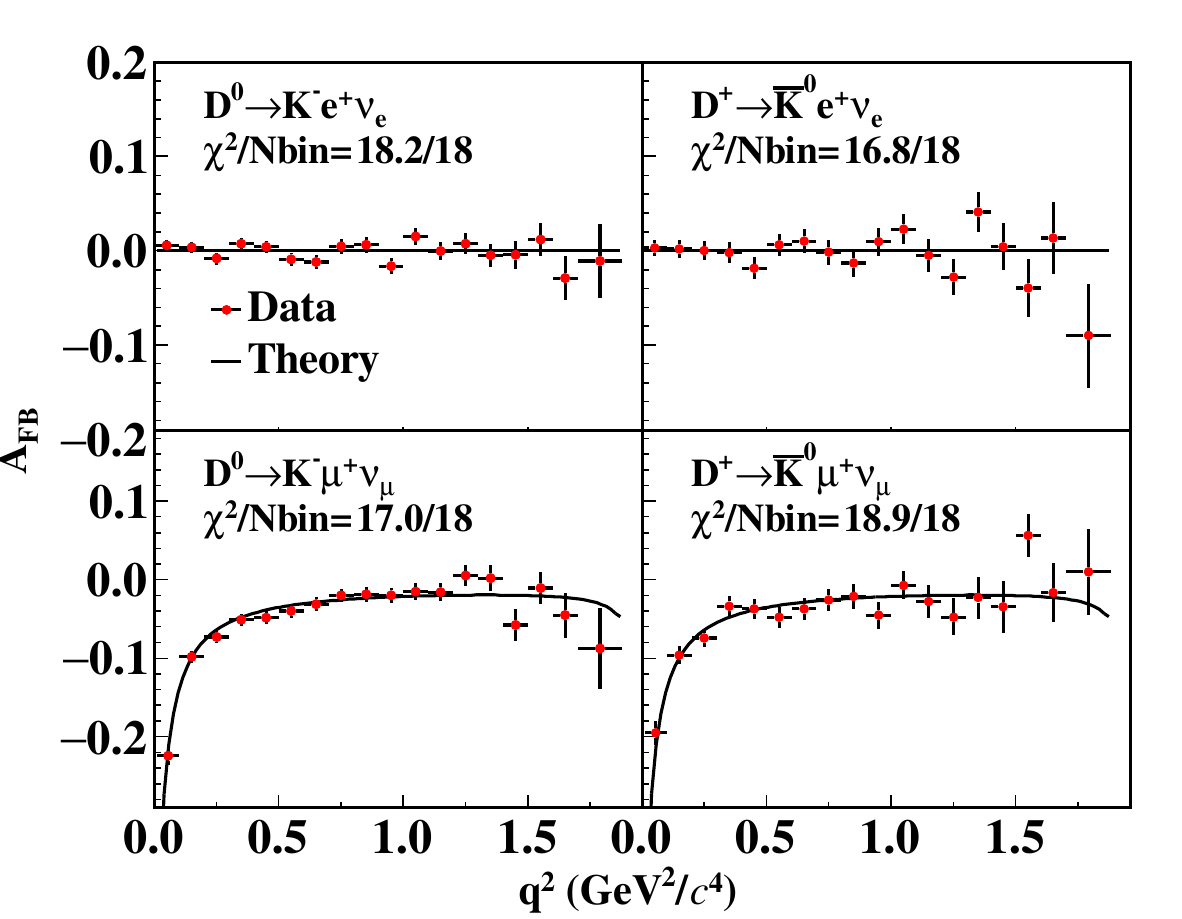}
		\caption{The measured $q^2$-binned forward and backward asymmetries. The red points with error bars are data. The black solid lines denote the theoretical curves.}
		\label{fig:afb_q2}
	\end{center}
\end{figure}

Table~\ref{tab:kenu_AFB_q2_stat} summarizes the measured $q^2$-binned forward-backward asymmetries of $\kenu$, and those of the other three channels are listed in Tables~\ref{tab:kmunu_AFB_q2_stat}$-$\ref{tab:ksmunu_AFB_q2_stat} of the \hyperref[appendix]{Appendix}. Figure~\ref{fig:afb_q2} shows the measured $q^2$-binned forward-backward asymmetries, which are found to be in good agreement with the theoretical predictions.

\begin{table}
	\caption{The numbers of observed forward/backward events $N_{\rm DT}$ in different $q^2$ intervals and the $q^2$-binned forward-backward asymmetries of $\kenu$, where the first uncertainties are statistical and the second are systematic.}
	\scalebox{0.85}{
		\centering
		\begin{tabular}{c|c|c|cc}
			\hline
			\hline
			$q^2({\rm GeV}/c^4)$&$N_{\rm DT}^{\rm Forward}$&$N_{\rm DT}^{\rm Backward}$&$A_{\rm FB}$\\ \hline
			$(0.0,0.1)$&  $27827\pm178$&$27482\pm186$&$0.006\pm0.005\pm0.002$ \\
			$(0.1,0.2)$&  $25812\pm171$&$25316\pm178$&$0.004\pm0.006\pm0.002$ \\
			$(0.2,0.3)$&  $23978\pm166$&$23913\pm173$&$-0.008\pm0.006\pm0.002$ \\
			$(0.3,0.4)$&  $22356\pm161$&$21894\pm163$&$0.007\pm0.006\pm0.002$ \\
			$(0.4,0.5)$&  $20925\pm158$&$20703\pm157$&$0.004\pm0.007\pm0.002$ \\
			$(0.5,0.6)$&  $19054\pm151$&$19473\pm151$&$-0.009\pm0.007\pm0.002$ \\
			$(0.6,0.7)$&  $17350\pm144$&$17868\pm143$&$-0.012\pm0.007\pm0.002$ \\
			$(0.7,0.8)$&  $15833\pm139$&$15992\pm135$&$0.005\pm0.008\pm0.002$ \\
			$(0.8,0.9)$&  $14155\pm131$&$14357\pm127$&$0.006\pm0.008\pm0.002$ \\
			$(0.9,1.0)$&  $12414\pm121$&$13074\pm121$&$-0.016\pm0.008\pm0.002$ \\
			$(1.0,1.1)$&  $11160\pm114$&$11210\pm112$&$0.015\pm0.009\pm0.002$ \\
			$(1.1,1.2)$&  $9303\pm105$&$9593\pm104$&$-0.000\pm0.010\pm0.002$ \\
			$(1.2,1.3)$&  $7846\pm96$&$7957\pm95$&$0.008\pm0.011\pm0.002$ \\
			$(1.3,1.4)$&  $6011\pm84$&$6233\pm84$&$-0.005\pm0.012\pm0.003$ \\
			$(1.4,1.5)$&  $4381\pm72$&$4546\pm72$&$-0.005\pm0.014\pm0.003$ \\
			$(1.5,1.6)$&  $3055\pm61$&$3090\pm60$&$0.012\pm0.017\pm0.004$ \\
			$(1.6,1.7)$&  $1727\pm46$&$1850\pm47$&$-0.029\pm0.023\pm0.005$ \\
			$>1.7$&  $677\pm31$&$732\pm30$&$-0.011\pm0.039\pm0.008$ \\
			\hline
			\hline
		\end{tabular}
	}
	\label{tab:kenu_AFB_q2_stat}
\end{table}

\section{\boldmath Constraint on scalar current in $c\to s\ell^+\nu_{\ell}$ transition}

As shown in Eq.~\ref{eq:mod_hel_amp}, the presence of potential scalar current in the $c\to s\ell^+\nu_{\ell}$ transition results in a modified $h_t(q^2)$. Hence, a simultaneous fit to the $q^2$-binned partial decay rates and forward-backward asymmetries of $\klnu$ is performed for a stringent constraint on the parameter space of $c^\ell_s$. This fit is performed using the least $\chi^2$ method with the objective function constructed as
\begin{equation}
\chi^2 = \chi^2_{\Delta \Gamma} + \sum{\chi^2_{A_{\rm FB,\klnu}}}.
\end{equation}
Here, the decay rate part $\chi^2_{\Delta \Gamma}$ is identical to that defined in Sec.~\ref{sec:ddr_smfit}. The forward-backward asymmetry part $\sum{\chi^2_{A_{\rm FB,\klnu}}}$ sums over all four decay channels, with the $\chi^2_{A_{\rm FB}}$ defined as
\begin{equation}
\sum_{i,j}{(A_{{\rm FB},i}^{\rm msr}- A_{{\rm FB},i}^{\rm fit})(C^{-1}_{\rm FB})_{ij}( A_{{\rm FB},j}^{\rm msr}- A_{{\rm FB},j}^{\rm fit})},
\end{equation}
where the small correlations between different decay channels due to the correlated systematic uncertainties are neglected.

In this fit, in addition to the parameters $f_+(0)|V_{cs}|$ and $r_1(t_0)$, the real and imaginary parts of the complex coefficient $c^\mu_{S}$ of the muon channels are also treated as free parameters. For the positron channels, the partial decay rate is only sensitive to the modulus $|c_S^{e}|$, while the forward-backward asymmetry is insensitive to the scalar current contribution. Therefore, only a single parameter $|c_S^{e}|$ is included in the fit. The different sensitivities of positron and muon channels  can be understood as follows. Under the limitation $m_\ell\to0$ for the positron channels, the forward-backward asymmetries in Eq.~\ref{eq:afb} are expected to be always zero, and hence can not constrain the real part of $c^{e}_S$. For the decay rates, the $|h_t(q^2)|^2$ in Eq.~\ref{ffk_function1} is proportional to $|c^{e}_S|^2$, so only the $|c^{e}_S|$ can be determined from the fit.

\begin{figure*}[htbp]
	\centering
	\setlength{\abovecaptionskip}{0.cm}
	\includegraphics[width=18.0cm]{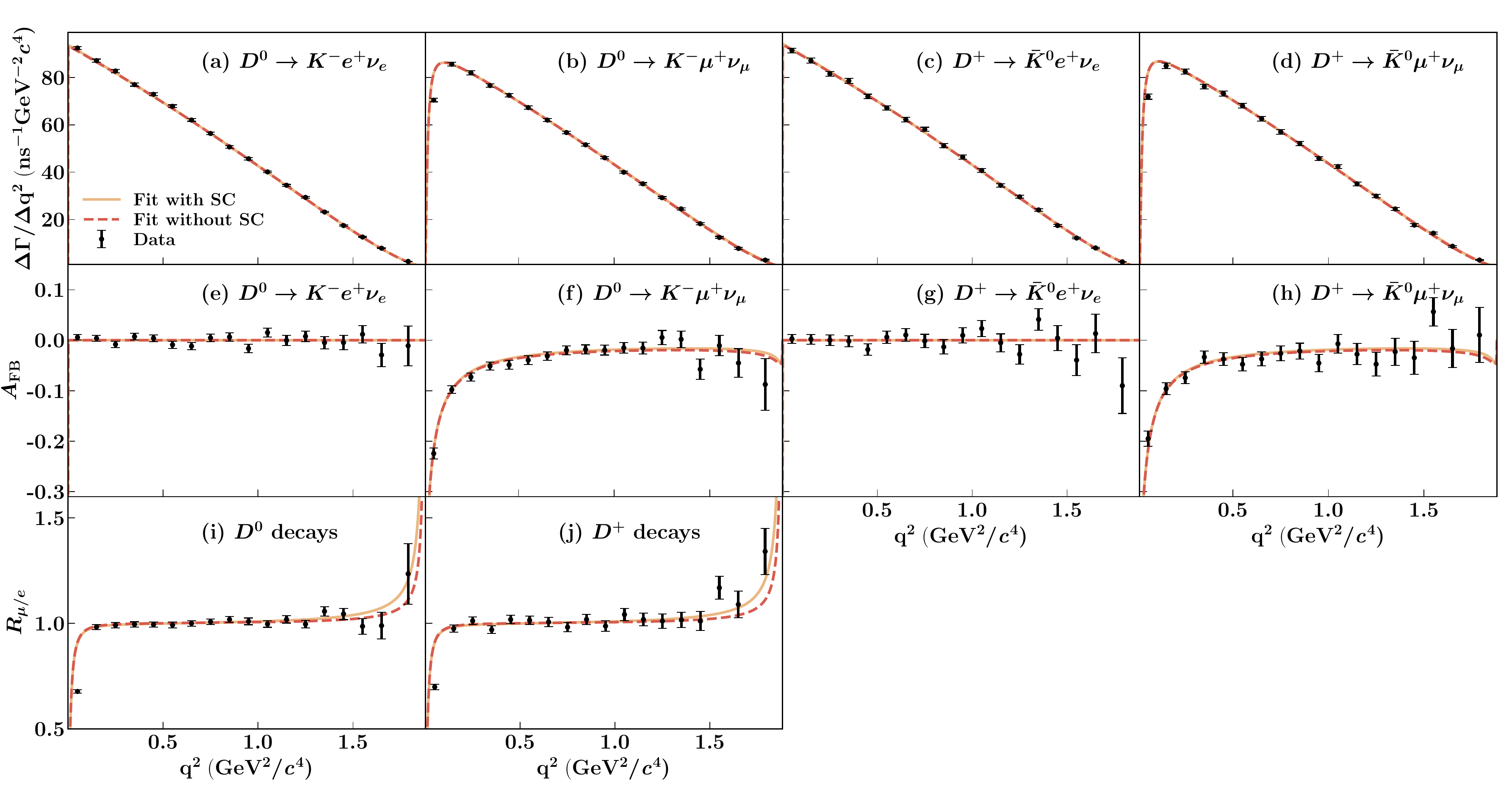}
	\caption{The simultaneous fit to the measured partial decay rates~(top) and forward-backward asymmetries~(middle) of $\klnu$, and the fit projections on the ratios of differential decay rates $\mathcal{R}_{\mu/e}$~(bottom). The black points with red error bars are data, the solid lines are the fit projections with scalar current~(SC) contribution included, and the dashed lines are the fit projections without scalar current contribution. The fit qualities are available in Table~\ref{tab:fited_pars}.}
	\label{fig:combined_fit}
\end{figure*}

\begin{figure*}[htbp]
	\centering
	\setlength{\abovecaptionskip}{0.cm}
	\includegraphics[width=18.6cm]{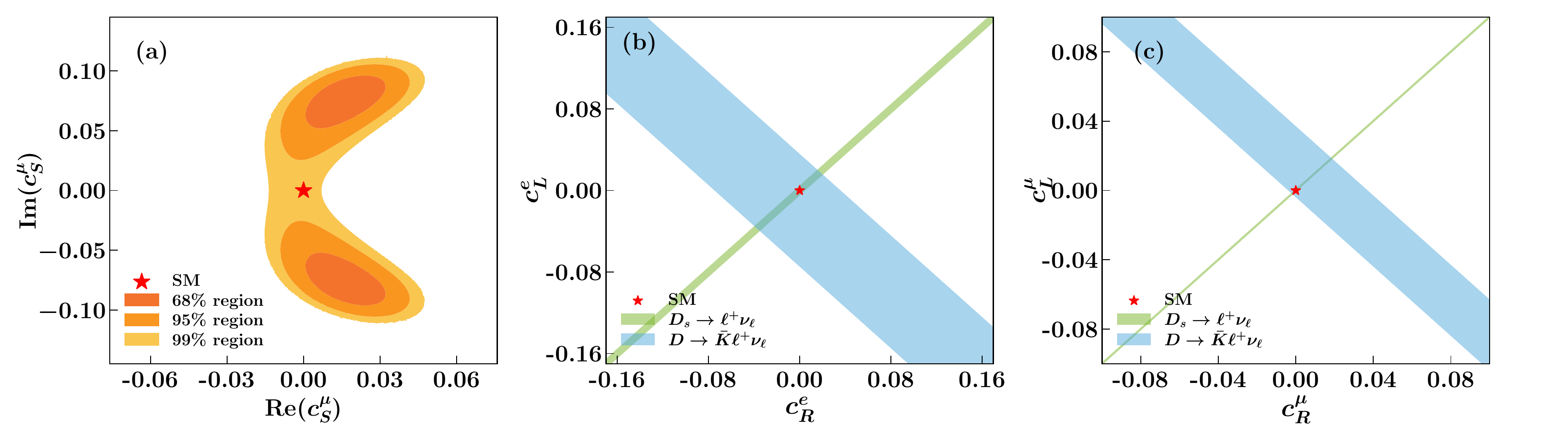}
	\caption{The confidence regions of (a) the scalar combination of complex Wilson coefficients $c_S^\mu$ with probabilities of 68\%, 95\%, and 99\% from $\klnu$, as well as (b)(c) the right- and left-handed real Wilson coefficients $c^\ell_{R}$ and $c^\ell_{L}$ with probabilities of 95\% by combining with $D_s^+\to\ell^+\nu_\ell$. The red dot indicates the SM value.}
	\label{fig:contour_muon}
\end{figure*}

First, two different fits are performed with and without the
scalar current contribution, with the $c_S^\ell$ shown in
Eq.~\ref{eq:mod_hel_amp} allowed to float and fixed to zero, respectively. Corresponding fit projections are shown in Fig.~\ref{fig:combined_fit},
and the fitted parameters are summarized in Table~\ref{tab:fited_pars}. No significant deviation from the SM is found
in the positron channels with the fitted $|c^e_S|$ consistent with zero. For the muon channels, $c^\mu_S$ is found to deviate from zero as shown in Fig.~\ref{fig:contour_muon}(a), especially on its imaginary part.

Given the LFU may not hold for the scalar current with possible dependence on lepton mass, its coupling with muon could be significantly larger than that of the positron. Therefore, to estimate the significance of non-zero $c^\mu_S$ alone, the third fit is performed by floating $c^\mu_S$ and fixing $|c^e_S|=0$ as the alternative hypothesis, which yields $\chi^2/{\rm ndf}=135.9/140$. Compared to the null hypothesis without a scalar current with $\chi^2/{\rm ndf}=143.5/142$, the changes $\Delta\chi^2=7.6$ and $\Delta {\rm ndf}=2$ suggest a significance of 2.3$\sigma$ based on the Wilks' theorem.

%Two separate fits are performed under the hypotheses with and without
%scalar current contribution, respectively. The $c_S^\ell$ shown in
%Eq.~\ref{eq:mod_hel_amp} is fixed to zero in the fit without scalar
%current, and is allowed to float in the fit with scalar current. The
%corresponding fit projections are shown in Fig.~\ref{fig:combined_fit}
%and the fitted parameters are summarized in
%Table~\ref{tab:fited_pars}. Based on the fit with scalar current
%contribution included, no significant deviation from the SM
%expectation is found in the positron channels, while a small
%non-zero $c^\mu_S$ is observed for the muon channels, as shown in
%Fig.~\ref{fig:contour_muon}(a). The significance of the scalar current
%in the $c\to s\mu^+\nu_\mu$ transition is estimated by performing an
%alternative fit, in which the scalar current is included only for the
%muon channels. The resultant fit quality $\chi^2/{\rm ndf}=133.0/140$
%suggests a significance 1.9$\sigma$, by making comparison to the null
%hypothesis without scalar current with $\chi^2/{\rm ndf}=138.5/142$.

\begin{table}[htbp]
	\centering
	\setlength\tabcolsep{6pt}
	\caption{Fitted parameters and overall fit qualities of fits with and without scalar current~(SC) contribution shown in Fig.~\ref{fig:combined_fit}, where the first uncertainties are statistical and the second are systematic.}
	\scalebox{0.92}{
	\begin{footnotesize}
		\begin{tabular}{ccc}
			\hline
			\hline
			Variable & With SC & Without SC \\ \hline
			$f_{+}(0)|V_{cs}|$ & $0.7193\pm0.0008\pm0.0014$ &  $0.7183\pm0.0007\pm0.0014$ \\
			$r_1$ & $-2.23\pm0.04\pm0.02$ &  $-2.28\pm0.02\pm0.02$\\
			$|c_S^{e}|$ & $0.02\pm0.02\pm0.02$ & ---\\
			${\rm Re}(c_S^{\mu})$  & $0.017\pm0.008\pm0.006$ & ---\\
			${\rm Im}(c_S^{\mu})$  & $\pm(0.077\pm0.011\pm0.009)$ & --- \\
			$\chi^2/{\rm ndf}$ & $135.8/139$ & $143.5/142$ \\
			\hline
			\hline
		\end{tabular}
	\end{footnotesize}
	}
	\label{tab:fited_pars}
\end{table}

Further constraints on the right- and left-handed components of the
scalar current are obtained by combining with the world average rates of
$D_s^+\to\ell^+\nu_\ell$, which are sensitive to the pseudoscalar
combination of the Wilson coefficients
$c^\ell_P=c^{\ell}_R-c^{\ell}_L$ as~\cite{Fajfer:2015ixa}
\begin{equation} \begin{split}
\mathcal{B}(D_s^+\to\ell^+\nu_\ell)&=\tau_{D_s}\frac{m_{D_s}}{8\pi}f^2_{D_s}\left(1-\frac{m_\ell^2}{m^2_{D_s}}\right)^2
G^2_F \\ &\times|V_{cs}|^2 m^2_{\ell}
|1-c^\ell_P\frac{m^2_{D_s}}{(m_c+m_s)m_\ell}|^2.  \end{split}
\end{equation} Here, the measured branching fractions
$\mathcal{B}(D_s^+\to \ell^+\nu_\ell)$ are taken from the
PDG~\cite{ParticleDataGroup:2024cfk}, and the inputs of
$f_{D_s}=(249.9\pm0.5)$ MeV from LQCD
calculations~\cite{Bazavov:2017lyh} and $|V_{cs}|=0.97349\pm0.00016$
from the global SM fit~\cite{ParticleDataGroup:2024cfk} are
used. Since the $D_s^+\to\ell^+\nu_\ell$ decay rates can not constrain
the real and imaginary parts of $c^\ell_P$ simultaneously, the Wilson
coefficient $c^\ell_{R(L)}$ is assumed to be real with T-parity
conserved here. As shown in Figs.~\ref{fig:contour_muon}(b)
and~\ref{fig:contour_muon}(c), both the right- and left-handed Wilson
coefficients $c^\ell_{R}$ and $c^\ell_{L}$ are consistent with zero.

\section{Summary}

In summary, using $e^+e^-$ collision data corresponding to an
integrated luminosity of 20.3 fb$^{-1}$, collected at
$\sqrt{s}=3.773$~GeV with the BESIII detector, we perform precise
measurements of the semileptonic decays $\klnu$. The absolute
branching fractions of $\kenu$, $\kmunu$, $\koenu$, and $\komunu$ are
determined to be $(3.548\pm0.006_{\rm
	stat}\pm0.017_{\rm syst}) \%$, $(3.445\pm0.007_{\rm
	stat}\pm0.017_{\rm syst}) \%$, $(8.928\pm0.025_{\rm
	stat}\pm0.050_{\rm syst}) \%$, and $(8.770\pm0.029_{\rm
	stat}\pm0.053_{\rm syst}) \%$,
respectively. The ratios of the BFs between semi-muonic and
semi-electronic $D$ decays are determined to be
$\mathcal{R}^{D^0}_{\mu/e}=0.971\pm0.003_{\rm stat}\pm0.004_{\rm
syst}$ and $\mathcal{R}^{D^+}_{\mu/e}=0.982\pm0.004_{\rm
stat}\pm0.002_{\rm syst}$, which exhibit a slight difference with the
LFU prediction of $0.975\pm0.001$~\cite{Riggio:2017zwh} up to
1.5$\sigma$.

\begin{figure}[htpb]
	\centering
	\setlength{\abovecaptionskip}{-2pt}
	\setlength{\belowcaptionskip}{-3pt}
	\includegraphics[width=0.49\textwidth]{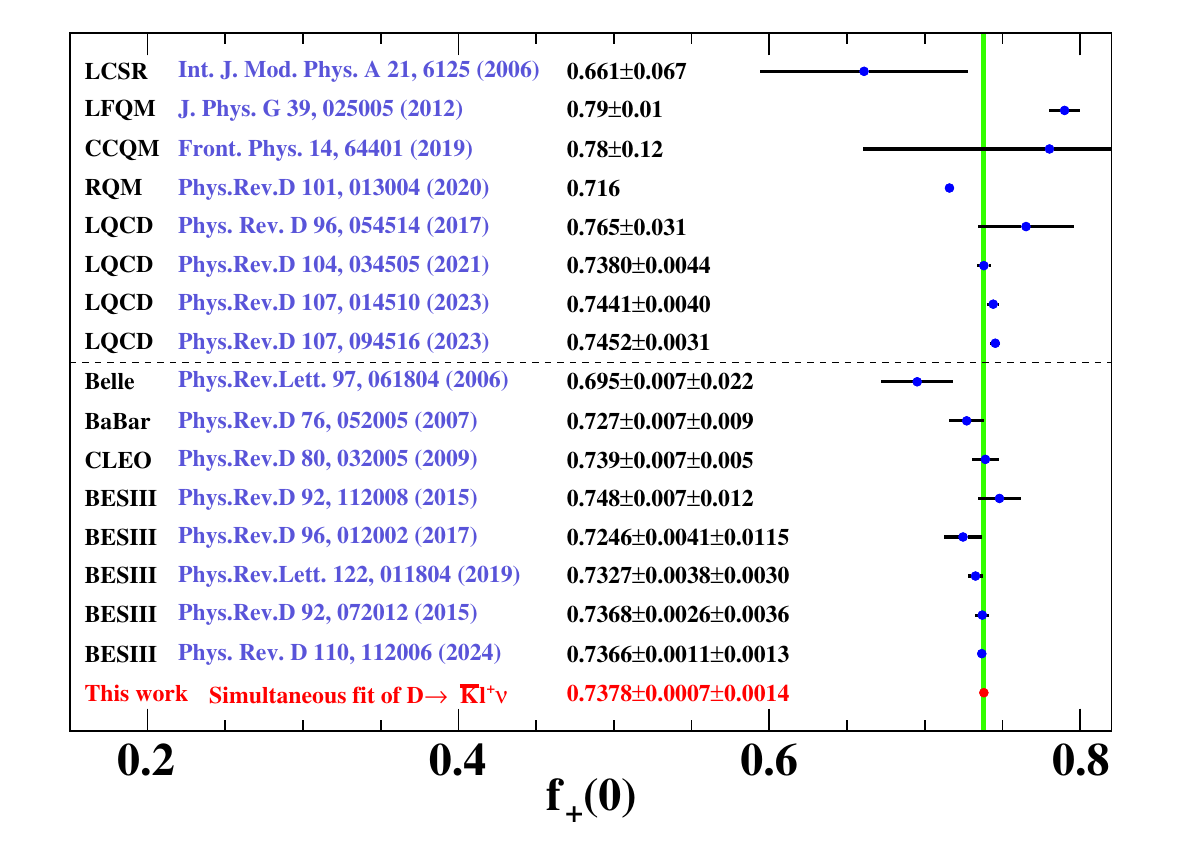}
	\caption{Comparison of the form factor $f_+(0)$ measured in this work with the theoretical and experimental calculations.
		The first and second uncertainties are statistical and systematic, respectively.  The green band corresponds to the $\pm1\sigma$ limit of the form factor calculated in this work.
		\label{compare_ff_klnu}
	}
\end{figure}

Based on the simultaneous fit to the partial decay rates of $\kenu$,
$\kmunu$, $\koenu$, and $\komunu$, the product of the hadronic form
factor $\ffK$ and the modulus of the CKM matrix element $|V_{cs}|$ is
determined to be $\ffK|V_{cs}|=0.7183\pm0.0007_{\rm
stat}\pm0.0014_{\rm syst}$. Taking the value of $|V_{cs}| =
0.97349\pm0.00016$ given by the PDG~\cite{ParticleDataGroup:2024cfk}
as input, the hadronic form factor is obtained to be
$\ffK=0.7383\pm0.0007_{\rm stat}\pm0.0014_{\rm syst}$. A comparison
of $\ffK$ values measured in different experiments and calculated in
various theoretical approaches is shown in Fig.~\ref{compare_ff_klnu}.
This work yields improved precision compared to the previous BESIII
measurements~\cite{BESIII:2015tql,BESIII:2018ccy,BESIII:2017ylw,BESIII:2015jmz,BESIII:2024slx},
providing an important benchmark for testing different theoretical
calculations. Conversely, taking the LQCD determination
$f_+(0)=0.7452\pm0.0031$~\cite{FermilabLattice:2022gku}, the CKM
matrix element is determined to be $|V_{cs}| = 0.9639\pm0.0009_{\rm
stat}\pm0.0018_{\rm syst}\pm0.0040_{\rm LQCD}$, which enables a
stringent test of CKM unitarity via direct measurement.

Furthermore, the overall and $q^2$-binned angular observables of forward-backward asymmetries $A_{\rm FB}$ are measured for $\klnu$ for the first time. Through a simultaneous fit to the $q^2$-binned partial decay rates and forward-backward asymmetries, the scalar current contribution in the $c\to s\ell^+\nu_{\ell}$ transition is searched for. The first experimental constraint on the scalar combination of complex Wilson coefficients is obtained to be ${\rm Re}(c_S^{\mu})=0.017\pm0.008_{\rm stat}\pm0.006_{\rm syst}$ and ${\rm Im}(c_S^{\mu})=\pm(0.077\pm0.011_{\rm stat}\pm0.009_{\rm syst})$, corresponding to a deviation from the SM with a significance of $2.3\sigma$. This result provides an important constraint on the scalar current contribution in the semileptonic $D^{0(+)}$ decays, offers new insights into the anomalies observed in the beauty sector~\cite{BaBar:2012obs,LHCb:2023zxo,LHCb:2023uiv,LHCb:2024jll,Belle:2015qfa,LHCb:2013ghj,Belle:2016fev,CMS:2017rzx,ATLAS:2018gqc,LHCb:2024onj}, and strengthens the search for new scalar bosons beyond the SM in various new-physics scenarios.

%The complex coupling of scalar current in weak interaction also serves as a potential new source of CP violation.

%The stringent constraint on the scalar combination of complex Wilson coefficients $c^\ell_{S}=c^\ell_{R}+c^\ell_{L}$ is obtained and no significant excess of the SM is seen beyond 2$\sigma$ region. Alongside the decay rate of $D_s^+\to\ell^+\nu_\ell$ and assuming a real $c^\ell_{R(L)}$, the right and left-handed components of scalar current in the $c\to s\ell^+\nu_\ell$ transition are further constrained, both of which are consistent with the SM.

\begin{acknowledgments}

The BESIII Collaboration thanks the staff of BEPCII (https://cstr.cn/31109.02.BEPC) and the IHEP computing center for their strong support. This work is supported in part by National Key R\&D Program of China under Contracts Nos. 2023YFA1606000, 2023YFA1606704; National Natural Science Foundation of China (NSFC) under Contracts Nos. 11635010, 11935015, 11935016, 11935018, 12025502, 12035009, 12035013, 12061131003, 12192260, 12192261, 12192262, 12192263, 12192264, 12192265, 12221005, 12225509, 12235017, 12342502, 12361141819; the Chinese Academy of Sciences (CAS) Large-Scale Scientific Facility Program; the Strategic Priority Research Program of Chinese Academy of Sciences under Contract No. XDA0480600; CAS under Contract No. YSBR-101; 100 Talents Program of CAS; The Institute of Nuclear and Particle Physics (INPAC) and Shanghai Key Laboratory for Particle Physics and Cosmology; ERC under Contract No. 758462; German Research Foundation DFG under Contract No. FOR5327; Istituto Nazionale di Fisica Nucleare, Italy; Knut and Alice Wallenberg Foundation under Contracts Nos. 2021.0174, 2021.0299, 2023.0315; Ministry of Development of Turkey under Contract No. DPT2006K-120470; National Research Foundation of Korea under Contract No. NRF-2022R1A2C1092335; National Science and Technology fund of Mongolia; Polish National Science Centre under Contract No. 2024/53/B/ST2/00975; STFC (United Kingdom); Swedish Research Council under Contract No. 2019.04595; U. S. Department of Energy under Contract No. DE-FG02-05ER41374.

\end{acknowledgments}

\clearpage

\bibliography{reference.bib}

@article{Lubicz:2017syv,
    author = "Lubicz, V. and Riggio, L. and Salerno, G. and Simula, S. and Tarantino, C.",
    collaboration = "ETM Collaboration",
    //title = "{Scalar and vector form factors of $D \to \pi(K) \ell \nu$ decays with $N_f=2+1+1$ twisted fermions}",
    //eprint = "1706.03017",
    archivePrefix = "arXiv",
    primaryClass = "hep-lat",
    reportNumber = "PREPRINT-RM3-TH-17-6, preprint RM3-TH/17-6",
    doi = "10.1103/PhysRevD.96.054514",
    journal = "Phys. Rev. D",
    volume = "96",
    number = "5",
    pages = "054514",
    year = "2017",
    //note = "[Erratum: Phys.Rev.D 99, 099902 (2019), Erratum: Phys.Rev.D 100, 079901 (2019)]"
}

@article{Chakraborty:2021qav,
    author = "Chakraborty, Bipasha and Parrott, W. G. and Bouchard, C. and Davies, C. T. H. and Koponen, J. and Lepage, G. P.",
    collaboration = "HPQCD Collaboration",
    //title = "{Improved Vcs determination using precise lattice QCD form factors for D\textrightarrow{}K\ensuremath{\ell}\ensuremath{\nu}}",
    //eprint = "2104.09883",
    archivePrefix = "arXiv",
    primaryClass = "hep-lat",
    doi = "10.1103/PhysRevD.104.034505",
    journal = "Phys. Rev. D",
    volume = "104",
    number = "3",
    pages = "034505",
    year = "2021"
}

@article{Parrott:2022rgu,
    author = "Parrott, W. G. and Bouchard, C. and Davies, C. T. H.",
    collaboration = "HPQCD Collaboration",
    //title = "{B\textrightarrow{}K and D\textrightarrow{}K form factors from fully relativistic lattice QCD}",
    //eprint = "2207.12468",
    archivePrefix = "arXiv",
    primaryClass = "hep-lat",
    doi = "10.1103/PhysRevD.107.014510",
    journal = "Phys. Rev. D",
    volume = "107",
    number = "1",
    pages = "014510",
    year = "2023"
}

@article{Wu:2006rd,
    author = "Wu, Yue-Liang and Zhong, Ming and Zuo, Ya-Bing",
    //title = "{B(s), D(s) ---\ensuremath{>} pi, K, eta, rho, K*, omega, phi Transition Form Factors and Decay Rates with Extraction of the CKM parameters |V(ub)|, |V(cs)|, |V(cd)|}",
    //eprint = "hep-ph/0604007",
    archivePrefix = "arXiv",
    doi = "10.1142/S0217751X06033209",
    journal = "Int. J. Mod. Phys. A",
    volume = "21",
    pages = "6125--6172",
    year = "2006"
}

@article{Verma:2011yw,
    author = "Verma, R. C.",
    //title = "{Decay constants and form factors of s-wave and p-wave mesons in the covariant light-front quark model}",
    //eprint = "1103.2973",
    archivePrefix = "arXiv",
    primaryClass = "hep-ph",
    doi = "10.1088/0954-3899/39/2/025005",
    journal = "J. Phys. G",
    volume = "39",
    pages = "025005",
    year = "2012"
}

@article{Ivanov:2019nqd,
    author = {Ivanov, Mikhail A. and K\"orner, J\"urgen G. and Pandya, Jignesh N. and Santorelli, Pietro and Soni, Nakul R. and Tran, Chien-Thang},
    //title = "{Exclusive semileptonic decays of D and D$_{s}$ mesons in the covariant confining quark model}",
    //eprint = "1904.07740",
    archivePrefix = "arXiv",
    primaryClass = "hep-ph",
    doi = "10.1007/s11467-019-0908-1",
    journal = "Front. Phys. (Beijing)",
    volume = "14",
    number = "6",
    pages = "64401",
    year = "2019"
}

@article{Ke:2023qzc,
    author = "Ke, Bai-Cian and Koponen, Jonna and Li, Hai-Bo and Zheng, Yangheng",
    //title = "{Recent Progress in Leptonic and Semileptonic Decays of Charmed Hadrons}",
    //eprint = "2310.05228",
    archivePrefix = "arXiv",
    primaryClass = "hep-ex",
    doi = "10.1146/annurev-nucl-110222-044046",
    journal = "Ann. Rev. Nucl. Part. Sci.",
    volume = "73",
    pages = "285--314",
    year = "2023"
}

@article{BaBar:2012obs,
    author = "Lees, J. P. and others",
    collaboration = "BaBar Collaboration",
    //title = "{Evidence for an excess of $\bar{B} \to D^{(*)} \tau^-\bar{\nu}_\tau$ decays}",
    //eprint = "1205.5442",
    archivePrefix = "arXiv",
    primaryClass = "hep-ex",
    reportNumber = "BABAR-PUB-12-012, SLAC-PUB-15028",
    doi = "10.1103/PhysRevLett.109.101802",
    journal = "Phys. Rev. Lett.",
    volume = "109",
    pages = "101802",
    year = "2012"
}

@article{Belle:2015qfa,
    author = "Huschle, M. and others",
    collaboration = "Belle Collaboration",
    //title = "{Measurement of the branching ratio of $\bar{B} \to D^{(\ast)} \tau^- \bar{\nu}_\tau$ relative to $\bar{B} \to D^{(\ast)} \ell^- \bar{\nu}_\ell$ decays with hadronic tagging at Belle}",
    //eprint = "1507.03233",
    archivePrefix = "arXiv",
    primaryClass = "hep-ex",
    reportNumber = "KEK-REPORT-2015-18",
    doi = "10.1103/PhysRevD.92.072014",
    journal = "Phys. Rev. D",
    volume = "92",
    number = "7",
    pages = "072014",
    year = "2015"
}

@article{LHCb:2024onj,
    author = "Aaij, Roel and others",
    collaboration = "LHCb Collaboration",
    //title = "{Comprehensive analysis of local and nonlocal amplitudes in the B$^{0}${\textrightarrow} K$^{*0}${\ensuremath{\mu}}$^{+}${\ensuremath{\mu}}$^{−}$ decay}",
    archivePrefix = "arXiv",
    primaryClass = "hep-ex",
    reportNumber = "LHCb-PAPER-2024-011, CERN-EP-2024-122",
    doi = "10.1007/JHEP09(2024)026",
    journal = "JHEP",
    volume = "09",
    pages = "026",
    year = "2024"
}

@article{LHCb:2013ghj,
    author = "Aaij, R and others",
    collaboration = "LHCb Collaboration",
    //title = "{Measurement of Form-Factor-Independent Observables in the Decay $B^{0} \to K^{*0} \mu^+ \mu^-$}",
    //eprint = "1308.1707",
    archivePrefix = "arXiv",
    primaryClass = "hep-ex",
    reportNumber = "LHCB-PAPER-2013-037, CERN-PH-EP-2013-146",
    doi = "10.1103/PhysRevLett.111.191801",
    journal = "Phys. Rev. Lett.",
    volume = "111",
    pages = "191801",
    year = "2013"
}

@article{Belle:2016fev,
    author = "Wehle, S. and others",
    collaboration = "Belle Collaboration",
    //title = "{Lepton-Flavor-Dependent Angular Analysis of $B\to K^\ast \ell^+\ell^-$}",
    //eprint = "1612.05014",
    archivePrefix = "arXiv",
    primaryClass = "hep-ex",
    reportNumber = "BELLE-PR//eprint-2016-15, KEK-PR//eprint-2016-54",
    doi = "10.1103/PhysRevLett.118.111801",
    journal = "Phys. Rev. Lett.",
    volume = "118",
    number = "11",
    pages = "111801",
    year = "2017"
}

@article{ATLAS:2018gqc,
    author = "Aaboud, Morad and others",
    collaboration = "ATLAS Collaboration",
    //title = "{Angular analysis of $B^0_d \rightarrow K^{*}\mu^+\mu^-$ decays in $pp$ collisions at $\sqrt{s}= 8$ TeV with the ATLAS detector}",
    //eprint = "1805.04000",
    archivePrefix = "arXiv",
    primaryClass = "hep-ex",
    reportNumber = "CERN-EP-2017-161",
    doi = "10.1007/JHEP10(2018)047",
    journal = "JHEP",
    volume = "10",
    pages = "047",
    year = "2018"
}

@article{CMS:2017rzx,
    author = "Sirunyan, Albert M and others",
    collaboration = "CMS Collaboration",
    //title = "{Measurement of angular parameters from the decay $\mathrm{B}^0 \to \mathrm{K}^{*0} \mu^+ \mu^-$ in proton-proton collisions at $\sqrt{s} = $ 8 TeV}",
    //eprint = "1710.02846",
    archivePrefix = "arXiv",
    primaryClass = "hep-ex",
    reportNumber = "CMS-BPH-15-008, CERN-EP-2017-240",
    doi = "10.1016/j.physletb.2018.04.030",
    journal = "Phys. Lett. B",
    volume = "781",
    pages = "517--541",
    year = "2018"
}

@article{Iguro:2024hyk,
    author = "Iguro, Syuhei and Kitahara, Teppei and Watanabe, Ryoutaro",
    //title = "{Global fit to b\textrightarrow{}c\ensuremath{\tau}\ensuremath{\nu} anomalies as of Spring 2024}",
    //eprint = "2405.06062",
    archivePrefix = "arXiv",
    primaryClass = "hep-ph",
    reportNumber = "KEK-TH-2614, P3H-24-024, TTP24-008, CHIBA-EP-262",
    doi = "10.1103/PhysRevD.110.075005",
    journal = "Phys. Rev. D",
    volume = "110",
    number = "7",
    pages = "075005",
    year = "2024"
}

@article{Iguro:2022uzz,
    author = "Iguro, Syuhei",
    //title = "{Revival of H- interpretation of RD(*) anomaly and closing low mass window}",
    //eprint = "2201.06565",
    archivePrefix = "arXiv",
    primaryClass = "hep-ph",
    reportNumber = "P3H-22-010, TTP22-004",
    doi = "10.1103/PhysRevD.105.095011",
    journal = "Phys. Rev. D",
    volume = "105",
    number = "9",
    pages = "095011",
    year = "2022"
}

@article{Blanke:2022pjy,
    author = "Blanke, Monika and Iguro, Syuhei and Zhang, Hantian",
    //title = "{Towards ruling out the charged Higgs interpretation of the $ {R}_{D^{\left(\ast \right)}} $ anomaly}",
    //eprint = "2202.10468",
    archivePrefix = "arXiv",
    primaryClass = "hep-ph",
    reportNumber = "P3H-22-018, TTP22-010",
    doi = "10.1007/JHEP06(2022)043",
    journal = "JHEP",
    volume = "06",
    pages = "043",
    year = "2022"
}

@article{Sakaki:2013bfa,
    author = "Sakaki, Yasuhito and Tanaka, Minoru and Tayduganov, Andrey and Watanabe, Ryoutaro",
    //title = "{Testing leptoquark models in $\bar B \to D^{(*)} \tau \bar\nu$}",
    //eprint = "1309.0301",
    archivePrefix = "arXiv",
    primaryClass = "hep-ph",
    reportNumber = "OU-HET-791, KEK-TH-1660, OU-HET 791",
    doi = "10.1103/PhysRevD.88.094012",
    journal = "Phys. Rev. D",
    volume = "88",
    number = "9",
    pages = "094012",
    year = "2013"
}

@article{Becirevic:2018afm,
    author = "Be\v{c}irevi\'c, Damir and Dor\v{s}ner, Ilja and Fajfer, Svjetlana and Ko\v{s}nik, Nejc and Faroughy, Darius A. and Sumensari, Olcyr",
    //title = "{Scalar leptoquarks from grand unified theories to accommodate the $B$-physics anomalies}",
    //eprint = "1806.05689",
    archivePrefix = "arXiv",
    primaryClass = "hep-ph",
    reportNumber = "LPT-Orsay-18-76, LPT-ORSAY-18-76",
    doi = "10.1103/PhysRevD.98.055003",
    journal = "Phys. Rev. D",
    volume = "98",
    number = "5",
    pages = "055003",
    year = "2018"
}

@article{Fajfer:2015ixa,
    author = "Fajfer, Svjetlana and Nisandzic, Ivan and Rojec, Ursa",
    //title = "{Discerning new physics in charm meson leptonic and semileptonic decays}",
    //eprint = "1502.07488",
    archivePrefix = "arXiv",
    primaryClass = "hep-ph",
    doi = "10.1103/PhysRevD.91.094009",
    journal = "Phys. Rev. D",
    volume = "91",
    number = "9",
    pages = "094009",
    year = "2015"
}

@article{Barranco:2013tba,
    author = "Barranco, J. and Delepine, D. and Gonzalez Macias, V. and Lopez-Lozano, L.",
    //title = "{Constraining New Physics with D meson decays}",
    //eprint = "1303.3896",
    archivePrefix = "arXiv",
    primaryClass = "hep-ph",
    doi = "10.1016/j.physletb.2014.02.008",
    journal = "Phys. Lett. B",
    volume = "731",
    pages = "36--42",
    year = "2014"
}

@article{Zhang:2018jtm,
    author = "Zhang, Jian and Yue, Chong-Xing and Li, Chun-Hua",
    //title = "{Determinations of the form factors of semileptonic $D\rightarrow K$ decays and leptoquark constraints}",
    //eprint = "1805.00700",
    archivePrefix = "arXiv",
    primaryClass = "hep-ph",
    doi = "10.1140/epjc/s10052-018-6184-3",
    journal = "Eur. Phys. J. C",
    volume = "78",
    number = "9",
    pages = "695",
    year = "2018"
}

@article{Barranco:2016njc,
    author = "Barranco, J. and Delepine, D. and Gonzalez Macias, V. and Lopez-Lozano, L.",
    //title = "{Two Higgs doublet model and leptoquarks constraints from D meson decays}",
    doi = "10.1088/0954-3899/43/11/115004",
    journal = "J. Phys. G",
    volume = "43",
    number = "11",
    pages = "115004",
    year = "2016"
}

@article{ParticleDataGroup:2024cfk,
    author = "Navas, S. and others",
    collaboration = "Particle Data Group",
    //title = "{Review of particle physics}",
    doi = "10.1103/PhysRevD.110.030001",
    journal = "Phys. Rev. D",
    volume = "110",
    number = "3",
    pages = "030001",
    year = "2024"
}

@article{Faustov:2019mqr,
    author = "Faustov, R. N. and Galkin, V. O. and Kang, Xian-Wei",
    //title = "{Semileptonic decays of $D$ and $D_s$ mesons in the relativistic quark model}",
    //eprint = "1911.08209",
    archivePrefix = "arXiv",
    primaryClass = "hep-ph",
    doi = "10.1103/PhysRevD.101.013004",
    journal = "Phys. Rev. D",
    volume = "101",
    number = "1",
    pages = "013004",
    year = "2020"
}

@article{BES:2004rav,
    author = "Ablikim, M. and others",
    collaboration = "BES Collaboration",
    //title = "{Direct measurements of the branching fractions for D0 ---\ensuremath{>} K- e+ nu(e) and D0 ---\ensuremath{>} pi- e+ nu(e) and determinations of the form-factors f**K(+)(0) and f**pi(+)(0)}",
    //eprint = "hep-ex/0406028",
    archivePrefix = "arXiv",
    doi = "10.1016/j.physletb.2004.07.004",
    journal = "Phys. Lett. B",
    volume = "597",
    pages = "39--46",
    year = "2004"
}

@article{BES:2004obp,
    author = "Ablikim, M. and others",
    collaboration = "BES Collaboration",
    //title = "{Direct measurement of the branching fraction for the decay of D+ ---\ensuremath{>} anti-K0 e+ nu(e) and determination of Gamma (D0 ---\ensuremath{>} K- e+ nu(e)) / Gamma (D+ ---\ensuremath{>} anti-K0 e+ nu(e))}",
    //eprint = "hep-ex/0410030",
    archivePrefix = "arXiv",
    doi = "10.1016/j.physletb.2004.12.040",
    journal = "Phys. Lett. B",
    volume = "608",
    pages = "24--30",
    year = "2005"
}

@article{BES:2006kzp,
    author = "Ablikim, M. and others",
    collaboration = "BES Collaboration",
    //title = "{Direct measurement of the absolute branching fraction for D+ ---\ensuremath{>} anti-K0 mu+ nu(mu) and determination of Gamma (D0 ---\ensuremath{>} K- mu+ nu(mu))/ Gamma (D+ ---\ensuremath{>} anti-K0 mu+ nu(mu))}",
    //eprint = "hep-ex/0610020",
    archivePrefix = "arXiv",
    doi = "10.1016/j.physletb.2006.11.026",
    journal = "Phys. Lett. B",
    volume = "644",
    pages = "20--24",
    year = "2007"
}

@article{BaBar:2007zgf,
    author = "Aubert, Bernard and others",
    editor = "Aug\'e, Etienne and Pietrzyk, Bolek and Tr\^an Thanh V\^an, Jean",
    collaboration = "BaBar Collaboration",
    //title = "{Measurement of the hadronic form-factor in $D^0 \to K^{-} e^{+} \nu_{e}$ 1}",
    //eprint = "0704.0020",
    archivePrefix = "arXiv",
    primaryClass = "hep-ex",
    reportNumber = "SLAC-PUB-12417, BABAR-PUB-07-015",
    doi = "10.1103/PhysRevD.76.052005",
    journal = "Phys. Rev. D",
    volume = "76",
    pages = "052005",
    year = "2007"
}

@article{CLEO:2005rxg,
    author = "Huang, G. S. and others",
    collaboration = "CLEO Collaboration",
    //title = "{Absolute branching fraction measurements of exclusive D+ semileptonic decays}",
    //eprint = "hep-ex/0506053",
    archivePrefix = "arXiv",
    reportNumber = "CLNS-05-1915, CLEO-05-07",
    doi = "10.1103/PhysRevLett.95.181801",
    journal = "Phys. Rev. Lett.",
    volume = "95",
    pages = "181801",
    year = "2005"
}

@article{CLEO:2005cuk,
    author = "Coan, T. E. and others",
    collaboration = "CLEO Collaboration",
    //title = "{Absolute branching fraction measurements of exclusive D0 semileptonic decays}",
    //eprint = "hep-ex/0506052",
    archivePrefix = "arXiv",
    reportNumber = "CLNS-05-1906, CLEO-05-01",
    doi = "10.1103/PhysRevLett.95.181802",
    journal = "Phys. Rev. Lett.",
    volume = "95",
    pages = "181802",
    year = "2005"
}

@article{CLEO:2007ntr,
    author = "Dobbs, S. and others",
    collaboration = "CLEO Collaboration",
    title = "{A Study of the semileptonic charm decays D0 ---\ensuremath{>}pi- e+ nu(e), D+ ---\ensuremath{>} pi0 e+ nu(e), D0 ---\ensuremath{>} K- e+ nu(e), and D+ ---\ensuremath{>} anti-K0 e+ nu(e)}",
    //eprint = "0712.1020",
    //archivePrefix = "arXiv",
    primaryClass = "hep-ex",
    reportNumber = "CLNS-06-1968, CLEO-06-13",
    doi = "10.1103/PhysRevD.77.112005",
    journal = "Phys. Rev. D",
    volume = "77",
    pages = "112005",
    year = "2008"
}

@article{CLEO:2009svp,
    author = "Besson, D. and others",
    collaboration = "CLEO Collaboration",
    //title = "{Improved measurements of D meson semileptonic decays to pi and K mesons}",
    //eprint = "0906.2983",
    archivePrefix = "arXiv",
    primaryClass = "hep-ex",
    reportNumber = "CLNS-09-2049, CLEO-09-02",
    doi = "10.1103/PhysRevD.80.032005",
    journal = "Phys. Rev. D",
    volume = "80",
    pages = "032005",
    year = "2009"
}

@article{BESIII:2021mfl,
    author = "Ablikim, Medina and others",
    collaboration = "BESIII Collaboration",
    //title = "{Determination of the absolute branching fractions of $D^0\to K^-e^+\nu_e$ and $D^+\to \bar K^0 e^+\nu_e$}",
    //eprint = "2104.08081",
    archivePrefix = "arXiv",
    primaryClass = "hep-ex",
    reportNumber = "BAM-00426",
    doi = "10.1103/PhysRevD.104.052008",
    journal = "Phys. Rev. D",
    volume = "104",
    number = "5",
    pages = "052008",
    year = "2021"
}

@article{BESIII:2015tql,
    author = "Ablikim, M. and others",
    collaboration = "BESIII Collaboration",
    //title = "{Study of Dynamics of $D^0 \to K^- e^+ \nu_{e}$ and $D^0\to\pi^- e^+ \nu_{e}$ Decays}",
    //eprint = "1508.07560",
    archivePrefix = "arXiv",
    primaryClass = "hep-ex",
    doi = "10.1103/PhysRevD.92.072012",
    journal = "Phys. Rev. D",
    volume = "92",
    number = "7",
    pages = "072012",
    year = "2015"
}

@article{BESIII:2018ccy,
    author = "Ablikim, Medina and others",
    collaboration = "BESIII Collaboration",
    //title = "{Study of the $D^0\to K^-\mu^+\nu_\mu$ dynamics and test of lepton flavor universality with $D^0\to K^-\ell^+\nu_\ell$ decays}",
    //eprint = "1810.03127",
    archivePrefix = "arXiv",
    primaryClass = "hep-ex",
    doi = "10.1103/PhysRevLett.122.011804",
    journal = "Phys. Rev. Lett.",
    volume = "122",
    number = "1",
    pages = "011804",
    year = "2019"
}

@article{BESIII:2017ylw,
    author = "Ablikim, M. and others",
    collaboration = "BESIII Collaboration",
    //title = "{Analysis of $D^+\to\bar K^0e^+\nu_e$ and $D^+\to\pi^0e^+\nu_e$ semileptonic decays}",
    //eprint = "1703.09084",
    archivePrefix = "arXiv",
    primaryClass = "hep-ex",
    doi = "10.1103/PhysRevD.96.012002",
    journal = "Phys. Rev. D",
    volume = "96",
    number = "1",
    pages = "012002",
    year = "2017"
}

@article{BESIII:2016hko,
    author = "Ablikim, Medina and others",
    collaboration = "BESIII Collaboration",
    //title = "{Measurement of the absolute branching fraction of $D^{+}\rightarrow\bar K^0 e^{+}\nu_{e}$ via $\bar K^0\to\pi^0\pi^0$}",
    //eprint = "1605.00208",
    archivePrefix = "arXiv",
    primaryClass = "hep-ex",
    doi = "10.1088/1674-1137/40/11/113001",
    journal = "Chin. Phys. C",
    volume = "40",
    number = "11",
    pages = "113001",
    year = "2016"
}

@article{BESIII:2015jmz,
    author = "Ablikim, M. and others",
    collaboration = "BESIII Collaboration",
    //title = "{Study of decay dynamics and $CP$ asymmetry in $D^+ \to K^0_L e^+ \nu_e$ decay}",
    //eprint = "1510.00308",
    archivePrefix = "arXiv",
    primaryClass = "hep-ex",
    doi = "10.1103/PhysRevD.92.112008",
    journal = "Phys. Rev. D",
    volume = "92",
    number = "11",
    pages = "112008",
    year = "2015"
}

@article{BESIII:2016gbw,
    author = "Ablikim, Medina and others",
    collaboration = "BESIII Collaboration",
    //title = "{Improved measurement of the absolute branching fraction of $D^{+}\rightarrow \bar{K}^0 \mu ^{+}\nu _{\mu }$}",
    //eprint = "1605.00068",
    archivePrefix = "arXiv",
    primaryClass = "hep-ex",
    doi = "10.1140/epjc/s10052-016-4198-2",
    journal = "Eur. Phys. J. C",
    volume = "76",
    number = "7",
    pages = "369",
    year = "2016"
}

@article{Belle:2006idb,
    author = "Widhalm, L. and others",
    collaboration = "Belle Collaboration",
    //title = "{Measurement of D0 ---\ensuremath{>} pi l nu (Kl nu) Form Factors and Absolute Branching Fractions}",
    //eprint = "hep-ex/0604049",
    archivePrefix = "arXiv",
    reportNumber = "BELLE-PREPRINT-2006-12",
    doi = "10.1103/PhysRevLett.97.061804",
    journal = "Phys. Rev. Lett.",
    volume = "97",
    pages = "061804",
    year = "2006"
}

@article{BESIII:2024lbn,
    author = "Ablikim, Medina and others",
    collaboration = "BESIII Collaboration",
    //title = "{Measurement of integrated luminosity of data collected at 3.773 GeV by BESIII from 2021 to 2024*}",
    //eprint = "2406.05827",
    archivePrefix = "arXiv",
    primaryClass = "hep-ex",
    doi = "10.1088/1674-1137/ad70a0",
    journal = "Chin. Phys. C",
    volume = "48",
    number = "12",
    pages = "123001",
    year = "2024"
}

@article{BESIII:2024slx,
    author = "Ablikim, Medina and others",
    collaboration = "BESIII Collaboration",
    //title = "{Improved measurements of D0\textrightarrow{}K-\ensuremath{\ell}+\ensuremath{\nu}\ensuremath{\ell} and D+\textrightarrow{}K\textasciimacron{}0\ensuremath{\ell}+\ensuremath{\nu}\ensuremath{\ell}}",
    //eprint = "2408.09087",
    archivePrefix = "arXiv",
    primaryClass = "hep-ex",
    doi = "10.1103/PhysRevD.110.112006",
    journal = "Phys. Rev. D",
    volume = "110",
    number = "11",
    pages = "112006",
    year = "2024"
}

@article{BESIII:2009fln,
    author = "Ablikim, M. and others",
    collaboration = "BESIII Collaboration",
    //title = "{Design and Construction of the BESIII Detector}",
    //eprint = "0911.4960",
    archivePrefix = "arXiv",
    primaryClass = "physics.ins-det",
    doi = "10.1016/j.nima.2009.12.050",
    journal = "Nucl. Instrum. Meth. A",
    volume = "614",
    pages = "345--399",
    year = "2010"
}

@inproceedings{Yu:2016cof,
    author = "Yu, Chenghui and others",
    title = "{BEPCII Performance and Beam Dynamics Studies on Luminosity}",
    booktitle = "{7th International Particle Accelerator Conference}",
    doi = "10.18429/JACoW-IPAC2016-TUYA01",
    pages = "TUYA01",
    year = "2016"
}

@article{BESIII:2020nme,
    author = "Ablikim, M. and others",
    collaboration = "BESIII Collaboration",
    //title = "{Future Physics Programme of BESIII}",
    //eprint = "1912.05983",
    archivePrefix = "arXiv",
    primaryClass = "hep-ex",
    reportNumber = "HEP-Physics-Report-BESIII-2019-12-13",
    doi = "10.1088/1674-1137/44/4/040001",
    journal = "Chin. Phys. C",
    volume = "44",
    number = "4",
    pages = "040001",
    year = "2020"
}

@article{Li:2021iwf,
    author = "Li, Hai-Bo and Lyu, Xiao-Rui",
    //title = "{Study of the standard model with weak decays of charmed hadrons at BESIII}",
    //eprint = "2103.00908",
    archivePrefix = "arXiv",
    primaryClass = "hep-ex",
    doi = "10.1093/nsr/nwab181",
    journal = "Natl. Sci. Rev.",
    volume = "8",
    number = "11",
    pages = "nwab181",
    year = "2021"
}

@article{Li:2017jpg,
    author = "Li, Xin and others",
    //title = "{Study of MRPC technology for BESIII endcap-TOF upgrade}",
    doi = "10.1007/s41605-017-0014-2",
    journal = "Radiat. Detect. Technol. Methods",
    volume = "1",
    pages = "13",
    year = "2017"
}

@article{Guo:2017sjt,
    author = "Guo, Ying-Xiao and others",
    //title = "{The study of time calibration for upgraded end cap TOF of BESIII}",
    doi = "10.1007/s41605-017-0012-4",
    journal = "Radiat. Detect. Technol. Methods",
    volume = "1",
    pages = "15",
    year = "2017"
}

@article{Cao:2020ibk,
    author = "Cao, P. and others",
    //title = "{Design and construction of the new BESIII endcap Time-of-Flight system with MRPC Technology}",
    doi = "10.1016/j.nima.2019.163053",
    journal = "Nucl. Instrum. Meth. A",
    volume = "953",
    pages = "163053",
    year = "2020"
}

@article{Huang:2022wuo,
    author = "Huang, Kai-Xuan and Li, Zhi-Jun and Qian, Zhen and Zhu, Jiang and Li, Hao-Yuan and Zhang, Yu-Mei and Sun, Sheng-Sen and You, Zheng-Yun",
    //title = "{Method for detector description transformation to Unity and application in BESIII}",
    //eprint = "2206.10117",
    archivePrefix = "arXiv",
    primaryClass = "physics.ins-det",
    doi = "10.1007/s41365-022-01133-8",
    journal = "Nucl. Sci. Tech.",
    volume = "33",
    number = "11",
    pages = "142",
    year = "2022"
}

@article{GEANT4:2002zbu,
    author = "Agostinelli, S. and others",
    collaboration = "GEANT4 Collaboration",
    //title = "{GEANT4 - A Simulation Toolkit}",
    reportNumber = "SLAC-PUB-9350, FERMILAB-PUB-03-339, CERN-IT-2002-003",
    doi = "10.1016/S0168-9002(03)01368-8",
    journal = "Nucl. Instrum. Meth. A",
    volume = "506",
    pages = "250--303",
    year = "2003"
}

@article{Jadach:1999vf,
    author = "Jadach, S. and Ward, B. F. L. and Was, Z.",
    //title = "{The Precision Monte Carlo event generator K K for two fermion final states in e+ e- collisions}",
    //eprint = "hep-ph/9912214",
    archivePrefix = "arXiv",
    reportNumber = "DESY-99-106, CERN-TH-99-235, UTHEP-99-08-01",
    doi = "10.1016/S0010-4655(00)00048-5",
    journal = "Comput. Phys. Commun.",
    volume = "130",
    pages = "260--325",
    year = "2000"
}

@article{Lange:2001uf,
    author = "Lange, D. J.",
    editor = "Erhan, S. and Schlein, P. and Rozen, Y.",
    //title = "{The EvtGen particle decay simulation package}",
    doi = "10.1016/S0168-9002(01)00089-4",
    journal = "Nucl. Instrum. Meth. A",
    volume = "462",
    pages = "152--155",
    year = "2001"
}

@article{Ping:2008zz,
    author = "Ping, Rong-Gang",
    //title = "{Event generators at BESIII}",
    doi = "10.1088/1674-1137/32/8/001",
    journal = "Chin. Phys. C",
    volume = "32",
    pages = "599",
    year = "2008"
}

@article{Chen:2000tv,
    author = "Chen, J. C. and Huang, G. S. and Qi, X. R. and Zhang, D. H. and Zhu, Y. S.",
    //title = "{Event generator for J / psi and psi (2S) decay}",
    doi = "10.1103/PhysRevD.62.034003",
    journal = "Phys. Rev. D",
    volume = "62",
    pages = "034003",
    year = "2000"
}

@article{Yang:2014vra,
    author = "Yang, Rui-Ling and Ping, Rong-Gang and Chen, Hong",
    //title = "{Tuning and Validation of the Lundcharm Model with $J/\psi$ Decays}",
    doi = "10.1088/0256-307X/31/6/061301",
    journal = "Chin. Phys. Lett.",
    volume = "31",
    pages = "061301",
    year = "2014"
}

@article{Golonka:2005pn,
    author = "Golonka, Piotr and Was, Zbigniew",
    title = "{PHOTOS Monte Carlo: A Precision tool for QED corrections in $Z$ and $W$ decays}",
    //eprint = "hep-ph/0506026",
    archivePrefix = "arXiv",
    reportNumber = "IFJPAN-V-05-01, CERN-PH-TH-2005-091",
    doi = "10.1140/epjc/s2005-02396-4",
    journal = "Eur. Phys. J. C",
    volume = "45",
    pages = "97--107",
    year = "2006"
}

@article{Becher:2005bg,
    author = "Becher, Thomas and Hill, Richard J.",
    //title = "{Comment on form-factor shape and extraction of |V(ub)| from B ---\ensuremath{>} pi l nu}",
    //eprint = "hep-ph/0509090",
    archivePrefix = "arXiv",
    reportNumber = "FERMILAB-PUB-05-385-T, SLAC-PUB-11468",
    doi = "10.1016/j.physletb.2005.11.063",
    journal = "Phys. Lett. B",
    volume = "633",
    pages = "61--69",
    year = "2006"
}

@article{Bazavov:2017lyh,
    author = "Bazavov, A. and others",
    //title = "{$B$- and $D$-meson leptonic decay constants from four-flavor lattice QCD}",
    //eprint = "1712.09262",
    archivePrefix = "arXiv",
    primaryClass = "hep-lat",
    reportNumber = "FERMILAB-PUB-17/491-T, FERMILAB-PUB-17-491-T",
    doi = "10.1103/PhysRevD.98.074512",
    journal = "Phys. Rev. D",
    volume = "98",
    number = "7",
    pages = "074512",
    year = "2018"
}

@article{FermilabLattice:2022gku,
    author = "Bazavov, Alexei and others",
    collaboration = "Fermilab Lattice, MILC",
    //title = "{D-meson semileptonic decays to pseudoscalars from four-flavor lattice QCD}",
    //eprint = "2212.12648",
    archivePrefix = "arXiv",
    primaryClass = "hep-lat",
    reportNumber = "MIT-CTP/5513, FERMILAB-PUB-22-943-T",
    doi = "10.1103/PhysRevD.107.094516",
    journal = "Phys. Rev. D",
    volume = "107",
    number = "9",
    pages = "094516",
    year = "2023"
}

@article{Riggio:2017zwh,
    author = "Riggio, L. and Salerno, G. and Simula, S.",
    //title = "{Extraction of $|V_{cd}|$ and $|V_{cs}|$ from experimental decay rates using lattice QCD $D \to \pi(K) \ell \nu$ form factors}",
    //eprint = "1706.03657",
    archivePrefix = "arXiv",
    primaryClass = "hep-lat",
    reportNumber = "PREPRINT-RM3-TH-17-7, preprint RM3-TH/17-7",
    doi = "10.1140/epjc/s10052-018-5943-5",
    journal = "Eur. Phys. J. C",
    volume = "78",
    number = "6",
    pages = "501",
    year = "2018"
}

@article{BESIII:2014rtm,
    author = "Ablikim, M. and others",
    collaboration = "BESIII Collaboration",
    //title = "{Measurement of the $D\to K^-\pi^+$ strong phase difference in $\psi(3770)\to D^0\overline{D}{}^0$}",
    //eprint = "1404.4691",
    archivePrefix = "arXiv",
    primaryClass = "hep-ex",
    doi = "10.1016/j.physletb.2014.05.071",
    journal = "Phys. Lett. B",
    volume = "734",
    pages = "227--233",
    year = "2014"
}

@article{ARGUS:1990hfq,
    author = "Albrecht, H. and others",
    collaboration = "ARGUS Collaboration",
    //title = "{Search for Hadronic $b \to u$ Decays}",
    reportNumber = "DESY-90-008",
    doi = "10.1016/0370-2693(90)91293-K",
    journal = "Phys. Lett. B",
    volume = "241",
    pages = "278--282",
    year = "1990"
}

@article{Boyd:1995sq,
    author = "Boyd, C. Glenn and Grinstein, Benjamin and Lebed, Richard F.",
    //title = "{Model independent determinations of anti-B ---\ensuremath{>} D (lepton), D* (lepton) anti-neutrino form-factors}",
    //eprint = "hep-ph/9508211",
    archivePrefix = "arXiv",
    reportNumber = "UCSD-PTH-95-11",
    doi = "10.1016/0550-3213(95)00653-2",
    journal = "Nucl. Phys. B",
    volume = "461",
    pages = "493--511",
    year = "1996"
}

@article{MARK-III:1985hbd,
    author = "Baltrusaitis, R. M. and others",
    collaboration = "MARK-III Collaboration",
    title = "{Direct Measurements of Charmed d Meson Hadronic Branching Fractions}",
    reportNumber = "SLAC-PUB-3861",
    doi = "10.1103/PhysRevLett.56.2140",
    journal = "Phys. Rev. Lett.",
    volume = "56",
    pages = "2140",
    year = "1986"
}

@article{PRL_draft,
    author = "Ablikim, M. and others",
    collaboration = "BESIII Collaboration",
    //title = "{supplemental material}",
    //eprint = "2312.07572",
    archivePrefix = "arXiv",
    primaryClass = "hep-ex",
    doi = "10.1103/PhysRevLett.xxx.xxxxxx",
    journal = "Phys. Rev. Lett",
    volume = "xxx",
    number = "x",
    pages = "xxxxxx",
    year = "2025",
    note = "The PRL version of Draft"
}

@article{LHCb:2023zxo,
    author = "Aaij, Roel and others",
    collaboration = "LHCb Collaboration",
    title = "{Measurement of the ratios of branching fractions $\mathcal{R}(D^{*})$ and $\mathcal{R}(D^{0})$}",
    //eprint = "2302.02886",
    archivePrefix = "arXiv",
    primaryClass = "hep-ex",
    reportNumber = "LHCb-PAPER-2022-039, CERN-EP-2022-284",
    doi = "10.1103/PhysRevLett.131.111802",
    journal = "Phys. Rev. Lett.",
    volume = "131",
    pages = "111802",
    year = "2023"
}

@article{LHCb:2023uiv,
    author = "Aaij, Roel and others",
    collaboration = "LHCb Collaboration",
    title = "{Test of lepton flavor universality using B0{\textrightarrow}D*-{\ensuremath{\tau}}+{\ensuremath{\nu}}{\ensuremath{\tau}} decays with hadronic {\ensuremath{\tau}} channels}",
    //eprint = "2305.01463",
    archivePrefix = "arXiv",
    primaryClass = "hep-ex",
    reportNumber = "LHCb-PAPER-2022-052, CERN-EP-2023-062",
    doi = "10.1103/PhysRevD.108.012018",
    journal = "Phys. Rev. D",
    volume = "108",
    number = "1",
    pages = "012018",
    year = "2023",
    note = "[Erratum: Phys.Rev.D 109, 119902 (2024)]"
}

@article{LHCb:2024jll,
    author = "Aaij, Roel and others",
    collaboration = "LHCb Collaboration",
    title = "{Measurement of the Branching Fraction Ratios R(D+) and R(D*+) Using Muonic {\ensuremath{\tau}} Decays}",
    //eprint = "2406.03387",
    archivePrefix = "arXiv",
    primaryClass = "hep-ex",
    reportNumber = "LHCb-PAPER-2024-007, CERN-EP-2024-125",
    doi = "10.1103/PhysRevLett.134.061801",
    journal = "Phys. Rev. Lett.",
    volume = "134",
    number = "6",
    pages = "061801",
    year = "2025"
}

\newpage
\onecolumngrid
\appendix
\section*{Appendix}
\label{appendix}

Tables~\ref{kenu_effmatrix}, \ref{kmunu_effmatrix}, \ref{ksenu_effmatrix}, and \ref{ksmunu_effmatrix} report the elements of the weighted efficiency matrices for $\kenu$, $\kmunu$, $\koenu$, and $\komunu$, respectively.

\begin{figure*}[htbp]
\begin{center}
\subfigure{\includegraphics[width=0.98\textwidth]{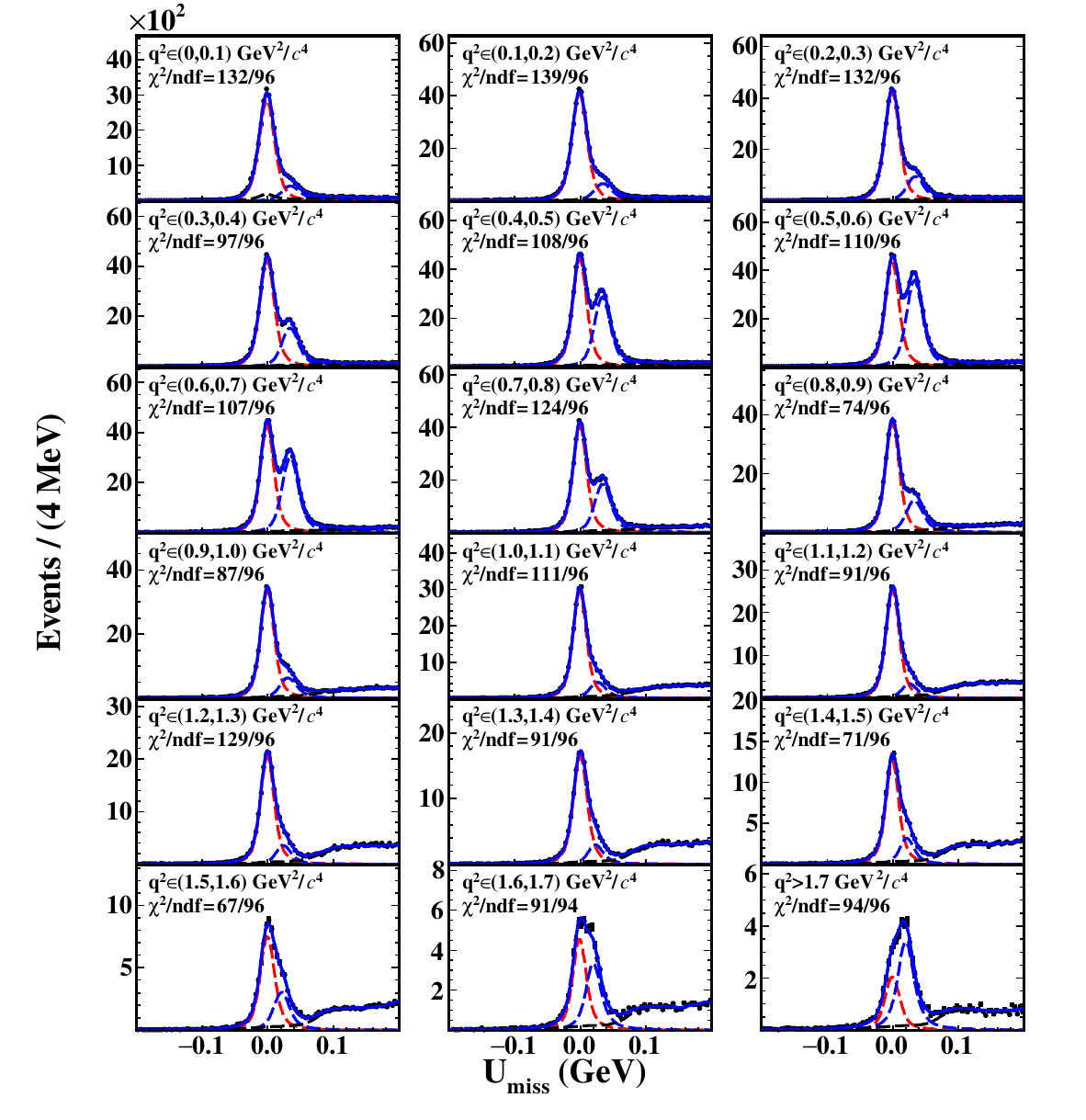}}
\caption{The $U_{\rm miss}$ distributions of the accepted candidate events for  $\kmunu$ in individual $q^2$ intervals in data, with fit results overlaid.
The points with  error bars are data. The blue solid curves are the fit results.
The violet dotted curves are the signal shapes.
The black dash-dotted curves are the peaking backgrounds.
The red dashed curves are the fitted combinatorial background shapes.
\label{kmunu_umissq2}
}
\end{center}
\end{figure*}

\begin{figure*}[htbp]
\begin{center}
\subfigure{\includegraphics[width=0.98\textwidth]{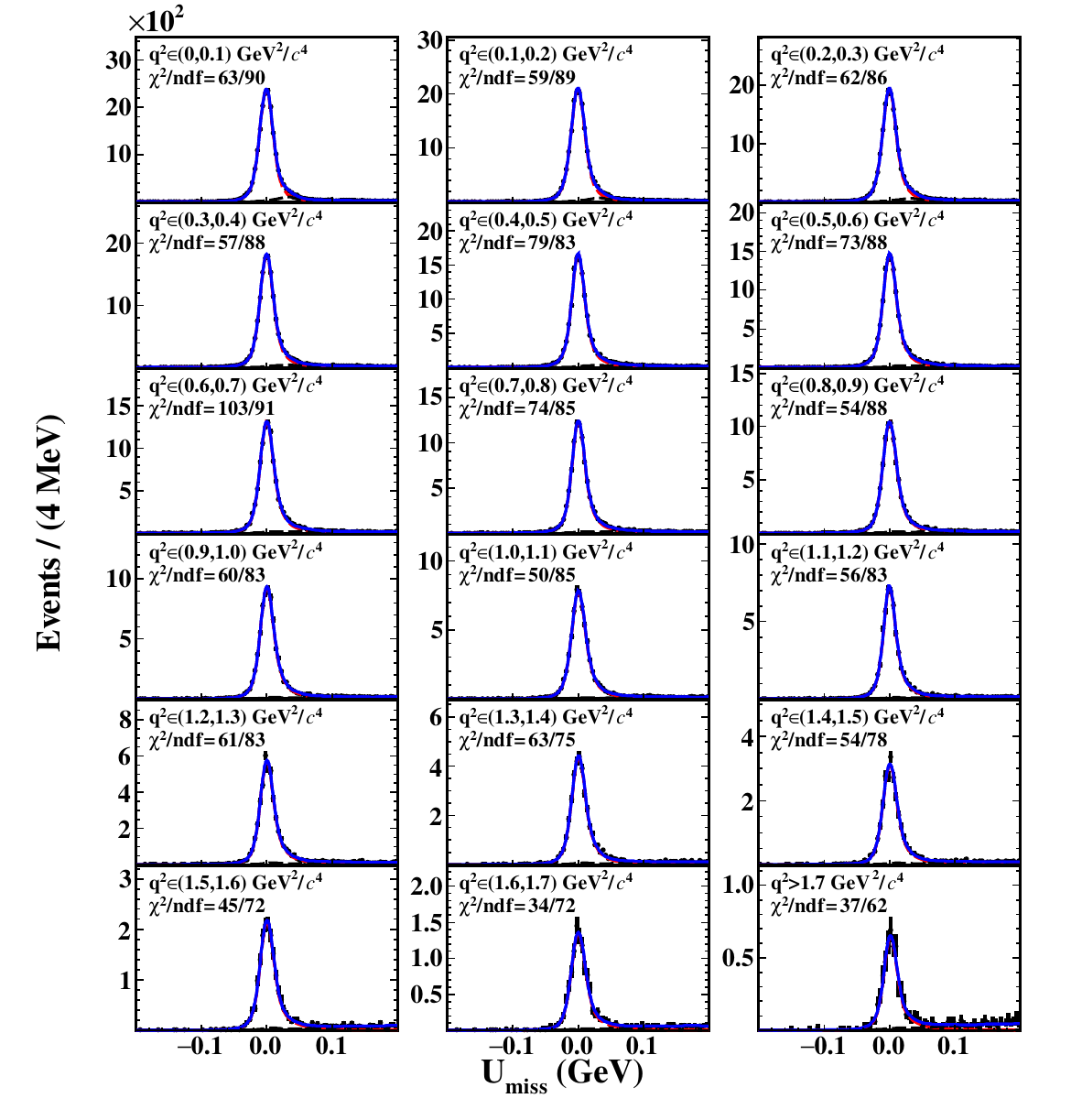}}
\caption{The $U_{\rm miss}$ distributions of the accepted candidate events for  $\ksenu$ in individual $q^2$ intervals in data. The curves and symbols are defined the same as in Fig.~\ref{kmunu_umissq2}.
%, with fit results overlaid. The points with error bars are data. The blue solid curves are the fit results. The violet dotted curves are the signal shapes. The red dashed curves are the fitted combinatorial background shapes.
\label{ksenu_umissq2}
}
\end{center}
\end{figure*}

\begin{figure*}[htbp]
\begin{center}
\subfigure{\includegraphics[width=0.98\textwidth]{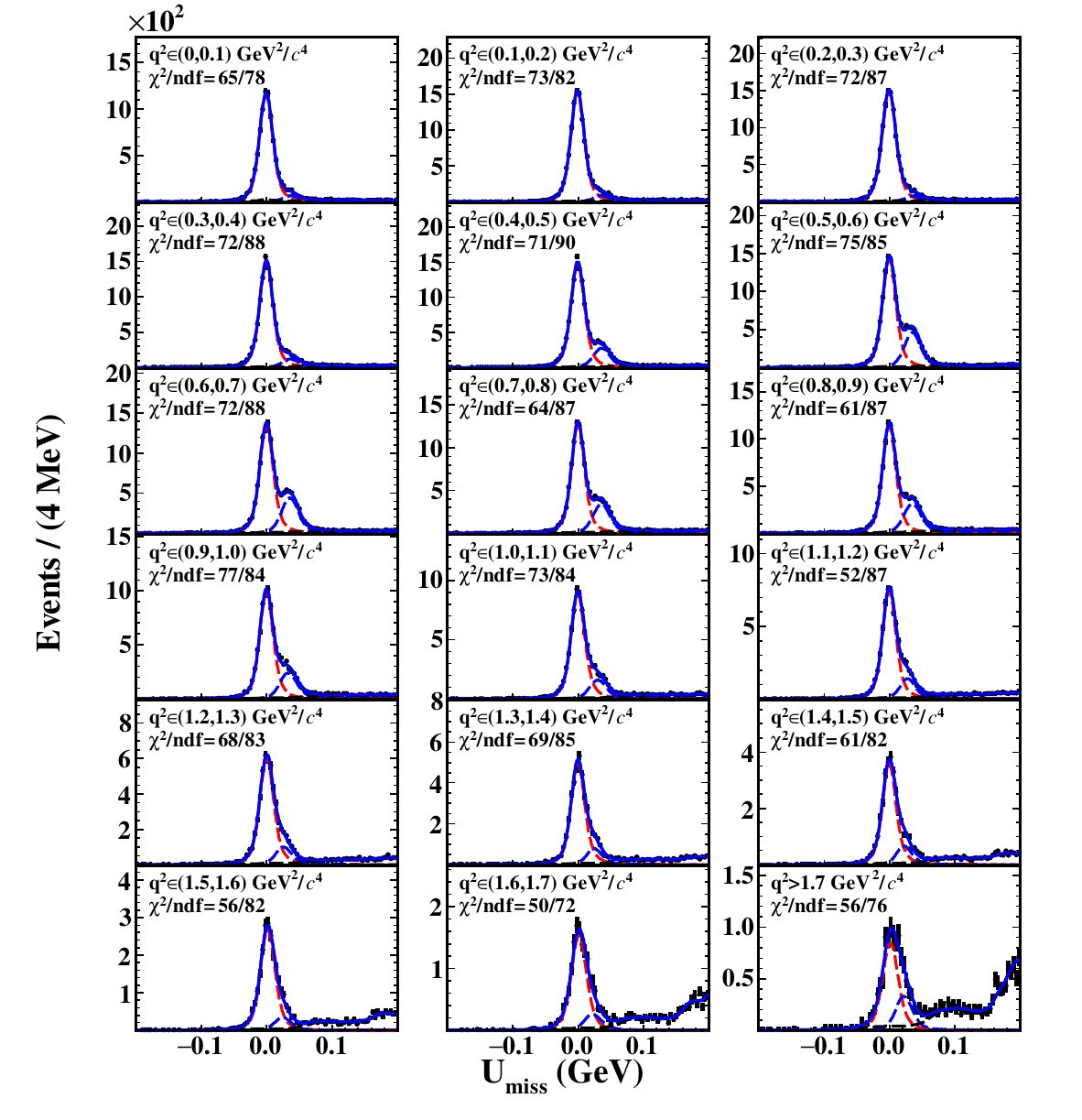}}
\caption{The $U_{\rm miss}$ distributions of the accepted candidate events for  $\ksmunu$ in individual $q^2$ intervals in data. The curves and symbols are defined the same as in Fig.~\ref{kmunu_umissq2}.
%, with fit results overlaid. The blue solid curves are the fit results. The violet dotted curves are the signal shapes. The black dash-dotted curves are the peaking backgrounds. The red dashed curves are the fitted combinatorial background shapes.
\label{ksmunu_umissq2}
}
\end{center}
\end{figure*}

Figures~\ref{kmunu_umissq2}, \ref{ksenu_umissq2}, and~\ref{ksmunu_umissq2} show the results of the fits to the $U_{\rm miss}$ distributions in the reconstructed $q^{2}$ intervals for $\kmunu$, $\koenu$, and $\komunu$, respectively.

Tables~\ref{tab:kmunu_decayrate}, \ref{tab:ksenu_decayrate}, and \ref{tab:ksmunu_decayrate} list the $q^{2}$ ranges, the fitted DT yields in data~($N_{\rm DT}$), the numbers of produced events ($N_{\rm prd}$) calculated by the weighted efficiency matrices and the decay rates ($\Delta\Gamma$) of $\kmunu$, $\koenu$, and $\komunu$ in individual $q^2$ intervals.

Tables~\ref{tab:kenu_statmatrix}, \ref{tab:kmunu_statmatrix}, \ref{tab:ksenu_statmatrix}, and \ref{tab:ksmunu_statmatrix} give the elements of the statistical covariance matrices for $\kenu$, $\kmunu$, $\koenu$, and $\komunu$, respectively.

Tables~\ref{tab:kmunu_sysq2}, \ref{tab:ksenu_sysq2}, and \ref{tab:ksmunu_sysq2} summarize the systematic uncertainties $\sigma_{\rm syst}$ of $\kmunu$, $\koenu$, and $\komunu$ in different $q^2$ intervals.

Tables~\ref{tab:kenu_sysmatrix}, \ref{tab:kmunu_sysmatrix}, \ref{tab:ksenu_sysmatrix} and \ref{tab:ksmunu_sysmatrix} present the elements of the systematic covariance matrices for $\kenu$, $\kmunu$, $\koenu$, and $\komunu$, respectively.

Tables~\ref{klnumatrix_cov1},~\ref{klnumatrix_cov2},~\ref{klnumatrix_cov3}, and~\ref{klnumatrix_cov4} provide the elements of the covariance matrix $\rho_{ij}$ ($i, j \in [1, 72]$) for the simultaneous fit.

\begin{figure*}[htbp]
	\begin{center}	
		\subfigure{\includegraphics[width=0.49\textwidth]{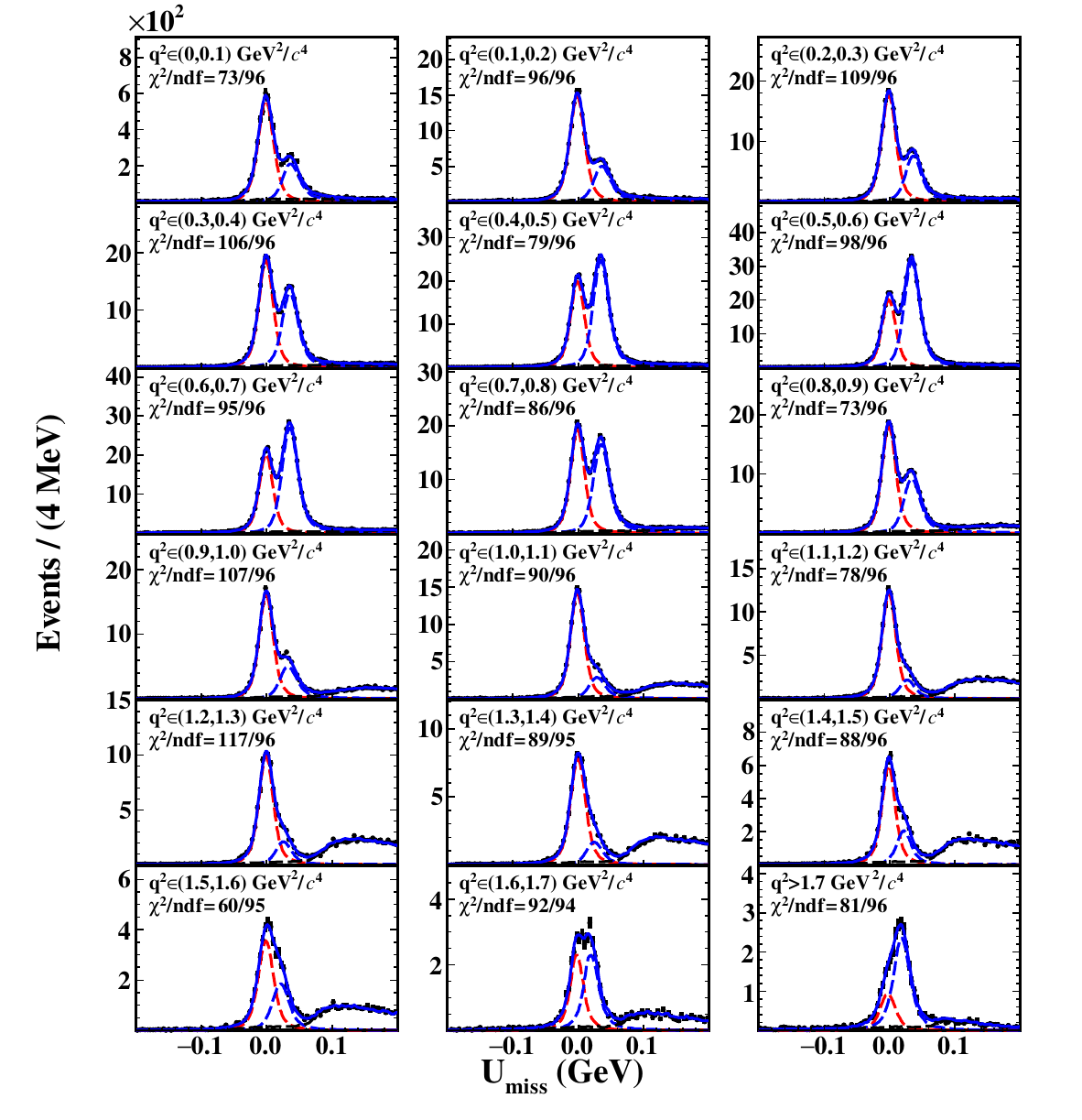}}
		\subfigure{\includegraphics[width=0.49\textwidth]{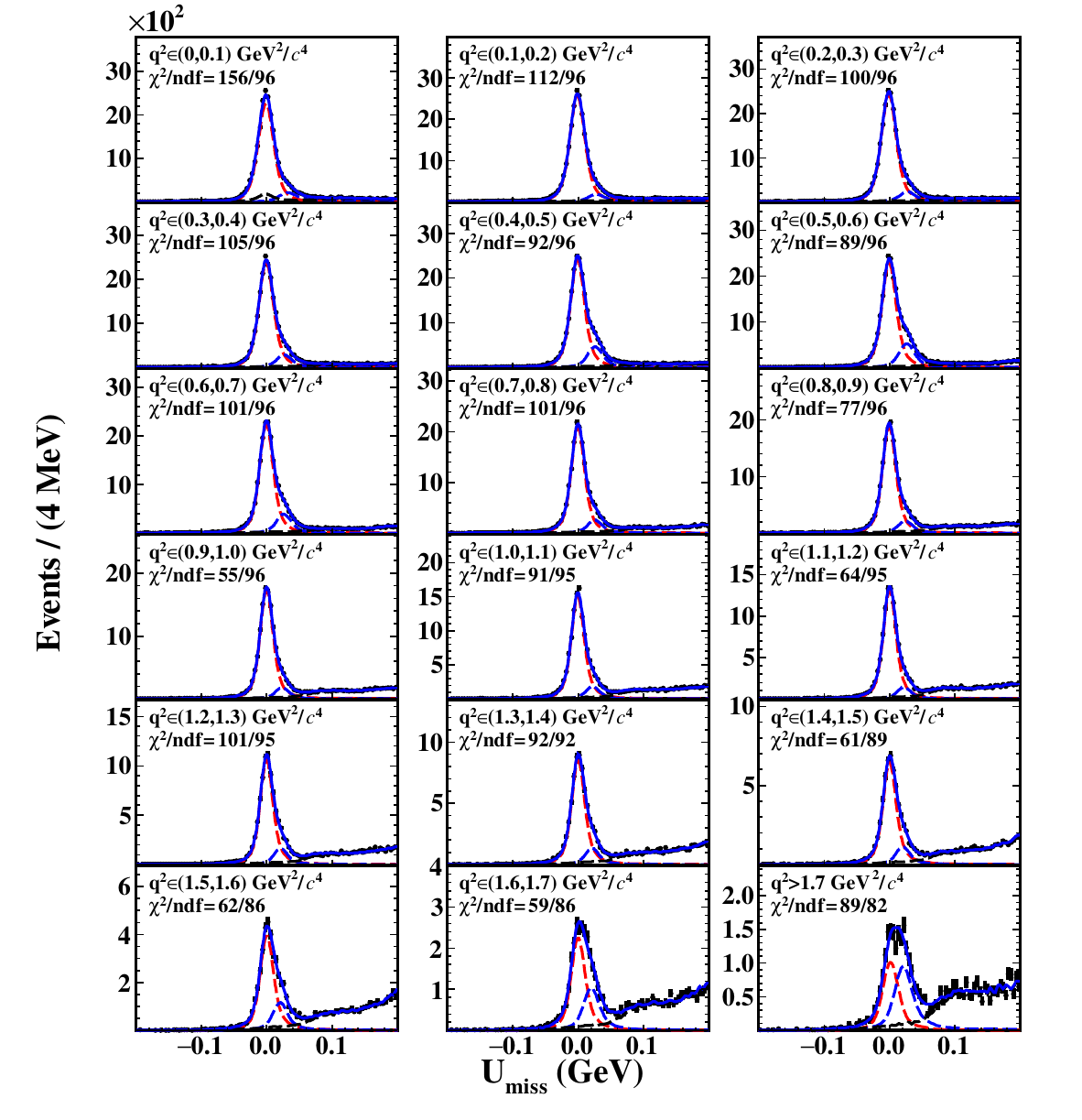}}
		\caption{The $U_{\rm miss}$ distributions of the accepted forward (left) and backward (right) candidate events in individual $q^2$ intervals for $\kmunu$ in data, with fit results overlaid. The blue solid curves are the fit results. The red dashed curves are the signal shapes, and the black dashed curves are the fitted combinatorial background shapes.}
		\label{kmunu_umiss_AFB_q2}
	\end{center}
\end{figure*}

\begin{figure*}[htbp]
	\begin{center}	
		\subfigure{\includegraphics[width=0.49\textwidth]{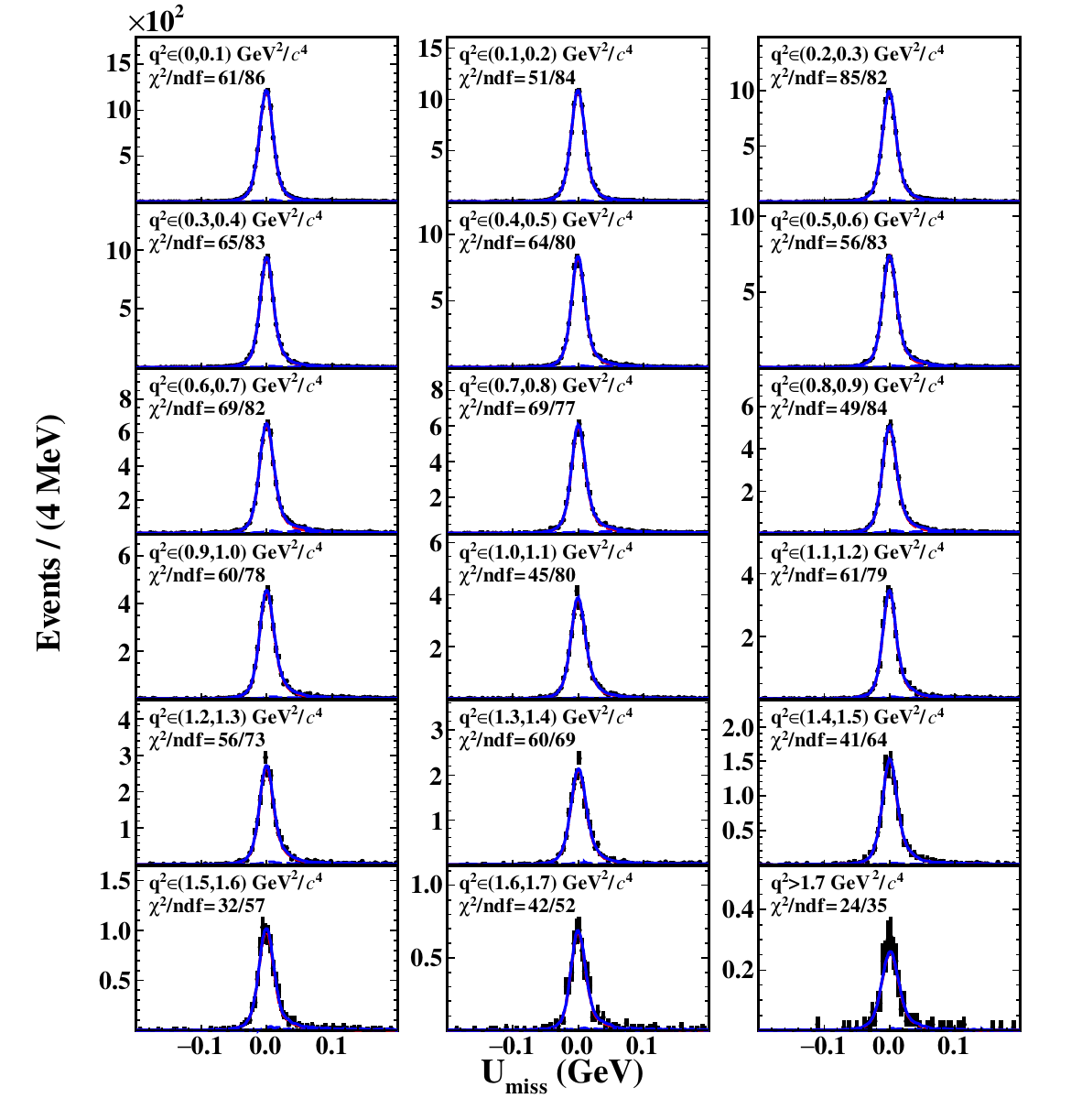}}
		\subfigure{\includegraphics[width=0.49\textwidth]{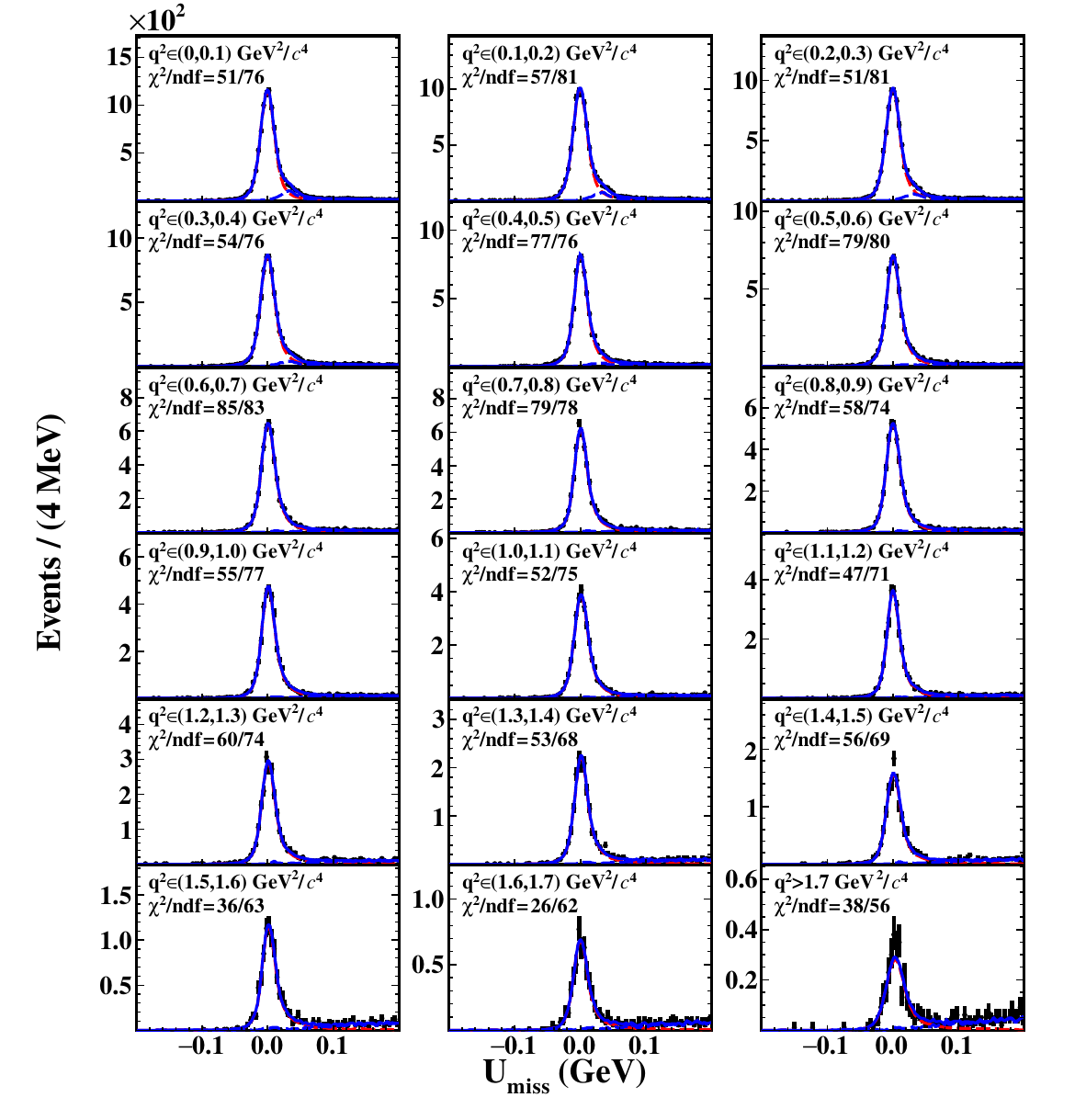}}
		\caption{%The $U_{\rm miss}$ distributions of the accepted forward (left) and backward (right) candidate events in different $q^2$ bins for $\ksenu$ in data, with fit results overlaid. The blue solid curves are the fit results. The red dashed curves are the signal shapes, and the black dashed curves are the fitted combinatorial background shapes.
		The $U_{\rm miss}$ distributions of the forward (left) and backward (right) events for $\ksenu$. The curves and symbols are defined the same as in Fig.~\ref{kmunu_umiss_AFB_q2}.
		}
		\label{ksenu_umiss_AFB_q2}
	\end{center}
\end{figure*}

\begin{figure*}[htbp]
	\begin{center}	
		\subfigure{\includegraphics[width=0.49\textwidth]{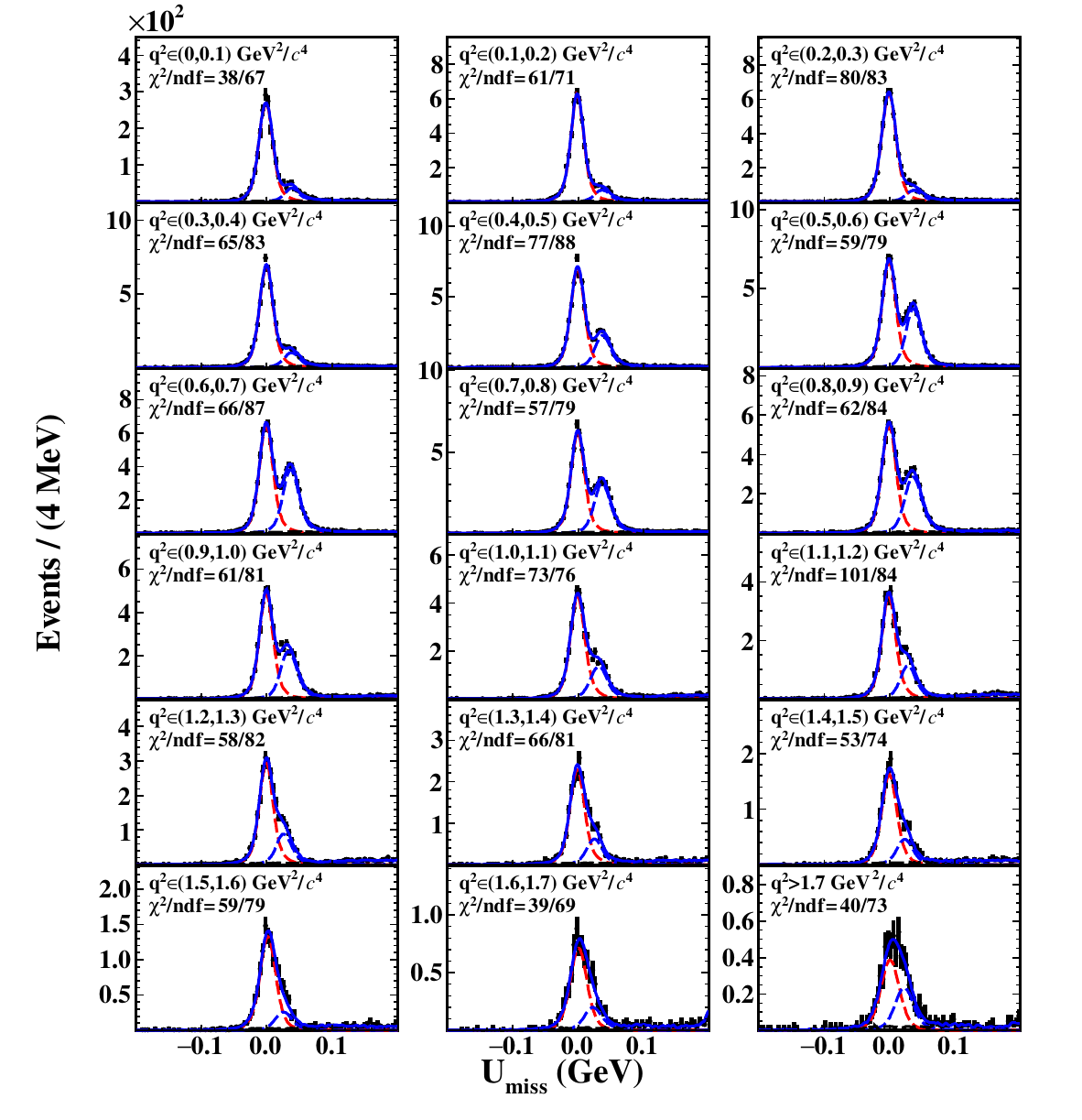}}
		\subfigure{\includegraphics[width=0.49\textwidth]{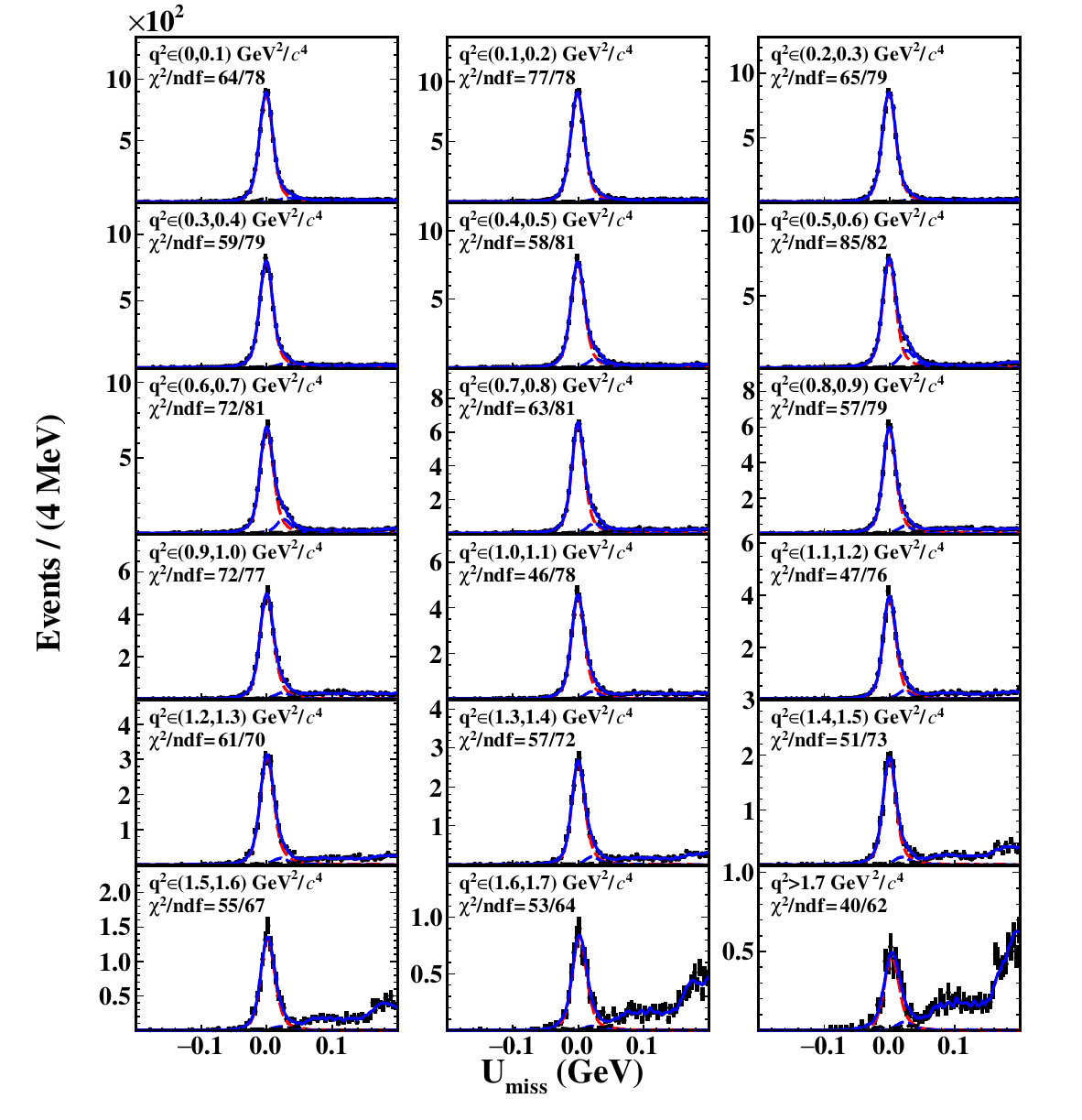}}
		\caption{%The $U_{\rm miss}$ distributions of the accepted forward (left) and backward (right) candidate events in different $q^2$ bins for $\ksmunu$ in data, with fit results overlaid. The blue solid curves are the fit results. The red dashed curves are the signal shapes, and the black dashed curves are the fitted combinatorial background shapes.
		The $U_{\rm miss}$ distributions of the forward (left) and backward (right) events for $\ksmunu$. The curves and symbols are defined the same as in Fig.~\ref{kmunu_umiss_AFB_q2}.
		}
		\label{ksmunu_umiss_AFB_q2}
	\end{center}
\end{figure*}

Figures~\ref{kmunu_umiss_AFB_q2}, ~\ref{ksenu_umiss_AFB_q2}, and~\ref{ksmunu_umiss_AFB_q2} provide the results of the fits to the $U_{\rm miss}$ distributions of forward/backward events in the reconstructed $q^2$ intervals of $\kmunu$, $\ksenu$, and $\ksmunu$, respectively.

Tables~\ref{tab:kmunu_AFB_q2_stat}, ~\ref{tab:ksenu_AFB_q2_stat}, and~\ref{tab:ksmunu_AFB_q2_stat} exhibit the fitted forward/backward DT yields in data and the determined $q^2$-binned forward-backward asymmetries of $\kmunu$, $\ksenu$, and $\ksmunu$, respectively.

\begin{table*}[htbp]
\caption{%The weighted efficiency matrix (in \%) for $\kenu$, where $i$ denotes the reconstructed bin, and $j$ represents the produced bin.
The weighted efficiency matrix $\varepsilon_{ij}$ (in \%) for $\kenu$, where the indices $i$ and $j$ denote the reconstructed and the produced $q^2$ intervals, respectively.
\label{kenu_effmatrix}}
\centering
%\scalebox{0.85}{
\resizebox{1.0\textwidth}{!}{
\begin{tabular}{c|cccccccccccccccccc}
\hline
\hline
$\varepsilon_{ij}$&1&2&3&4&5&6&7&8&9&10&11&12&13&14&15&16&17&18\\ \hline
1&67.76&4.08&0.33&0.12&0.03&0.01&0.00&0.00&0.00&0.00&0.00&0.00&0.00&0.00&0.00&0.00&0.00&0.00\\
2&2.60&62.57&5.03&0.41&0.14&0.02&0.01&0.01&0.00&0.00&0.00&0.00&0.00&0.00&0.00&0.00&0.00&0.00\\
3&0.09&3.26&60.63&5.45&0.43&0.12&0.03&0.01&0.01&0.00&0.00&0.00&0.00&0.00&0.00&0.00&0.00&0.00\\
4&0.03&0.13&3.65&59.30&5.68&0.44&0.12&0.03&0.01&0.01&0.01&0.00&0.00&0.00&0.00&0.00&0.00&0.00\\
5&0.01&0.04&0.16&3.89&58.72&5.75&0.46&0.11&0.03&0.02&0.01&0.00&0.00&0.00&0.00&0.00&0.00&0.00\\
6&0.01&0.02&0.06&0.19&4.08&58.14&5.70&0.44&0.10&0.03&0.02&0.01&0.00&0.00&0.00&0.00&0.00&0.00\\
7&0.01&0.01&0.03&0.06&0.22&4.26&57.89&5.67&0.42&0.10&0.04&0.02&0.01&0.00&0.00&0.00&0.00&0.00\\
8&0.01&0.01&0.01&0.03&0.07&0.24&4.27&57.51&5.64&0.40&0.11&0.04&0.02&0.01&0.00&0.00&0.00&0.00\\
9&0.00&0.01&0.01&0.02&0.03&0.08&0.26&4.26&57.38&5.39&0.40&0.10&0.04&0.01&0.00&0.00&0.00&0.00\\
10&0.00&0.00&0.01&0.01&0.02&0.03&0.08&0.27&4.30&57.00&5.24&0.37&0.09&0.03&0.01&0.01&0.00&0.00\\
11&0.00&0.00&0.00&0.00&0.01&0.02&0.04&0.08&0.28&4.18&57.04&5.09&0.35&0.09&0.03&0.01&0.00&0.00\\
12&0.00&0.00&0.00&0.00&0.00&0.01&0.02&0.04&0.09&0.31&4.06&56.46&4.79&0.32&0.09&0.03&0.00&0.00\\
13&0.00&0.00&0.00&0.00&0.00&0.00&0.01&0.01&0.04&0.08&0.31&3.95&55.87&4.51&0.28&0.06&0.01&0.00\\
14&0.00&0.00&0.00&0.00&0.00&0.00&0.00&0.01&0.01&0.03&0.08&0.32&3.72&54.89&4.19&0.21&0.03&0.01\\
15&0.00&0.00&0.00&0.00&0.00&0.00&0.00&0.00&0.00&0.01&0.02&0.07&0.30&3.51&53.46&3.86&0.18&0.02\\
16&0.00&0.00&0.00&0.00&0.00&0.00&0.00&0.00&0.00&0.00&0.01&0.02&0.06&0.24&3.12&51.56&3.48&0.10\\
17&0.00&0.00&0.00&0.00&0.00&0.00&0.00&0.00&0.00&0.00&0.00&0.00&0.01&0.03&0.16&2.58&48.40&2.76\\
18&0.00&0.00&0.00&0.00&0.00&0.00&0.00&0.00&0.00&0.00&0.00&0.00&0.00&0.00&0.01&0.08&1.84&37.29\\
\hline
\hline
\end{tabular}
	}
\end{table*}

\begin{table*}[htbp]
\caption{%The weighted efficiency matrix (in \%) for $\kmunu$, where $i$ denotes the reconstructed bin, and $j$ represents the produced bin.
The weighted efficiency matrix for $\kmunu$, defined as in Table~\ref{kenu_effmatrix}.
\label{kmunu_effmatrix}}
\centering
\resizebox{1.0\textwidth}{!}{
\begin{tabular}{c|cccccccccccccccccc}
\hline
\hline
$\varepsilon_{ij}$&1&2&3&4&5&6&7&8&9&10&11&12&13&14&15&16&17&18\\ \hline
1&44.06&1.39&0.02&0.00&0.00&0.00&0.00&0.00&0.00&0.00&0.00&0.00&0.00&0.00&0.00&0.00&0.00&0.00\\
2&1.86&43.59&2.13&0.05&0.01&0.00&0.00&0.00&0.00&0.00&0.00&0.00&0.00&0.00&0.00&0.00&0.00&0.00\\
3&0.06&2.05&45.33&2.72&0.08&0.02&0.01&0.00&0.00&0.00&0.00&0.00&0.00&0.00&0.00&0.00&0.00&0.00\\
4&0.02&0.08&2.58&47.81&3.15&0.11&0.03&0.01&0.00&0.00&0.00&0.00&0.00&0.00&0.00&0.00&0.00&0.00\\
5&0.01&0.03&0.11&3.09&50.24&3.51&0.15&0.05&0.02&0.00&0.00&0.00&0.00&0.00&0.00&0.00&0.00&0.00\\
6&0.01&0.01&0.04&0.14&3.44&52.79&3.76&0.18&0.06&0.03&0.01&0.00&0.00&0.00&0.00&0.00&0.00&0.00\\
7&0.01&0.01&0.02&0.05&0.17&3.80&55.00&4.00&0.21&0.07&0.03&0.02&0.00&0.00&0.00&0.00&0.00&0.00\\
8&0.00&0.00&0.01&0.02&0.06&0.21&4.03&56.58&4.05&0.22&0.08&0.04&0.02&0.01&0.00&0.00&0.00&0.00\\
9&0.00&0.00&0.01&0.01&0.03&0.07&0.23&4.13&57.57&4.13&0.23&0.08&0.03&0.01&0.00&0.00&0.00&0.00\\
10&0.00&0.00&0.00&0.01&0.01&0.03&0.08&0.25&4.17&57.95&4.00&0.26&0.08&0.03&0.01&0.00&0.00&0.00\\
11&0.00&0.00&0.00&0.00&0.01&0.02&0.03&0.08&0.27&4.11&57.81&3.83&0.25&0.09&0.03&0.01&0.00&0.00\\
12&0.00&0.00&0.00&0.00&0.00&0.01&0.02&0.03&0.09&0.29&3.91&57.46&3.75&0.24&0.07&0.02&0.00&0.00\\
13&0.00&0.00&0.00&0.00&0.00&0.00&0.01&0.01&0.03&0.08&0.27&3.77&56.58&3.49&0.23&0.06&0.01&0.00\\
14&0.00&0.00&0.00&0.00&0.00&0.00&0.00&0.00&0.01&0.03&0.08&0.29&3.56&56.00&3.33&0.18&0.03&0.01\\
15&0.00&0.00&0.00&0.00&0.00&0.00&0.00&0.00&0.00&0.01&0.02&0.07&0.28&3.35&54.79&3.11&0.13&0.02\\
16&0.00&0.00&0.00&0.00&0.00&0.00&0.00&0.00&0.00&0.00&0.01&0.02&0.05&0.20&2.95&52.39&2.85&0.07\\
17&0.00&0.00&0.00&0.00&0.00&0.00&0.00&0.00&0.00&0.00&0.00&0.00&0.01&0.03&0.16&2.45&48.92&2.11\\
18&0.00&0.00&0.00&0.00&0.00&0.00&0.00&0.00&0.00&0.00&0.00&0.00&0.00&0.01&0.01&0.06&1.77&37.01\\
\hline
\hline
\end{tabular}
	}
\end{table*}

\begin{table*}[htbp]
\caption{%The weighted efficiency matrix (in \%) for $\ksenu$, where $i$ denotes the reconstructed bin, and $j$ represents the produced bin.
The weighted efficiency matrix for $\ksenu$, defined as in Table~\ref{kenu_effmatrix}.
\label{ksenu_effmatrix}}
\centering
\resizebox{1.0\textwidth}{!}{
\begin{tabular}{c|cccccccccccccccccc}
\hline
\hline
$\varepsilon_{ij}$&1&2&3&4&5&6&7&8&9&10&11&12&13&14&15&16&17&18\\ \hline
1&48.49&2.69&0.19&0.07&0.01&0.00&0.00&0.00&0.00&0.00&0.00&0.00&0.00&0.00&0.00&0.00&0.00&0.00\\
2&1.49&44.44&3.34&0.22&0.07&0.01&0.00&0.00&0.00&0.00&0.00&0.00&0.00&0.00&0.00&0.00&0.00&0.00\\
3&0.02&1.85&42.32&3.62&0.21&0.05&0.00&0.00&0.00&0.00&0.00&0.00&0.00&0.00&0.00&0.00&0.00&0.00\\
4&0.00&0.03&2.04&41.02&3.69&0.21&0.04&0.00&0.00&0.00&0.00&0.00&0.00&0.00&0.00&0.00&0.00&0.00\\
5&0.00&0.01&0.05&2.16&39.99&3.72&0.19&0.02&0.00&0.00&0.00&0.00&0.00&0.00&0.00&0.00&0.00&0.00\\
6&0.00&0.00&0.01&0.06&2.24&39.11&3.68&0.18&0.02&0.00&0.00&0.00&0.00&0.00&0.00&0.00&0.00&0.00\\
7&0.00&0.00&0.00&0.01&0.06&2.30&38.44&3.64&0.15&0.01&0.00&0.00&0.00&0.00&0.00&0.00&0.00&0.00\\
8&0.00&0.00&0.00&0.00&0.01&0.06&2.35&37.80&3.51&0.14&0.01&0.00&0.00&0.00&0.00&0.00&0.00&0.00\\
9&0.00&0.00&0.00&0.00&0.00&0.01&0.07&2.35&37.26&3.38&0.11&0.01&0.00&0.00&0.00&0.00&0.00&0.00\\
10&0.00&0.00&0.00&0.00&0.00&0.00&0.01&0.07&2.33&36.87&3.18&0.09&0.00&0.00&0.00&0.00&0.00&0.00\\
11&0.00&0.00&0.00&0.00&0.00&0.00&0.00&0.01&0.08&2.28&36.43&3.02&0.08&0.00&0.00&0.00&0.00&0.00\\
12&0.00&0.00&0.00&0.00&0.00&0.00&0.00&0.00&0.01&0.08&2.26&36.15&2.88&0.06&0.00&0.00&0.00&0.00\\
13&0.00&0.00&0.00&0.00&0.00&0.00&0.00&0.00&0.00&0.01&0.08&2.19&35.78&2.68&0.05&0.00&0.00&0.00\\
14&0.00&0.00&0.00&0.00&0.00&0.00&0.00&0.00&0.00&0.00&0.01&0.08&2.12&35.31&2.50&0.03&0.00&0.00\\
15&0.00&0.00&0.00&0.00&0.00&0.00&0.00&0.00&0.00&0.00&0.00&0.01&0.08&1.93&35.06&2.33&0.02&0.00\\
16&0.00&0.00&0.00&0.00&0.00&0.00&0.00&0.00&0.00&0.00&0.00&0.00&0.01&0.08&1.78&34.70&2.13&0.02\\
17&0.00&0.00&0.00&0.00&0.00&0.00&0.00&0.00&0.00&0.00&0.00&0.00&0.00&0.01&0.06&1.55&34.27&1.70\\
18&0.00&0.00&0.00&0.00&0.00&0.00&0.00&0.00&0.00&0.00&0.00&0.00&0.00&0.00&0.01&0.05&1.24&32.54\\
\hline
\hline
\end{tabular}
	}
\end{table*}

\begin{table*}[htbp]
\caption{%The weighted efficiency matrix (in \%) for $\ksmunu$, where $i$ denotes the reconstructed bin, and $j$ represents the produced bin.
The weighted efficiency matrix for $\ksmunu$, defined as in Table~\ref{kenu_effmatrix}.
\label{ksmunu_effmatrix}}
\centering
\resizebox{1.0\textwidth}{!}{
\begin{tabular}{c|cccccccccccccccccc}
\hline
\hline
$\varepsilon_{ij}$&1&2&3&4&5&6&7&8&9&10&11&12&13&14&15&16&17&18\\ \hline
1&33.03&0.97&0.00&0.00&0.00&0.00&0.00&0.00&0.00&0.00&0.00&0.00&0.00&0.00&0.00&0.00&0.00&0.00\\
2&1.15&32.04&1.45&0.01&0.00&0.00&0.00&0.00&0.00&0.00&0.00&0.00&0.00&0.00&0.00&0.00&0.01&0.00\\
3&0.01&1.24&32.64&1.80&0.02&0.00&0.00&0.00&0.00&0.00&0.00&0.00&0.00&0.00&0.00&0.01&0.01&0.01\\
4&0.00&0.02&1.52&33.71&2.06&0.02&0.00&0.00&0.00&0.00&0.00&0.00&0.00&0.00&0.00&0.01&0.00&0.01\\
5&0.00&0.00&0.02&1.76&34.89&2.28&0.03&0.00&0.00&0.00&0.00&0.00&0.00&0.01&0.00&0.01&0.01&0.01\\
6&0.00&0.00&0.00&0.04&1.92&35.93&2.43&0.03&0.00&0.00&0.00&0.00&0.01&0.00&0.01&0.01&0.01&0.01\\
7&0.00&0.00&0.00&0.01&0.04&2.10&36.77&2.48&0.03&0.01&0.00&0.00&0.01&0.01&0.01&0.01&0.02&0.01\\
8&0.00&0.00&0.00&0.00&0.01&0.05&2.19&37.18&2.52&0.03&0.01&0.01&0.01&0.01&0.01&0.01&0.01&0.01\\
9&0.00&0.00&0.00&0.00&0.00&0.01&0.06&2.25&37.28&2.46&0.03&0.01&0.01&0.00&0.01&0.01&0.01&0.00\\
10&0.00&0.00&0.00&0.00&0.00&0.00&0.01&0.07&2.26&36.87&2.40&0.03&0.01&0.01&0.01&0.01&0.01&0.00\\
11&0.00&0.00&0.00&0.00&0.00&0.01&0.01&0.01&0.07&2.20&36.54&2.28&0.03&0.01&0.01&0.01&0.00&0.00\\
12&0.00&0.00&0.00&0.00&0.00&0.01&0.00&0.01&0.02&0.08&2.14&36.33&2.19&0.02&0.01&0.01&0.00&0.00\\
13&0.00&0.00&0.00&0.00&0.00&0.00&0.01&0.00&0.01&0.02&0.07&2.08&35.81&2.05&0.02&0.00&0.00&0.00\\
14&0.00&0.00&0.00&0.00&0.00&0.00&0.01&0.01&0.01&0.01&0.02&0.07&1.98&35.69&1.90&0.00&0.00&0.00\\
15&0.00&0.00&0.00&0.00&0.00&0.00&0.01&0.01&0.01&0.01&0.01&0.02&0.07&1.84&35.09&1.88&0.01&0.00\\
16&0.00&0.00&0.00&0.00&0.00&0.01&0.01&0.01&0.01&0.01&0.01&0.01&0.01&0.07&1.70&35.08&1.68&0.00\\
17&0.00&0.00&0.00&0.01&0.00&0.00&0.01&0.01&0.01&0.00&0.00&0.00&0.00&0.01&0.05&1.46&34.20&1.28\\
18&0.00&0.01&0.01&0.01&0.01&0.01&0.00&0.00&0.00&0.00&0.00&0.00&0.00&0.00&0.01&0.04&1.17&32.74\\
\hline
\hline
\end{tabular}
	}
\end{table*}

\begin{table}[htbp]
\caption{ The fitted DT yields ($N_{\rm DT}^i$), the produced yields ($N_{\rm prd}^i$) and the determined partial decay rates ($\Delta\Gamma$) of $\kmunu$ in different $q^2$ intervals of data, where the uncertainties are statistical only. \label{tab:kmunu_decayrate}}
\centering
\resizebox{0.49\textwidth}{!}{
\begin{tabular}{cccc}
\hline
\hline
$q^2~({\rm GeV^2}/c^4)$&$N_{\rm DT}^i$&$N_{\rm prd}^i$&$\Delta\Gamma~({\rm ns^{-1}})$\\ \hline
$(0.0,0.1)$&  $23870\pm183$&$51906\pm417$&$6.255\pm0.050$ \\
$(0.1,0.2)$&  $33402\pm209$&$71004\pm481$&$8.556\pm0.058$ \\
$(0.2,0.3)$&  $34109\pm214$&$68012\pm475$&$8.195\pm0.057$ \\
$(0.3,0.4)$&  $34171\pm220$&$63532\pm464$&$7.655\pm0.056$ \\
$(0.4,0.5)$&  $34333\pm230$&$60112\pm463$&$7.243\pm0.056$ \\
$(0.5,0.6)$&  $33721\pm236$&$55816\pm453$&$6.726\pm0.055$ \\
$(0.6,0.7)$&  $32588\pm224$&$51439\pm413$&$6.198\pm0.050$ \\
$(0.7,0.8)$&  $30747\pm212$&$47083\pm380$&$5.673\pm0.046$ \\
$(0.8,0.9)$&  $28447\pm202$&$42776\pm356$&$5.154\pm0.043$ \\
$(0.9,1.0)$&  $25562\pm193$&$38241\pm338$&$4.608\pm0.041$ \\
$(1.0,1.1)$&  $22134\pm186$&$33171\pm327$&$3.997\pm0.039$ \\
$(1.1,1.2)$&  $19180\pm182$&$29121\pm320$&$3.509\pm0.039$ \\
$(1.2,1.3)$&  $15714\pm174$&$24253\pm312$&$2.922\pm0.038$ \\
$(1.3,1.4)$&  $12893\pm174$&$20318\pm314$&$2.448\pm0.038$ \\
$(1.4,1.5)$&  $9396\pm154$&$15124\pm284$&$1.822\pm0.034$ \\
$(1.5,1.6)$&  $6102\pm99$&$10320\pm190$&$1.243\pm0.023$ \\
$(1.6,1.7)$&  $3540\pm85$&$6467\pm176$&$0.779\pm0.021$ \\
$>1.7$&  $1710\pm71$&$4285\pm194$&$0.516\pm0.023$ \\
\hline
\hline
\end{tabular}
	}
\end{table}

\begin{table}[htbp]
\caption{ %The fitted DT yields ($N_{\rm DT}^i$), the produced yields ($N_{\rm prd}^i$) and the determined partial decay rates ($\Delta\Gamma$) of $\ksenu$ in different $q^2$ intervals of data, where the uncertainties are statistical only. 
Same as Table~\ref{tab:kmunu_decayrate}, but for $\ksenu$.
\label{tab:ksenu_decayrate}}
\centering
\resizebox{0.49\textwidth}{!}{
\begin{tabular}{cccc}
\hline
\hline
$q^2~({\rm GeV^2}/c^4)$&$N_{\rm DT}^i$&$N_{\rm prd}^i$&$\Delta\Gamma~({\rm ns^{-1}})$\\ \hline
$(0.0,0.1)$&  $17837\pm144$&$100514\pm859$&$9.139\pm0.078$ \\
$(0.1,0.2)$&  $16363\pm138$&$95738\pm906$&$8.705\pm0.082$ \\
$(0.2,0.3)$&  $14902\pm131$&$89666\pm906$&$8.153\pm0.082$ \\
$(0.3,0.4)$&  $13977\pm127$&$86390\pm906$&$7.855\pm0.082$ \\
$(0.4,0.5)$&  $12609\pm120$&$79104\pm879$&$7.192\pm0.080$ \\
$(0.5,0.6)$&  $11532\pm114$&$73772\pm857$&$6.708\pm0.078$ \\
$(0.6,0.7)$&  $10540\pm109$&$68410\pm835$&$6.220\pm0.076$ \\
$(0.7,0.8)$&  $9637\pm104$&$63872\pm810$&$5.807\pm0.074$ \\
$(0.8,0.9)$&  $8403\pm97$&$56229\pm764$&$5.113\pm0.069$ \\
$(0.9,1.0)$&  $7485\pm91$&$51026\pm726$&$4.639\pm0.066$ \\
$(1.0,1.1)$&  $6466\pm85$&$44755\pm681$&$4.069\pm0.062$ \\
$(1.1,1.2)$&  $5432\pm78$&$37865\pm630$&$3.443\pm0.057$ \\
$(1.2,1.3)$&  $4577\pm71$&$32532\pm585$&$2.958\pm0.053$ \\
$(1.3,1.4)$&  $3652\pm64$&$26470\pm527$&$2.407\pm0.048$ \\
$(1.4,1.5)$&  $2626\pm54$&$19210\pm450$&$1.747\pm0.041$ \\
$(1.5,1.6)$&  $1801\pm45$&$13404\pm378$&$1.219\pm0.034$ \\
$(1.6,1.7)$&  $1146\pm36$&$8800\pm305$&$0.800\pm0.028$ \\
$>1.7$&  $528\pm25$&$4330\pm218$&$0.394\pm0.020$ \\
\hline
\hline
\end{tabular}
	}
\end{table}

\begin{table}[htbp]
\caption{ %The fitted DT yields ($N_{\rm DT}^i$), the produced yields ($N_{\rm prd}^i$) and the determined partial decay rates ($\Delta\Gamma$) of $\ksmunu$ in different $q^2$ intervals of data, where the uncertainties are statistical only. 
Same as Table~\ref{tab:kmunu_decayrate}, but for $\ksmunu$.
\label{tab:ksmunu_decayrate}}
\centering
\resizebox{0.49\textwidth}{!}{
\begin{tabular}{cccc}
\hline
\hline
$q^2~({\rm GeV^2}/c^4)$&$N_{\rm DT}^i$&$N_{\rm prd}^i$&$\Delta\Gamma~({\rm ns^{-1}})$\\ \hline
$(0.0,0.1)$&  $8335\pm97$&$70192\pm852$&$6.382\pm0.078$ \\
$(0.1,0.2)$&  $11083\pm112$&$93318\pm1012$&$8.485\pm0.092$ \\
$(0.2,0.3)$&  $11176\pm113$&$90718\pm1007$&$8.248\pm0.092$ \\
$(0.3,0.4)$&  $10832\pm112$&$83751\pm965$&$7.615\pm0.088$ \\
$(0.4,0.5)$&  $10835\pm114$&$80508\pm958$&$7.320\pm0.087$ \\
$(0.5,0.6)$&  $10436\pm117$&$74820\pm949$&$6.803\pm0.086$ \\
$(0.6,0.7)$&  $9864\pm114$&$68826\pm905$&$6.258\pm0.082$ \\
$(0.7,0.8)$&  $9113\pm108$&$62691\pm850$&$5.700\pm0.077$ \\
$(0.8,0.9)$&  $8327\pm105$&$57250\pm822$&$5.205\pm0.075$ \\
$(0.9,1.0)$&  $7283\pm100$&$50350\pm791$&$4.578\pm0.072$ \\
$(1.0,1.1)$&  $6606\pm96$&$46606\pm772$&$4.238\pm0.070$ \\
$(1.1,1.2)$&  $5465\pm93$&$38543\pm747$&$3.504\pm0.068$ \\
$(1.2,1.3)$&  $4561\pm89$&$32869\pm727$&$2.989\pm0.066$ \\
$(1.3,1.4)$&  $3696\pm81$&$26902\pm661$&$2.446\pm0.060$ \\
$(1.4,1.5)$&  $2650\pm72$&$19408\pm597$&$1.765\pm0.054$ \\
$(1.5,1.6)$&  $2092\pm51$&$15661\pm420$&$1.424\pm0.038$ \\
$(1.6,1.7)$&  $1253\pm40$&$9583\pm344$&$0.871\pm0.031$ \\
$>1.7$&  $714\pm34$&$5804\pm304$&$0.528\pm0.028$ \\
\hline
\hline
\end{tabular}
	}
\end{table}

\begin{table*}[htbp]
\caption{%The statistical covariance matrix for $\kenu$, where $i$ denotes the reconstructed bin, and $j$ represents the produced bin.
The statistical covariance matrix for $\kenu$. The indices $i$ and $j$ denote the $q^2$ intervals.
\label{tab:kenu_statmatrix}}
\centering
\resizebox{1.0\textwidth}{!}{
\begin{tabular}{c|cccccccccccccccccc}
\hline
\hline
$\rho_{ij}^{\rm stat}$&1&2&3&4&5&6&7&8&9&10&11&12&13&14&15&16&17&18\\ \hline
1&1.000&-0.102&0.004&-0.002&-0.000&-0.000&-0.000&-0.000&0.000&-0.000&0.000&0.000&0.000&0.000&-0.000&-0.000&-0.000&0.000\\
2&-0.102&1.000&-0.133&0.007&-0.002&-0.000&-0.000&-0.000&-0.000&0.000&-0.000&0.000&0.000&0.000&-0.000&0.000&-0.000&0.000\\
3&0.004&-0.133&1.000&-0.149&0.009&-0.002&-0.000&-0.000&-0.000&-0.000&0.000&-0.000&0.000&0.000&0.000&-0.000&0.000&-0.000\\
4&-0.002&0.007&-0.149&1.000&-0.159&0.010&-0.002&-0.000&-0.000&-0.000&-0.000&0.000&0.000&0.000&0.000&0.000&-0.000&0.000\\
5&-0.000&-0.002&0.009&-0.159&1.000&-0.165&0.011&-0.002&-0.000&-0.000&-0.000&0.000&-0.000&0.000&0.000&0.000&-0.000&0.000\\
6&-0.000&-0.000&-0.002&0.010&-0.165&1.000&-0.168&0.011&-0.002&-0.000&-0.000&-0.000&0.000&-0.000&0.000&0.000&-0.000&0.000\\
7&-0.000&-0.000&-0.000&-0.002&0.011&-0.168&1.000&-0.168&0.011&-0.002&-0.001&-0.000&-0.000&0.000&-0.000&0.000&-0.000&-0.000\\
\hline
\hline
\end{tabular}
	}
\end{table*}

\begin{table*}[htbp]
\caption{%The statistical covariance matrix for $\kmunu$, where $i$ denotes the reconstructed bin, and $j$ represents the produced bin.
Same as Table~\ref{tab:kenu_statmatrix}, but for $\kmunu$.
\label{tab:kmunu_statmatrix}}
\centering
\resizebox{1.0\textwidth}{!}{
\begin{tabular}{c|cccccccccccccccccc}
\hline
\hline
$\rho_{ij}^{\rm stat}$&1&2&3&4&5&6&7&8&9&10&11&12&13&14&15&16&17&18\\ \hline
1&1.000&-0.073&0.003&-0.000&-0.000&-0.000&-0.000&0.000&-0.000&0.000&0.000&0.000&0.000&-0.000&-0.000&-0.000&-0.000&0.000\\
2&-0.073&1.000&-0.094&0.005&-0.001&-0.000&-0.000&0.000&0.000&-0.000&0.000&0.000&0.000&-0.000&0.000&-0.000&0.000&0.000\\
3&0.003&-0.094&1.000&-0.113&0.007&-0.001&-0.000&-0.000&-0.000&0.000&-0.000&0.000&0.000&0.000&-0.000&0.000&-0.000&0.000\\
4&-0.000&0.005&-0.113&1.000&-0.127&0.008&-0.002&-0.000&-0.000&-0.000&0.000&-0.000&0.000&0.000&0.000&-0.000&0.000&-0.000\\
5&-0.000&-0.001&0.007&-0.127&1.000&-0.134&0.009&-0.002&-0.001&-0.000&-0.000&0.000&-0.000&0.000&-0.000&0.000&-0.000&-0.000\\
6&-0.000&-0.000&-0.001&0.008&-0.134&1.000&-0.139&0.008&-0.002&-0.001&-0.000&-0.000&0.000&-0.000&0.000&0.000&-0.000&0.000\\
7&-0.000&-0.000&-0.000&-0.002&0.009&-0.139&1.000&-0.143&0.008&-0.002&-0.000&-0.001&-0.000&0.000&0.000&0.000&-0.000&-0.000\\
\hline
\hline
\end{tabular}
	}
\end{table*}

\begin{table*}[htbp]
\caption{%The statistical covariance matrix for $\ksenu$, where $i$ denotes the reconstructed bin, and $j$ represents the produced bin.
Same as Table~\ref{tab:kenu_statmatrix}, but for $\ksenu$.
\label{tab:ksenu_statmatrix}}
\centering
\resizebox{1.0\textwidth}{!}{
\begin{tabular}{c|cccccccccccccccccc}
\hline
\hline
$\rho_{ij}^{\rm stat}$&1&2&3&4&5&6&7&8&9&10&11&12&13&14&15&16&17&18\\ \hline
1&1.000&-0.090&0.004&-0.001&0.000&0.000&-0.000&0.000&-0.000&0.000&-0.000&0.000&-0.000&0.000&-0.000&0.000&-0.000&0.000\\
2&-0.090&1.000&-0.118&0.007&-0.002&0.000&0.000&0.000&-0.000&0.000&-0.000&0.000&-0.000&0.000&-0.000&0.000&-0.000&0.000\\
3&0.004&-0.118&1.000&-0.134&0.009&-0.002&0.000&-0.000&0.000&-0.000&0.000&-0.000&0.000&-0.000&0.000&-0.000&0.000&-0.000\\
4&-0.001&0.007&-0.134&1.000&-0.142&0.010&-0.001&0.000&-0.000&0.000&-0.000&0.000&-0.000&0.000&-0.000&0.000&-0.000&0.000\\
5&0.000&-0.002&0.009&-0.142&1.000&-0.148&0.011&-0.001&0.000&-0.000&0.000&-0.000&0.000&-0.000&0.000&-0.000&0.000&-0.000\\
6&0.000&0.000&-0.002&0.010&-0.148&1.000&-0.152&0.012&-0.001&0.000&-0.000&0.000&-0.000&0.000&-0.000&0.000&-0.000&0.000\\
7&-0.000&0.000&0.000&-0.001&0.011&-0.152&1.000&-0.155&0.013&-0.001&0.000&-0.000&0.000&-0.000&0.000&-0.000&0.000&-0.000\\
\hline
\hline
\end{tabular}
	}
\end{table*}

\begin{table*}[htbp]
\caption{%The statistical covariance matrix for $\ksmunu$, where $i$ denotes the reconstructed bin, and $j$ represents the produced bin.
Same as Table~\ref{tab:kenu_statmatrix}, but for $\ksmunu$.
\label{tab:ksmunu_statmatrix}}
\centering
\resizebox{1.0\textwidth}{!}{
\begin{tabular}{c|cccccccccccccccccc}
\hline
\hline
$\rho_{ij}^{\rm stat}$&1&2&3&4&5&6&7&8&9&10&11&12&13&14&15&16&17&18\\ \hline
1&1.000&-0.065&0.004&-0.000&0.000&-0.000&0.000&-0.000&-0.000&0.000&0.000&0.000&0.000&-0.000&0.000&-0.000&0.000&0.000\\
2&-0.065&1.000&-0.083&0.005&-0.000&0.000&-0.000&0.000&0.000&-0.000&-0.000&-0.000&-0.000&0.000&-0.000&0.000&-0.000&-0.001\\
3&0.004&-0.083&1.000&-0.100&0.007&-0.000&0.000&-0.000&0.000&0.000&0.000&0.000&0.000&-0.000&0.000&-0.000&0.000&-0.001\\
4&-0.000&0.005&-0.100&1.000&-0.111&0.008&-0.001&0.000&-0.000&0.000&-0.000&-0.000&-0.000&0.000&0.000&-0.000&-0.001&-0.001\\
5&0.000&-0.000&0.007&-0.111&1.000&-0.118&0.009&-0.001&0.000&-0.000&0.000&0.000&0.000&-0.000&0.000&-0.000&0.000&-0.001\\
6&-0.000&0.000&-0.000&0.008&-0.118&1.000&-0.124&0.010&-0.001&0.000&-0.000&-0.000&-0.000&0.000&-0.000&-0.001&0.000&-0.001\\
7&0.000&-0.000&0.000&-0.001&0.009&-0.124&1.000&-0.126&0.010&-0.001&-0.000&0.000&-0.001&-0.000&-0.000&-0.000&-0.001&0.000\\
\hline
\hline
\end{tabular}
	}
\end{table*}

\begin{table*}[htbp]
\caption{Systematic uncertainties (in \%) of the partial decay rates for $\kmunu$ in individual $q^2$ intervals.
\label{tab:kmunu_sysq2}}
\centering
\resizebox{1.0\textwidth}{!}{
\begin{tabular}{ccccccccccccccccccc}
\hline
\hline
$i$-th $q^2$ bin&1&2&3&4&5&6&7&8&9&10&11&12&13&14&15&16&17&18\\ \hline
$N_{\rm tag}$&0.30&0.30&0.30&0.30&0.30&0.30&0.30&0.30&0.30&0.30&0.30&0.30&0.30&0.30&0.30&0.30&0.30&0.30\\
$D^0$ lifetime&0.24&0.24&0.24&0.24&0.24&0.24&0.24&0.24&0.24&0.24&0.24&0.24&0.24&0.24&0.24&0.24&0.24&0.24\\
MC stat.&0.13&0.11&0.11&0.11&0.11&0.11&0.11&0.11&0.12&0.12&0.13&0.14&0.16&0.18&0.20&0.25&0.33&0.54\\
$E_{\rm extra \gamma}^{\rm max}$ cut&0.10&0.10&0.10&0.10&0.10&0.10&0.10&0.10&0.10&0.10&0.10&0.10&0.10&0.10&0.10&0.10&0.10&0.10\\
$U_{\rm miss}$ fit&0.42&0.13&0.12&0.12&0.14&0.17&0.16&0.18&0.18&0.18&0.17&0.19&0.17&0.15&0.18&2.59&5.12&9.89\\
$K$ tracking&0.10&0.10&0.10&0.10&0.10&0.10&0.10&0.10&0.10&0.10&0.10&0.10&0.10&0.10&0.10&0.11&0.14&0.27\\
$K$ PID&0.10&0.10&0.10&0.10&0.10&0.10&0.10&0.10&0.10&0.10&0.10&0.10&0.10&0.10&0.10&0.10&0.10&0.10\\
$\mu$ tracking&0.10&0.10&0.10&0.10&0.10&0.10&0.10&0.10&0.10&0.10&0.10&0.10&0.10&0.10&0.10&0.10&0.10&0.10\\
$\mu$ PID&0.12&0.12&0.11&0.11&0.11&0.11&0.10&0.10&0.10&0.11&0.11&0.11&0.12&0.12&0.12&0.13&0.13&0.15\\
MC model&0.12&0.40&0.59&0.07&0.20&0.39&0.24&0.38&0.47&0.27&0.34&0.49&0.44&0.12&0.92&0.61&1.23&1.83\\
\hline
Total&0.64&0.63&0.76&0.48&0.52&0.63&0.55&0.62&0.68&0.56&0.60&0.70&0.67&0.52&1.06&2.71&5.30&10.08\\
\hline
\hline
\end{tabular}
	}
\end{table*}

\begin{table*}[htbp]
\caption{%The systematic uncertainties (in \%) of the measured decay rates of $\ksenu$ in different $q^2$ bins.
Same as Table~\ref{tab:kmunu_sysq2}, but for $\ksenu$.
\label{tab:ksenu_sysq2}}
\centering
\resizebox{1.0\textwidth}{!}{
\begin{tabular}{ccccccccccccccccccc}
\hline
\hline
$i$-th $q^2$ bin&1&2&3&4&5&6&7&8&9&10&11&12&13&14&15&16&17&18\\ \hline
$N_{\rm tag}$&0.30&0.30&0.30&0.30&0.30&0.30&0.30&0.30&0.30&0.30&0.30&0.30&0.30&0.30&0.30&0.30&0.30&0.30\\
$D^+$ lifetime&0.48&0.48&0.48&0.48&0.48&0.48&0.48&0.48&0.48&0.48&0.48&0.48&0.48&0.48&0.48&0.48&0.48&0.48\\
MC stat.&0.10&0.11&0.11&0.12&0.12&0.13&0.13&0.14&0.15&0.16&0.17&0.18&0.20&0.23&0.26&0.31&0.40&0.58\\
$E_{\rm extra \gamma}^{\rm max}$ cut&0.10&0.10&0.10&0.10&0.10&0.10&0.10&0.10&0.10&0.10&0.10&0.10&0.10&0.10&0.10&0.10&0.10&0.10\\
$U_{\rm miss}$ fit&0.07&0.07&0.06&0.06&0.06&0.06&0.06&0.05&0.07&0.05&0.04&0.06&0.07&0.05&0.08&0.08&0.09&0.09\\
$e^+$ tracking&0.10&0.10&0.10&0.10&0.10&0.10&0.10&0.10&0.10&0.10&0.10&0.10&0.10&0.10&0.10&0.10&0.10&0.10\\
$e^+$ PID&0.10&0.10&0.10&0.10&0.10&0.10&0.10&0.10&0.10&0.10&0.10&0.10&0.10&0.10&0.10&0.10&0.10&0.10\\
$K_S^0$ reconstruction&0.15&0.17&0.19&0.22&0.24&0.31&0.39&0.46&0.54&0.63&0.75&0.83&0.88&0.95&1.04&1.13&1.17&1.37\\
$\pi$ tracking&0.10&0.10&0.10&0.10&0.10&0.11&0.11&0.12&0.12&0.13&0.13&0.13&0.14&0.14&0.14&0.15&0.16&0.19\\
Quoted BF&0.07&0.07&0.07&0.07&0.07&0.07&0.07&0.07&0.07&0.07&0.07&0.07&0.07&0.07&0.07&0.07&0.07&0.07\\
MC model&0.15&0.40&0.27&0.42&0.41&0.21&0.54&0.29&0.66&0.74&0.69&0.47&0.14&0.33&0.87&1.24&1.52&0.54\\
\hline
Total&0.65&0.76&0.70&0.78&0.78&0.73&0.91&0.83&1.06&1.16&1.20&1.15&1.10&1.20&1.51&1.82&2.06&1.71\\
\hline
\hline
\end{tabular}
	}
\end{table*}

\begin{table*}[htbp]
\caption{%The systematic uncertainties (in \%) of the measured decay rates of $\ksmunu$ in different $q^2$ bins.
Same as Table~\ref{tab:kmunu_sysq2}, but for $\ksmunu$.
\label{tab:ksmunu_sysq2}}
\centering
\resizebox{1.0\textwidth}{!}{
\begin{tabular}{ccccccccccccccccccc}
\hline
\hline
$i$-th $q^2$ bin&1&2&3&4&5&6&7&8&9&10&11&12&13&14&15&16&17&18\\ \hline
$N_{\rm tag}$&0.30&0.30&0.30&0.30&0.30&0.30&0.30&0.30&0.30&0.30&0.30&0.30&0.30&0.30&0.30&0.30&0.30&0.30\\
$D^+$ lifetime&0.48&0.48&0.48&0.48&0.48&0.48&0.48&0.48&0.48&0.48&0.48&0.48&0.48&0.48&0.48&0.48&0.48&0.48\\
MC stat.&0.15&0.13&0.13&0.13&0.13&0.13&0.14&0.14&0.15&0.16&0.17&0.18&0.20&0.23&0.26&0.31&0.40&0.55\\
$E_{\rm extra \gamma}^{\rm max}$ cut&0.10&0.10&0.10&0.10&0.10&0.10&0.10&0.10&0.10&0.10&0.10&0.10&0.10&0.10&0.10&0.10&0.10&0.10\\
$U_{\rm miss}$ fit&0.04&0.03&0.04&0.05&0.04&0.06&0.06&0.06&0.06&0.07&0.07&0.06&0.09&0.08&0.11&0.84&1.14&2.52\\
$\mu^+$ tracking&0.10&0.10&0.10&0.10&0.10&0.10&0.10&0.10&0.10&0.10&0.10&0.10&0.10&0.10&0.10&0.10&0.10&0.10\\
$\mu^+$ PID&0.10&0.11&0.11&0.10&0.10&0.10&0.10&0.10&0.10&0.11&0.11&0.11&0.12&0.12&0.12&0.13&0.13&0.15\\
$K_S^0$ reconstruction&0.55&0.17&0.20&0.22&0.24&0.31&0.39&0.46&0.54&0.63&0.75&0.83&0.88&0.95&1.04&1.13&1.18&1.40\\
$\pi$ tracking&0.10&0.10&0.10&0.10&0.10&0.11&0.11&0.12&0.12&0.13&0.13&0.13&0.14&0.14&0.14&0.15&0.16&0.19\\
Quoted BF&0.07&0.07&0.07&0.07&0.07&0.07&0.07&0.07&0.07&0.07&0.07&0.07&0.07&0.07&0.07&0.07&0.07&0.07\\
MC model&0.33&0.37&0.30&0.07&0.40&0.16&0.42&0.42&0.10&0.28&0.38&0.21&0.67&0.50&1.11&0.36&1.11&1.22\\
\hline
Total&0.90&0.74&0.72&0.66&0.78&0.72&0.85&0.89&0.84&0.94&1.06&1.07&1.29&1.26&1.67&1.61&2.12&3.25\\
\hline
\hline
\end{tabular}
	}
\end{table*}

\begin{table*}[htbp]
\caption{ %The systematic covariance matrix for $\kenu$, where $i$ denotes the reconstructed bin, and $j$ represents the produced bin.
The systematic covariance matrix for $\kenu$. The indices $i$ and $j$ denote the $q^2$ intervals.
\label{tab:kenu_sysmatrix}}
\centering
\resizebox{1.0\textwidth}{!}{
\begin{tabular}{c|cccccccccccccccccc}
\hline
\hline
$\rho_{ij}^{\rm syst}$&1&2&3&4&5&6&7&8&9&10&11&12&13&14&15&16&17&18\\ \hline
1&1.000&0.764&0.716&0.757&0.770&0.709&0.754&0.775&0.750&0.569&0.806&0.779&0.776&0.674&0.720&0.558&0.608&0.442\\
2&0.764&1.000&0.776&0.821&0.809&0.798&0.801&0.745&0.819&0.726&0.799&0.803&0.793&0.775&0.692&0.701&0.649&0.567\\
3&0.716&0.776&1.000&0.807&0.801&0.834&0.799&0.673&0.836&0.830&0.746&0.778&0.762&0.825&0.625&0.793&0.650&0.650\\
4&0.757&0.821&0.807&1.000&0.789&0.833&0.812&0.716&0.842&0.796&0.782&0.802&0.788&0.816&0.665&0.764&0.660&0.623\\
5&0.770&0.809&0.801&0.789&1.000&0.770&0.799&0.730&0.817&0.736&0.788&0.795&0.785&0.778&0.679&0.710&0.646&0.575\\
6&0.709&0.798&0.834&0.833&0.770&1.000&0.768&0.669&0.837&0.839&0.741&0.776&0.760&0.829&0.619&0.801&0.650&0.657\\
7&0.754&0.801&0.799&0.812&0.799&0.768&1.000&0.677&0.816&0.746&0.774&0.786&0.774&0.780&0.664&0.718&0.641&0.583\\
\hline
\hline
\end{tabular}
	}
\end{table*}

\begin{table*}[htbp]
\caption{ %The systematic covariance matrix for $\kmunu$, where $i$ denotes the reconstructed bin, and $j$ represents the produced bin.
Same as Table~\ref{tab:kenu_sysmatrix}, but for $\kmunu$.
\label{tab:kmunu_sysmatrix}}
\centering
\resizebox{1.0\textwidth}{!}{
\begin{tabular}{c|cccccccccccccccccc}
\hline
\hline
$\rho_{ij}^{\rm syst}$&1&2&3&4&5&6&7&8&9&10&11&12&13&14&15&16&17&18\\ \hline
1&1.000&0.579&0.542&0.643&0.640&0.587&0.621&0.585&0.557&0.605&0.591&0.546&0.563&0.603&0.440&0.137&0.091&0.062\\
2&0.579&1.000&0.904&0.733&0.833&0.885&0.843&0.876&0.883&0.838&0.862&0.875&0.874&0.725&0.840&0.232&0.187&0.140\\
3&0.542&0.904&1.000&0.630&0.781&0.881&0.805&0.870&0.895&0.806&0.847&0.891&0.880&0.653&0.904&0.240&0.205&0.157\\
4&0.643&0.733&0.630&1.000&0.799&0.721&0.785&0.719&0.672&0.762&0.733&0.657&0.684&0.786&0.494&0.161&0.099&0.065\\
5&0.640&0.833&0.781&0.799&1.000&0.794&0.834&0.809&0.786&0.816&0.811&0.774&0.788&0.779&0.665&0.197&0.142&0.101\\
6&0.587&0.885&0.881&0.721&0.794&1.000&0.801&0.853&0.856&0.816&0.838&0.848&0.848&0.711&0.810&0.224&0.180&0.134\\
7&0.621&0.843&0.805&0.785&0.834&0.801&1.000&0.794&0.803&0.811&0.814&0.791&0.801&0.754&0.704&0.204&0.152&0.110\\
\hline
\hline
\end{tabular}
	}
\end{table*}

\begin{table*}[htbp]
\caption{ %The systematic covariance matrix for $\ksenu$, where $i$ denotes the reconstructed bin, and $j$ represents the produced bin.
Same as Table~\ref{tab:kenu_sysmatrix}, but for $\ksenu$.
\label{tab:ksenu_sysmatrix}}
\centering
\resizebox{1.0\textwidth}{!}{
\begin{tabular}{c|cccccccccccccccccc}
\hline
\hline
$\rho_{ij}^{\rm syst}$&1&2&3&4&5&6&7&8&9&10&11&12&13&14&15&16&17&18\\ \hline
1&1.000&0.749&0.774&0.745&0.739&0.742&0.697&0.706&0.653&0.629&0.616&0.606&0.570&0.558&0.528&0.486&0.445&0.371\\
2&0.749&1.000&0.758&0.804&0.791&0.730&0.777&0.717&0.752&0.737&0.709&0.653&0.544&0.572&0.628&0.619&0.592&0.403\\
3&0.774&0.758&1.000&0.737&0.761&0.731&0.737&0.713&0.707&0.690&0.672&0.641&0.569&0.578&0.591&0.566&0.531&0.401\\
4&0.745&0.804&0.737&1.000&0.748&0.723&0.773&0.715&0.754&0.742&0.717&0.664&0.557&0.586&0.641&0.633&0.605&0.418\\
5&0.739&0.791&0.761&0.748&1.000&0.680&0.770&0.716&0.750&0.739&0.718&0.669&0.568&0.595&0.644&0.632&0.603&0.425\\
6&0.742&0.730&0.731&0.723&0.680&1.000&0.671&0.720&0.687&0.675&0.672&0.665&0.623&0.620&0.602&0.564&0.520&0.438\\
7&0.697&0.777&0.737&0.773&0.770&0.671&1.000&0.693&0.790&0.785&0.769&0.718&0.606&0.644&0.708&0.701&0.670&0.477\\
\hline
\hline
\end{tabular}
	}
\end{table*}

\begin{table*}[htbp]
\caption{ %The systematic covariance matrix for $\ksmunu$, where $i$ denotes the reconstructed bin, and $j$ represents the produced bin.
Same as Table~\ref{tab:kenu_sysmatrix}, but for $\ksmunu$.
\label{tab:ksmunu_sysmatrix}}
\centering
\resizebox{1.0\textwidth}{!}{
\begin{tabular}{c|cccccccccccccccccc}
\hline
\hline
$\rho_{ij}^{\rm syst}$&1&2&3&4&5&6&7&8&9&10&11&12&13&14&15&16&17&18\\ \hline
1&1.000&0.688&0.708&0.661&0.725&0.717&0.762&0.770&0.730&0.762&0.770&0.738&0.754&0.736&0.708&0.586&0.555&0.410\\
2&0.688&1.000&0.727&0.664&0.756&0.692&0.738&0.724&0.618&0.650&0.643&0.578&0.639&0.593&0.615&0.437&0.468&0.335\\
3&0.708&0.727&1.000&0.649&0.744&0.701&0.729&0.718&0.640&0.660&0.650&0.597&0.632&0.597&0.595&0.449&0.455&0.326\\
4&0.661&0.664&0.649&1.000&0.624&0.694&0.650&0.643&0.661&0.634&0.608&0.597&0.545&0.545&0.466&0.434&0.361&0.259\\
5&0.725&0.756&0.744&0.624&1.000&0.663&0.753&0.741&0.637&0.677&0.675&0.612&0.675&0.630&0.652&0.472&0.500&0.361\\
6&0.717&0.692&0.701&0.694&0.663&1.000&0.671&0.703&0.691&0.688&0.674&0.652&0.630&0.621&0.565&0.492&0.440&0.319\\
7&0.762&0.738&0.729&0.650&0.753&0.671&1.000&0.736&0.680&0.722&0.728&0.674&0.728&0.691&0.700&0.531&0.542&0.396\\
\hline
\hline
\end{tabular}
	}
\end{table*}

\begin{sidewaystable}[htbp]
\caption{%The elements of the covariance matrix $\rho_{ij}(i=0,1,2...,35,36; j=0,1,2...,35,36;)$ for the simultaneous fit.
Elements of the covariance matrix $\rho_{ij}$ ($i, j \in [1, 36]$) for the simultaneous fit.
\label{klnumatrix_cov1}}
\centering
\resizebox{1.0\textwidth}{!}{
\begin{tabular}{c|cccccccccccccccccccccccccccccccccccc}
\hline
\hline
$\rho_{ij}$&1&2&3&4&5&6&7&8&9&10&11&12&13&14&15&16&17&18&19&20&21&22&23&24&25&26&27&28&29&30&31&32&33&34&35&36\\ \hline
1&1.000&0.331&0.376&0.369&0.361&0.349&0.341&0.320&0.327&0.303&0.296&0.282&0.264&0.249&0.200&0.199&0.145&0.112&0.038&0.034&0.033&0.031&0.030&0.029&0.027&0.027&0.023&0.022&0.021&0.020&0.019&0.017&0.014&0.012&0.010&0.008\\
2&0.331&1.000&0.335&0.400&0.374&0.388&0.357&0.303&0.352&0.381&0.290&0.287&0.267&0.282&0.190&0.246&0.153&0.142&0.034&0.031&0.030&0.028&0.027&0.027&0.024&0.024&0.021&0.020&0.019&0.018&0.017&0.016&0.013&0.011&0.009&0.007\\
3&0.376&0.335&1.000&0.333&0.393&0.423&0.372&0.286&0.376&0.455&0.283&0.291&0.268&0.314&0.179&0.292&0.160&0.170&0.031&0.028&0.027&0.026&0.024&0.024&0.022&0.022&0.019&0.018&0.017&0.016&0.016&0.014&0.012&0.010&0.008&0.006\\
4&0.369&0.400&0.333&1.000&0.271&0.403&0.353&0.285&0.355&0.409&0.278&0.281&0.260&0.291&0.179&0.263&0.152&0.153&0.032&0.028&0.028&0.026&0.025&0.025&0.022&0.022&0.020&0.019&0.018&0.017&0.016&0.014&0.012&0.010&0.008&0.006\\
5&0.361&0.374&0.393&0.271&1.000&0.263&0.342&0.278&0.331&0.363&0.269&0.268&0.248&0.267&0.175&0.235&0.143&0.136&0.032&0.028&0.028&0.026&0.025&0.025&0.022&0.022&0.020&0.018&0.017&0.017&0.016&0.014&0.012&0.010&0.008&0.006\\
6&0.349&0.388&0.423&0.403&0.263&1.000&0.245&0.275&0.355&0.435&0.265&0.274&0.253&0.299&0.168&0.278&0.151&0.163&0.029&0.026&0.025&0.024&0.023&0.023&0.020&0.020&0.018&0.017&0.016&0.015&0.015&0.013&0.011&0.009&0.008&0.006\\
7&0.341&0.357&0.372&0.353&0.342&0.245&1.000&0.143&0.325&0.353&0.254&0.254&0.236&0.257&0.165&0.229&0.137&0.132&0.030&0.026&0.026&0.024&0.023&0.023&0.021&0.021&0.018&0.017&0.016&0.016&0.015&0.014&0.011&0.009&0.008&0.006\\
\hline
\hline
\end{tabular}

	}
\end{sidewaystable}

\begin{sidewaystable}[htbp]
\caption{%The elements of the covariance matrix $\rho_{ij}(i=0,1,2...,35,36; j=37,38,39...,71,72;)$ for the simultaneous fit.
Elements of the covariance matrix $\rho_{ij}$ ($i \in [1, 36]$ and $j \in [37, 72]$) for the simultaneous fit.
\label{klnumatrix_cov2}}
\centering
\resizebox{1.0\textwidth}{!}{
\begin{tabular}{c|cccccccccccccccccccccccccccccccccccc}
\hline
\hline
$\rho_{ij}$&37&38&39&40&41&42&43&44&45&46&47&48&49&50&51&52&53&54&55&56&57&58&59&60&61&62&63&64&65&66&67&68&69&70&71&72\\ \hline
1&0.244&0.274&0.246&0.289&0.271&0.246&0.259&0.247&0.233&0.240&0.218&0.192&0.174&0.155&0.117&0.071&0.040&0.023&0.009&0.010&0.010&0.010&0.009&0.009&0.009&0.008&0.008&0.007&0.007&0.006&0.005&0.005&0.004&0.004&0.003&0.002\\
2&0.223&0.250&0.224&0.263&0.247&0.224&0.236&0.225&0.213&0.218&0.199&0.175&0.158&0.141&0.106&0.065&0.036&0.021&0.008&0.009&0.009&0.009&0.009&0.008&0.008&0.008&0.007&0.007&0.006&0.006&0.005&0.004&0.004&0.004&0.003&0.002\\
3&0.201&0.225&0.202&0.237&0.223&0.202&0.213&0.203&0.192&0.197&0.179&0.158&0.143&0.127&0.096&0.059&0.033&0.019&0.007&0.008&0.008&0.008&0.008&0.008&0.007&0.007&0.007&0.006&0.006&0.005&0.004&0.004&0.003&0.004&0.003&0.002\\
4&0.205&0.230&0.206&0.242&0.227&0.206&0.217&0.207&0.196&0.201&0.183&0.161&0.146&0.130&0.098&0.060&0.033&0.020&0.008&0.009&0.009&0.008&0.008&0.008&0.007&0.007&0.007&0.006&0.006&0.005&0.004&0.004&0.003&0.004&0.003&0.002\\
5&0.204&0.229&0.206&0.242&0.227&0.206&0.217&0.207&0.195&0.201&0.182&0.161&0.145&0.129&0.098&0.060&0.033&0.020&0.008&0.009&0.009&0.008&0.008&0.008&0.007&0.007&0.007&0.006&0.006&0.005&0.004&0.004&0.003&0.004&0.003&0.002\\
6&0.187&0.210&0.188&0.221&0.208&0.188&0.199&0.189&0.179&0.184&0.167&0.147&0.133&0.119&0.089&0.055&0.030&0.018&0.007&0.008&0.008&0.008&0.007&0.007&0.007&0.006&0.006&0.006&0.005&0.005&0.004&0.004&0.003&0.003&0.002&0.002\\
7&0.191&0.215&0.192&0.226&0.213&0.193&0.203&0.193&0.183&0.188&0.171&0.151&0.136&0.121&0.091&0.056&0.031&0.018&0.007&0.008&0.008&0.008&0.007&0.007&0.007&0.007&0.006&0.006&0.005&0.005&0.004&0.004&0.003&0.003&0.003&0.002\\
\hline
\hline
\end{tabular}
	}
\end{sidewaystable}

\begin{sidewaystable}[htbp]
\caption{%The elements of the covariance matrix $\rho_{ij}(i=37,38,39...,71,72; j=0,1,2...,35,36;)$ for the simultaneous fit.
Elements of the covariance matrix $\rho_{ij}$ ($i \in [37, 72]$ and $j \in [1, 36]$) for the simultaneous fit.
\label{klnumatrix_cov3}}
\centering
\resizebox{1.0\textwidth}{!}{
\begin{tabular}{c|cccccccccccccccccccccccccccccccccccc}
\hline
\hline
$\rho_{ij}$&1&2&3&4&5&6&7&8&9&10&11&12&13&14&15&16&17&18&19&20&21&22&23&24&25&26&27&28&29&30&31&32&33&34&35&36\\ \hline
37&0.244&0.223&0.201&0.205&0.204&0.187&0.191&0.193&0.180&0.144&0.175&0.162&0.153&0.131&0.121&0.096&0.081&0.054&0.009&0.008&0.008&0.007&0.007&0.007&0.006&0.006&0.005&0.005&0.005&0.005&0.004&0.004&0.003&0.003&0.002&0.002\\
38&0.274&0.250&0.225&0.230&0.229&0.210&0.215&0.216&0.201&0.161&0.196&0.182&0.171&0.147&0.135&0.108&0.091&0.060&0.010&0.009&0.008&0.008&0.008&0.008&0.007&0.007&0.006&0.006&0.005&0.005&0.005&0.004&0.004&0.003&0.003&0.002\\
39&0.246&0.224&0.202&0.206&0.206&0.188&0.192&0.194&0.181&0.145&0.176&0.163&0.154&0.132&0.121&0.097&0.082&0.054&0.009&0.008&0.008&0.007&0.007&0.007&0.006&0.006&0.005&0.005&0.005&0.005&0.004&0.004&0.003&0.003&0.002&0.002\\
40&0.289&0.263&0.237&0.242&0.242&0.221&0.226&0.227&0.212&0.170&0.207&0.192&0.180&0.155&0.143&0.114&0.096&0.063&0.010&0.009&0.009&0.008&0.008&0.008&0.007&0.007&0.006&0.006&0.006&0.005&0.005&0.005&0.004&0.003&0.003&0.002\\
41&0.271&0.247&0.223&0.227&0.227&0.208&0.213&0.214&0.199&0.160&0.194&0.180&0.170&0.145&0.134&0.107&0.090&0.060&0.010&0.009&0.008&0.008&0.008&0.007&0.007&0.007&0.006&0.006&0.005&0.005&0.005&0.004&0.004&0.003&0.003&0.002\\
42&0.246&0.224&0.202&0.206&0.206&0.188&0.193&0.194&0.181&0.145&0.176&0.163&0.154&0.132&0.122&0.097&0.082&0.054&0.009&0.008&0.008&0.007&0.007&0.007&0.006&0.006&0.005&0.005&0.005&0.005&0.004&0.004&0.003&0.003&0.002&0.002\\
43&0.259&0.236&0.213&0.217&0.217&0.199&0.203&0.204&0.191&0.153&0.186&0.172&0.162&0.139&0.128&0.102&0.086&0.057&0.009&0.008&0.008&0.008&0.007&0.007&0.006&0.006&0.006&0.005&0.005&0.005&0.005&0.004&0.004&0.003&0.002&0.002\\
\hline
\hline
\end{tabular}
	}
\end{sidewaystable}

\begin{sidewaystable}[htbp]
\caption{%The elements of the covariance matrix $\rho_{ij}(i=37,38,39...,71,72; j=37,38,39...,71,72;)$ for the simultaneous fit.
Elements of the covariance matrix $\rho_{ij}$ ($i \in [37, 72]$ and $j \in [37, 72]$) for the simultaneous fit.
\label{klnumatrix_cov4}}
\centering
\resizebox{1.0\textwidth}{!}{
\begin{tabular}{c|cccccccccccccccccccccccccccccccccccc}
\hline
\hline
$\rho_{ij}$&37&38&39&40&41&42&43&44&45&46&47&48&49&50&51&52&53&54&55&56&57&58&59&60&61&62&63&64&65&66&67&68&69&70&71&72\\ \hline
37&1.000&0.204&0.251&0.222&0.225&0.226&0.220&0.225&0.223&0.206&0.195&0.187&0.164&0.123&0.138&0.073&0.052&0.036&0.012&0.014&0.014&0.014&0.013&0.013&0.012&0.011&0.011&0.010&0.010&0.008&0.007&0.007&0.005&0.006&0.004&0.003\\
38&0.204&1.000&0.402&0.276&0.315&0.367&0.323&0.364&0.381&0.308&0.307&0.323&0.275&0.160&0.283&0.134&0.115&0.087&0.014&0.016&0.016&0.016&0.015&0.014&0.013&0.013&0.013&0.011&0.011&0.009&0.008&0.008&0.006&0.007&0.005&0.003\\
39&0.251&0.402&1.000&0.189&0.324&0.395&0.334&0.391&0.419&0.321&0.327&0.356&0.300&0.156&0.330&0.151&0.137&0.106&0.012&0.014&0.014&0.014&0.013&0.013&0.012&0.012&0.011&0.010&0.010&0.008&0.007&0.007&0.005&0.006&0.004&0.003\\
40&0.222&0.276&0.189&1.000&0.158&0.248&0.243&0.242&0.236&0.227&0.212&0.196&0.174&0.140&0.135&0.076&0.049&0.033&0.015&0.017&0.017&0.016&0.015&0.015&0.014&0.014&0.013&0.012&0.011&0.010&0.009&0.008&0.006&0.007&0.005&0.004\\
41&0.225&0.315&0.324&0.158&1.000&0.184&0.270&0.276&0.280&0.248&0.238&0.236&0.205&0.142&0.185&0.095&0.072&0.052&0.014&0.016&0.016&0.015&0.015&0.014&0.013&0.013&0.012&0.011&0.011&0.009&0.008&0.008&0.006&0.007&0.005&0.003\\
42&0.226&0.367&0.395&0.248&0.184&1.000&0.186&0.325&0.333&0.271&0.269&0.283&0.241&0.142&0.247&0.118&0.100&0.076&0.012&0.014&0.014&0.014&0.013&0.013&0.012&0.012&0.011&0.010&0.010&0.008&0.007&0.007&0.005&0.006&0.004&0.003\\
43&0.220&0.323&0.334&0.243&0.270&0.186&1.000&0.182&0.295&0.247&0.242&0.243&0.210&0.139&0.198&0.099&0.078&0.057&0.013&0.015&0.015&0.015&0.014&0.014&0.013&0.012&0.012&0.011&0.010&0.009&0.008&0.007&0.006&0.006&0.005&0.003\\
\hline
\hline
\end{tabular}
	}
\end{sidewaystable}

\begin{table}
	\caption{The fitted forward/backward DT yields $N_{\rm DT}^{\rm Forward/Backward}$ in individual $q^2$ intervals and the resulting forward-backward asymmetries ($A_{\rm FB}$) for $\kmunu$, where the first uncertainties are statistical and the second are systematic.}
	\scalebox{1.0}{
		\centering
		\begin{tabular}{c|c|c|cc}
			\hline
			\hline
			$q^2({\rm GeV}/c^4)$&$N_{\rm DT}^{\rm Forward}$&$N_{\rm DT}^{\rm Backward}$&$A_{\rm FB}$\\ \hline
			$(0.0,0.1)$&  $4704\pm80$&$19188\pm166$&$-0.225\pm0.011\pm0.003$ \\
			$(0.1,0.2)$&  $12133\pm127$&$21254\pm174$&$-0.098\pm0.007\pm0.002$ \\
			$(0.2,0.3)$&  $13941\pm137$&$20150\pm175$&$-0.073\pm0.008\pm0.002$ \\
			$(0.3,0.4)$&  $14802\pm146$&$19411\pm176$&$-0.051\pm0.008\pm0.002$ \\
			$(0.4,0.5)$&  $15226\pm155$&$18994\pm182$&$-0.049\pm0.008\pm0.002$ \\
			$(0.5,0.6)$&  $15493\pm161$&$18332\pm186$&$-0.039\pm0.009\pm0.002$ \\
			$(0.6,0.7)$&  $15398\pm155$&$17276\pm174$&$-0.031\pm0.009\pm0.002$ \\
			$(0.7,0.8)$&  $14857\pm150$&$15806\pm162$&$-0.020\pm0.009\pm0.002$ \\
			$(0.8,0.9)$&  $13990\pm143$&$14353\pm157$&$-0.018\pm0.009\pm0.002$ \\
			$(0.9,1.0)$&  $12614\pm136$&$12822\pm150$&$-0.020\pm0.010\pm0.002$ \\
			$(1.0,1.1)$&  $10953\pm127$&$11090\pm147$&$-0.015\pm0.011\pm0.002$ \\
			$(1.1,1.2)$&  $9444\pm124$&$9668\pm140$&$-0.016\pm0.012\pm0.002$ \\
			$(1.2,1.3)$&  $7865\pm120$&$7771\pm134$&$0.005\pm0.014\pm0.002$ \\
			$(1.3,1.4)$&  $6449\pm121$&$6425\pm130$&$0.002\pm0.016\pm0.003$ \\
			$(1.4,1.5)$&  $4489\pm100$&$4934\pm126$&$-0.058\pm0.020\pm0.003$ \\
			$(1.5,1.6)$&  $3024\pm72$&$3070\pm67$&$-0.011\pm0.019\pm0.006$ \\
			$(1.6,1.7)$&  $1735\pm60$&$1826\pm57$&$-0.045\pm0.028\pm0.007$ \\
			$>1.7$&  $782\pm54$&$906\pm47$&$-0.088\pm0.050\pm0.010$ \\
			\hline
			\hline
		\end{tabular}
	}
	\label{tab:kmunu_AFB_q2_stat}
\end{table}

\begin{table}
	\caption{%The fitted forward/backward DT yields  $N_{\rm DT}$ in different $q^2$ intervals and the $q^2$-binned forward-backward asymmetries of $\ksenu$, where the first uncertainties are statistical and the seconds are systematic.
	Same as Table~\ref{tab:kmunu_AFB_q2_stat}, but for $\ksenu$.
	}
	\scalebox{1.0}{
		\centering
		\begin{tabular}{c|c|c|cc}
			\hline
			\hline
			$q^2({\rm GeV}/c^4)$&$N_{\rm DT}^{\rm Forward}$&$N_{\rm DT}^{\rm Backward}$&$A_{\rm FB}$\\ \hline
			$(0.0,0.1)$&  $9023\pm99$&$8839\pm103$&$0.003\pm0.009\pm0.001$ \\
			$(0.1,0.2)$&  $8300\pm95$&$8041\pm99$&$0.002\pm0.010\pm0.001$ \\
			$(0.2,0.3)$&  $7581\pm91$&$7347\pm94$&$0.000\pm0.010\pm0.001$ \\
			$(0.3,0.4)$&  $7076\pm89$&$6909\pm90$&$-0.002\pm0.011\pm0.001$ \\
			$(0.4,0.5)$&  $6288\pm85$&$6331\pm85$&$-0.018\pm0.011\pm0.002$ \\
			$(0.5,0.6)$&  $5842\pm82$&$5698\pm80$&$0.006\pm0.012\pm0.002$ \\
			$(0.6,0.7)$&  $5343\pm79$&$5185\pm76$&$0.010\pm0.013\pm0.002$ \\
			$(0.7,0.8)$&  $4808\pm76$&$4825\pm73$&$-0.001\pm0.013\pm0.002$ \\
			$(0.8,0.9)$&  $4134\pm70$&$4245\pm68$&$-0.013\pm0.014\pm0.002$ \\
			$(0.9,1.0)$&  $3749\pm67$&$3735\pm64$&$0.010\pm0.015\pm0.002$ \\
			$(1.0,1.1)$&  $3270\pm62$&$3214\pm59$&$0.023\pm0.016\pm0.002$ \\
			$(1.1,1.2)$&  $2687\pm56$&$2750\pm55$&$-0.005\pm0.018\pm0.002$ \\
			$(1.2,1.3)$&  $2207\pm52$&$2358\pm51$&$-0.028\pm0.019\pm0.003$ \\
			$(1.3,1.4)$&  $1859\pm47$&$1796\pm45$&$0.041\pm0.021\pm0.003$ \\
			$(1.4,1.5)$&  $1310\pm39$&$1337\pm39$&$0.004\pm0.025\pm0.003$ \\
			$(1.5,1.6)$&  $857\pm32$&$948\pm33$&$-0.039\pm0.030\pm0.004$ \\
			$(1.6,1.7)$&  $567\pm26$&$586\pm26$&$0.013\pm0.038\pm0.005$ \\
			$>1.7$&  $243\pm16$&$302\pm19$&$-0.090\pm0.055\pm0.008$ \\
			
			\hline
			\hline
		\end{tabular}
	}
	\label{tab:ksenu_AFB_q2_stat}
\end{table}

\begin{table}
	\caption{%The fitted forward/backward DT yields  $N_{\rm DT}$ in different $q^2$ intervals and the $q^2$-binned forward-backward asymmetries of $\ksmunu$, where the first uncertainties are statistical and the seconds are systematic.
	Same as Table~\ref{tab:kmunu_AFB_q2_stat}, but for $\ksmunu$.
	}
	\scalebox{1.0}{
		\centering
		\begin{tabular}{c|c|c|cc}
			\hline
			\hline
			$q^2({\rm GeV}/c^4)$&$N_{\rm DT}^{\rm Forward}$&$N_{\rm DT}^{\rm Backward}$&$A_{\rm FB}$\\ \hline
			$(0.0,0.1)$&  $2013\pm47$&$6324\pm85$&$-0.195\pm0.015\pm0.002$ \\
			$(0.1,0.2)$&  $4389\pm70$&$6695\pm89$&$-0.096\pm0.011\pm0.002$ \\
			$(0.2,0.3)$&  $4857\pm75$&$6324\pm87$&$-0.074\pm0.012\pm0.002$ \\
			$(0.3,0.4)$&  $5043\pm76$&$5777\pm85$&$-0.034\pm0.012\pm0.002$ \\
			$(0.4,0.5)$&  $5118\pm78$&$5694\pm89$&$-0.037\pm0.013\pm0.002$ \\
			$(0.5,0.6)$&  $4934\pm80$&$5432\pm90$&$-0.048\pm0.013\pm0.002$ \\
			$(0.6,0.7)$&  $4778\pm79$&$5072\pm87$&$-0.037\pm0.014\pm0.002$ \\
			$(0.7,0.8)$&  $4530\pm76$&$4556\pm83$&$-0.026\pm0.014\pm0.002$ \\
			$(0.8,0.9)$&  $4164\pm76$&$4136\pm80$&$-0.021\pm0.015\pm0.002$ \\
			$(0.9,1.0)$&  $3576\pm70$&$3709\pm79$&$-0.045\pm0.017\pm0.002$ \\
			$(1.0,1.1)$&  $3311\pm70$&$3260\pm76$&$-0.007\pm0.018\pm0.002$ \\
			$(1.1,1.2)$&  $2660\pm68$&$2759\pm69$&$-0.028\pm0.021\pm0.002$ \\
			$(1.2,1.3)$&  $2189\pm63$&$2337\pm66$&$-0.047\pm0.023\pm0.003$ \\
			$(1.3,1.4)$&  $1789\pm62$&$1876\pm57$&$-0.023\pm0.027\pm0.003$ \\
			$(1.4,1.5)$&  $1298\pm55$&$1378\pm50$&$-0.035\pm0.033\pm0.003$ \\
			$(1.5,1.6)$&  $1088\pm38$&$1015\pm34$&$0.056\pm0.028\pm0.004$ \\
			$(1.6,1.7)$&  $626\pm30$&$641\pm27$&$-0.016\pm0.038\pm0.005$ \\
			$>1.7$&  $373\pm27$&$375\pm22$&$0.010\pm0.054\pm0.008$ \\
			\hline
			\hline
		\end{tabular}
	}
	\label{tab:ksmunu_AFB_q2_stat}
\end{table}

\end{document}